\documentclass{aa}  

\newcommand{\bzcat}{Roma-BZCAT}

\newcommand{\fer}{{\it Fermi}}

\newcommand{\wse}{{\it WISE}}

\usepackage{graphicx}
\usepackage{longtable}
\usepackage{mdwlist}

\usepackage{natbib}
\usepackage{multirow}
\usepackage{upgreek}
\usepackage[colorlinks=true, pdfstartview=FitV, linkcolor=blue,citecolor=blue, urlcolor=blue]{hyperref}
\usepackage{wasysym}
\usepackage{txfonts}
\usepackage{gensymb}
\usepackage{morefloats}
\usepackage{threeparttable}
\usepackage{rotating}
\usepackage{graphicx}

\begin{document} 

\title{Optical spectroscopic observations of $\gamma$-ray blazar candidates}
\subtitle{II. The 2013 KPNO campaign in the Northern Hemisphere}

\titlerunning{Optical spectroscopic observations of $\gamma$-ray blazar candidates II}
\authorrunning{Massaro et al. 2014}

\author{F. Massaro\inst{1,2}, M. Landoni\inst{3}, R. D'Abrusco\inst{4}, D. Milisavljevic\inst{4}, \\ A. Paggi\inst{4}, N. Masetti\inst{5}, Howard A. Smith\inst{4}, \and G. Tosti\inst{6}}

\institute{Dipartimento di Fisica, Universit\`a degli Studi di Torino, via Pietro Giuria 1, I-10125 Torino, Italy \\
                \email{f.massaro@unito.it}
                \and
                Yale Center for Astronomy and Astrophysics, Physics Department, Yale University, PO Box 208120, New Haven, CT 06520-8120, USA
                \and
                INAF-Osservatorio Astronomico di Brera, Via Emilio Bianchi 46, I-23807 Merate, Italy
                \and
                Harvard-Smithsonian Center for Astrophysics, 60 Garden Street, Cambridge, MA 02138, USA
                \and
                INAF - Istituto di Astrofisica Spaziale e Fisica Cosmica di Bologna, via Gobetti 101, 40129, Bologna, Italy
                \and
                Dipartimento di Fisica, Universit\`a degli Studi di Perugia, 06123, Perugia, Italy
                 }

\date{Received October 6, 2014; accepted  ... }

 \abstract
{We recently started a systematic search of low-energy counterparts of the unidentified $\gamma$-ray sources (UGSs) listed in the \fer-Large Area Telescope (LAT) First Source Catalog (1FGL) and the \fer-LAT 2-Year Source Catalog (2FGL).}
{The main goal of our investigation is to find active galaxies belonging to the blazar class 
that lie within the positional uncertainty region of the UGSs and thus could be their potential low-energy counterparts.}
{To achieve our aims, we first adopted several procedures based on the peculiar observational properties of blazars
in the radio and in the IR. Then we carried out a follow-up spectroscopic campaign in the optical band to verify the nature of the candidates selected
as potential counterparts of the UGSs. Here we present the results of the observations carried out in 2013 in the Northern Hemisphere at 
Kitt Peak National Observatory (KPNO). Optical spectroscopy is crucial to confirm the nature of the sources
and can be  used to estimate their redshifts; it will also allow us to test the robustness of our methods when 
the whole campaign is completed.}
{Here we present the optical spectroscopic observations of 39 sources. 
Within our sample we found that 6 sources are blazars, candidates to be low-energy counterparts of the UGSs listed in the 2FGL. 
We confirm that an additional 8 sources, previously classified 
as active galaxies of uncertain type and associated in the 2FGL, are also all BL Lac objects.
Moreover, we also present 20 new spectra for known blazars listed in the Multi-frequency Catalogue of Blazars
as having an uncertain redshift and/or being classified as BL Lac candidates.}
{We conclude that our methods for selecting $\gamma$-ray blazar candidates allows us to discover
new blazars and increase the list of potential low-energy counterparts for the \fer\ UGSs.}


\maketitle

\section{Introduction}
Within the various types of active galaxies (AGNs), blazars fit as radio-loud sources.
Their spectral energy distribution, dominated by nonthermal emission, extends from radio frequencies to $\gamma$-rays,
where they are the largest known population of extragalactic sources \citep[e.g.,][]{abdo10,nolan12}. 
Blazars also show very rapid variability at all frequencies, high and variable polarization, superluminal motion, 
and high luminosities \citep[e.g.,][]{urry95}, all features that are coupled with peculiar infrared (IR) colors \citep{paper1}.
In 1978, Blandford \& Rees proposed the well-entertained interpretation of blazar emission as arising from
particles accelerated in a relativistic jet closely aligned to the line of sight \citep{blandford78}.

Blazars are historically classified on the basis of their optical spectra.
According to the nomenclature proposed in the \bzcat: 
Multi-frequency Catalogue of Blazars\footnote{{\underline http://www.asdc.asi.it/bzcat/}} 
\citep[e.g.,][]{massaro09,massaro11}, there are BL Lac objects (i.e., BZBs) 
that have optical spectra with emission and/or absorption lines of
rest-frame equivalent width (EW $\leq$ 5$\textrm{\AA}$) \citep[e.g.,][]{stickel91,stoke91,laurent99}
and flat spectrum radio quasars, indicated as BZQs, that show a typical quasar-like optical spectrum with broad emission lines. 
The \bzcat\ also lists a small fraction of sources classified as BL Lac candidates. They were indicated as BZBs
in the literature, but the lack of a published optical spectra did not permit us to verify their classification \citep[see also][]{sdss}.

We recently carried out a systematic search of blazars
among the \fer\ unidentified $\gamma$-ray sources (UGSs) since they might be their potential low-energy counterparts.
To achieve this goal, we developed several procedures based on the blazar properties, such as
1) the use of their peculiar IR colors \citep{ugs2,ugs3}, discovered thanks to the all-sky survey of the Wide-Field Infrared Survey Explorer 
\citep[\wse;][]{wright10}, 2) the flat shape of their radio spectra maintained even at MHz frequencies  \citep{paper3,ugs6},
or 3) combining radio observations with the IR colors \citep{paper5}.
In addition, multifrequency analyses based on X-ray follow-up observations 
\citep[e.g.,][]{mirabal09,paggi13,takeuchi13,stroh13,acero13}
and radio campaigns \citep[e.g.,][]{petrov13} were also used to complete our search.
However, all the proposed methods, even if coupled with X-ray observations, cannot guarantee that the selected candidates are indeed blazars
and that the selection procedures are not contaminated by 
different source classes with blazar-like properties \citep[e.g.,][]{stern13}.
Removing any possible degeneracy in the selection methods requires a certain source classification that can only beachieved by optical spectroscopy
\citep[e.g.,][]{masetti13,shaw13a,shaw13b,paggi14,sdss}.
Moreover, several optical spectroscopic campaigns have recently
been carried out to obtain redshift estimates for a large sample of BZBs using medium-resolution spectroscopy;
some of them are now known as $\gamma$-ray emitters associated with \fer\ sources
\citep[e.g.,][]{sbarufatti06a,sbarufatti06b,sbarufatti09,landoni13,shaw13a,shaw13b}.

In this paper we report the results of the 2013 observations carried out at Kitt Peak National Observatory (KPNO)
with the Mayall 4m class telescope in the Northern Hemisphere. Preliminary results for our exploratory program in the Northern Hemisphere 
obtained with the Loiano telescope, the Multiple Mirror Telescope (MMT), and the Observatorio Astron\'omico Nacional (OAN) in San Pedro M\'artir (M\'exico) 
were already presented in Paggi et al. (2014).
In addition, the results of the complementary observations carried out in the Southern Hemisphere in 2013  
are reported by Landoni et al. (2014).
Our final goal is to extend this optical spectroscopic campaign to complete the observations 
for all the blazar-like sources selected according to our procedures as well as to those with uncertain nature reported in the \fer\ catalogs.

This paper is organized as follows: 
in Sect.~\ref{sec:sample} we describe our sample selection.
Details on the cross-matches with multifrequency databases and catalogs for the observed targets
are reported in Sect.~\ref{sec:crossmatches}.
Then, in Sect.~\ref{sec:obs} we present our dataset 
and discuss the data reduction procedures, while results of our analysis for different types of sources are presented in Sect.~\ref{sec:results}.
Finally, Sect.~\ref{sec:conclusions} is devoted to our summary and conclusions.
We use cgs units unless stated otherwise. Spectral indices, $\alpha$, 
are defined by flux density, S$_{\nu}\propto\nu^{-\alpha}$.

\section{Sample selection}
\label{sec:sample}
Our sample lists a total of 39 sources divided into the following five subsamples:
Seven sources are potential counterparts of UGSs selected according to our method based on the IR colors
and/or using their spectral shape at low radio frequencies \citep{ugs3,ugs6,lorcat}.
Then eight \fer\ sources are classified as active galaxies of uncertain type (AGUs) in the Second Catalog of Active Galactic Nuclei Detected by the Fermi Large Area Telescope \citep[2LAC;][]{ackermann11a}.
According to the 2LAC, these AGUs are radio sources with a flat radio spectrum and/or an X-ray counterpart that appear to have a multifrequency behavior
similar to active galaxies and in some cases, more specifically to blazars; the lack of their optical spectra did not allow a precise classification, however.
Moreover, as shown in Massaro et al. (2012a), a large portion of the AGUs appear to have IR colors similar to other \fer\ blazars, and 
thus they were observed as part of our spectroscopic campaign.
The 2LAC catalog lists 157 AGUs in the CLEAN sample, one third
of which are visible from the Northern Hemisphere. We selected our AGU targets considering those that 
were visible during semester 2013A from the KPNO site, at an
airmass lower than 1.30 and with an R-band magnitude of between 15 and 20, suitable for a 4m class telescope,
and with IR colors similar to those of known \fer\ blazars. Our future plan is to complete the whole sample of AGUs to verify their nature.

In addition to these eight AGUs, there were also four sources classified as BZBs in the 2LAC that are not listed in the \bzcat.
Because no optical spectra were available in the literature when the the \fer\ catalogs were prepared, 
we added these four targets to our sample to confirm their classification. This also allows us to consider them for future releases of the \bzcat.
 
The remaining 20 sources belong to the \bzcat\ version 4.1 \citep{massaro11}.
In particular, 14 of them are BL Lac candidates, 9 are also detected by \fer; for these sources 
no optical spectroscopic information was found in literature \citep{massaro11}, and, as already performed for others targets, we filled the gaps in our schedule to
obtain their correct classification as well. The remaining 6 targets were selected from the 82 sources listed in the \bzcat\ that have an uncertain redshift estimate.
There are 270 BL Lac candidates in the \bzcat\ version 4.1, 159 in the Northern Hemisphere, at declinations higher than zero, and in the sample of 950 BZBs, 82 
have an uncertain redshift estimate (76 in the northern sky and only 6 in the    Southern Hemisphere);
among these, we selected sources adopting the same criteria used for the AGUs: visibility, R magnitude range, and airmass constraints.
We included in our target list known BZBs that were not yet detected by \fer\ since they can be discovered as $\gamma$-ray emitters 
when the \fer\ all-sky survey is completed; thus the information reported here will be used for future releases of the \fer\ catalogs.
  
All the selected sources are listed in Table~\ref{tab:log}. 
For the first two subsamples described above we also performed a multifrequency analysis to verify possible 
additional information that might support the blazar-like behavior.
Sources that belong to the \bzcat\ and have been observed during our campaign 
are reported in Table~\ref{tab:log} with their proper \bzcat\ name.

We highlight that some of the sources observed during our campaign have also been observed at different observatories by other groups,
as reported in the following sections. However, we reobserved these targets for two main reasons: 1) at the time when our observations were performed,
these spectra were not yet published, and 2) as a result of the well-known variability of BL Lacs in the optical energy range, there is always the chance to
observe the source during a quiescent or low state and thus detect emission and/or absorption features that enable a
redshift measurement.

\section{Correlation with existing databases}
\label{sec:crossmatches}
We searched the following major radio, infrared, optical, and X-ray surveys 
and both the NASA Extragalactic Database (NED)\footnote{\underline{http://ned.ipac.caltech.edu/}} 
and the SIMBAD Astronomical Database\footnote{\underline{http://simbad.u-strasbg.fr/simbad/}}
to verify whether multifrequency information can help to confirm  the natures of uncertain counterparts and blazar candidates.

In Table~\ref{tab:log} we summarize the multifrequency notes for each source
with the exception of those already listed in the \bzcat\ to avoid duplicating information.
In the multifrequency notes in Table~\ref{tab:log} we also indicate whether the spectral energy distribution (SED) of the source
is shown in Takeuchi et al. (2013) and if the radio counterpart has a flat radio spectrum 
(marked as ``rf'' in the notes of Table~\ref{tab:log} whenever $\alpha<$0.5 in the radio band).

The surveys and catalogs used to search of the counterparts of our targets are
the VLA Low-Frequency Sky Survey Discrete Source Catalog \citep[VLSS;][- V]{cohen07} 
and the recent revision VLSSr\footnote{\underline{http://heasarc.gsfc.nasa.gov/W3Browse/all/vlssr.html}}
\citep{lane14}, the two Westerbork Northern Sky Survey \citep[WENSS;][- W]{rengelink97},
the Texas Survey of Radio Sources at 365 MHz \citep[TEXAS;][- T]{douglas96},
and the Low-frequency Radio Catalog of Flat-spectrum Sources \citep[LORCAT;][- L]{lorcat}.
For the radio counterparts at higher frequency, above $\sim$1GHz, we searched the NRAO VLA Sky Survey \citep[NVSS;][- N]{condon98}, 
the VLA Faint Images of the Radio Sky at Twenty-Centimeters \citep[FIRST;][- F]{becker95,white97},
the 87 Green Bank catalog of radio sources \citep[87GB;][- 87]{gregory91}, and the 
Green Bank 6-cm (GB6) Radio Source Catalog \citep[GB6;][- GB]{gregory96}.

In the infrared, we queried the \wse\ all-sky survey in the  
AllWISE source catalog\footnote{\underline{http://wise2.ipac.caltech.edu/docs/release/allwise/}} \citep[][- w]{wright10}
and the Two Micron All Sky Survey \citep[2MASS;][- M]{skrutskie06}
since each \wse\ source is automatically matched to the closest 2MASS potential counterpart \citep[see][for details]{cutri12}.
Then, we also searched for optical counterparts
in the Sloan Digital Sky Survey Data Release 9 \citep[SDSS DR9; e.g.][- s]{ahn12}, 
and in the USNO-B1 Catalog \citep[][- U]{monet03}.
In the ultraviolet energy range, 
we used the GALaxy Evolution eXplorer All-Sky Survey Source Catalog\footnote{http://galex.stsci.edu/GR6/} \citep[][- g]{seibert12}.
At high-energies, in the X-rays, we searched the ROSAT all-sky survey 
in both the ROSAT Bright Source Catalog \citep[RBSC;][- X]{voges99} 
and the ROSAT Faint Source Catalog \citep[RFSC;][- X]{voges00}.

To perform the search in all these surveys and catalogs, we considered
the 1$\sigma$ positional uncertainty, with the two exceptions of the \wse\ all-sky survey
and the SDSS DR9. In these two cases we searched the closest IR and optical counterpart 
within a maximum angular separation of 3\arcsec.3 and of 1\arcsec.8
for the AllWISE survey and the SDSS DR9, respectively.
These values were derived on the basis of the statistical analysis 
described in D'Abrusco et al. (2013) and Massaro et al. (2014a), which was 
developed following the approach presented in Maselli et al. (2010) and Stephen et al. (2010).

\begin{figure}[]
\begin{center}
\includegraphics[height=8.4cm,width=8.4cm,angle=0]{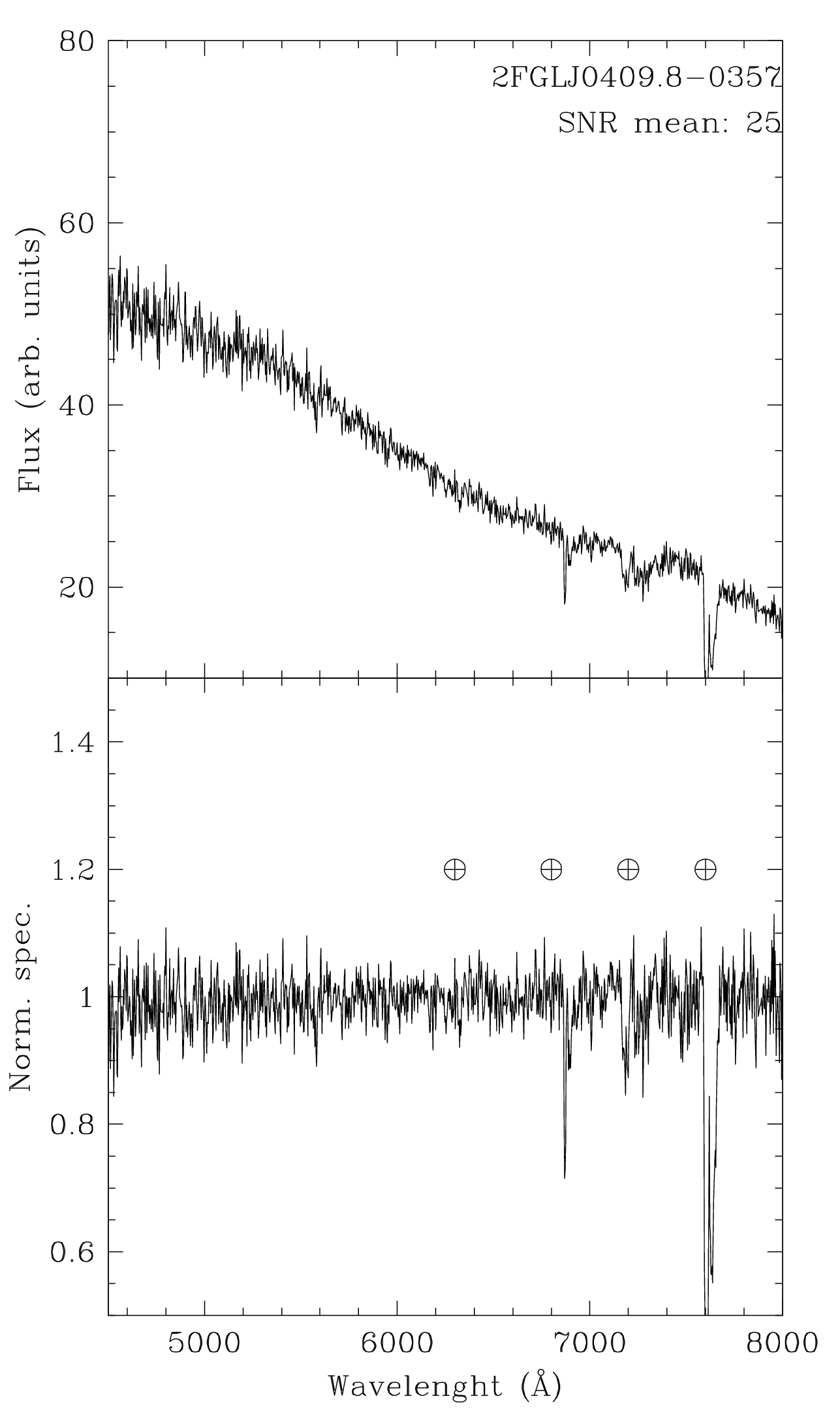}
\includegraphics[height=5.6cm,width=5.6cm,angle=0]{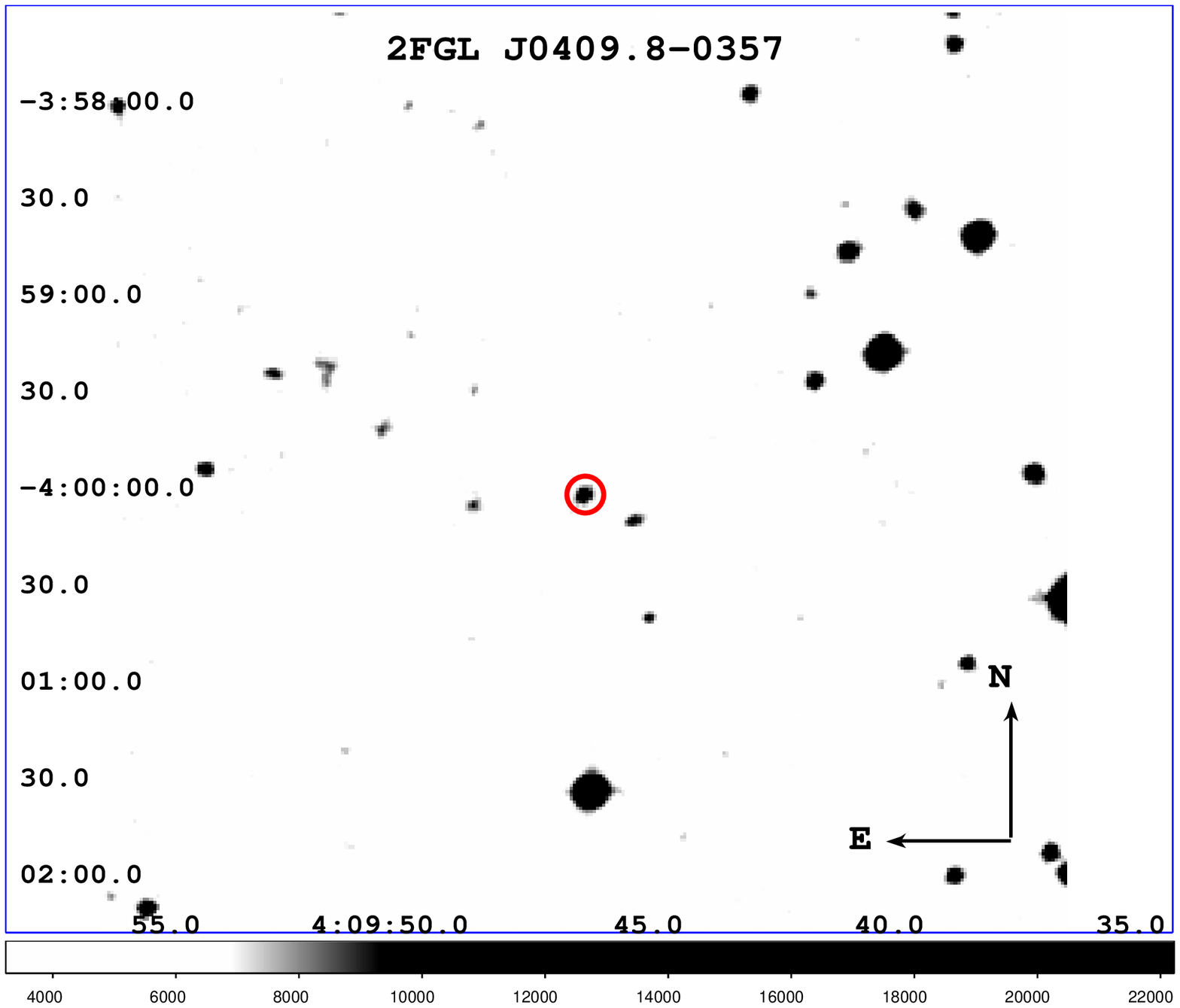}
\end{center}
\caption{Upper panel: optical spectra of WISE J040946.58-040003.5, potential counterpart of 
2FGL J0409.8-0357, classified as a BZB on the basis of its featureless continuum.
The average S/N.
Middle panel: normalized spectrum is shown here.
Lower panel: 5\arcmin\,x\,5\arcmin\ finding chart from the Digital Sky Survey (red filter). 
The potential counterpart of  2FGL J0409.8-0357
 is indicated by the red circle.}
\label{fig:J0409.8}
\end{figure}

\begin{figure}[]
\begin{center}
\includegraphics[height=12.2cm,width=12.2cm,angle=0]{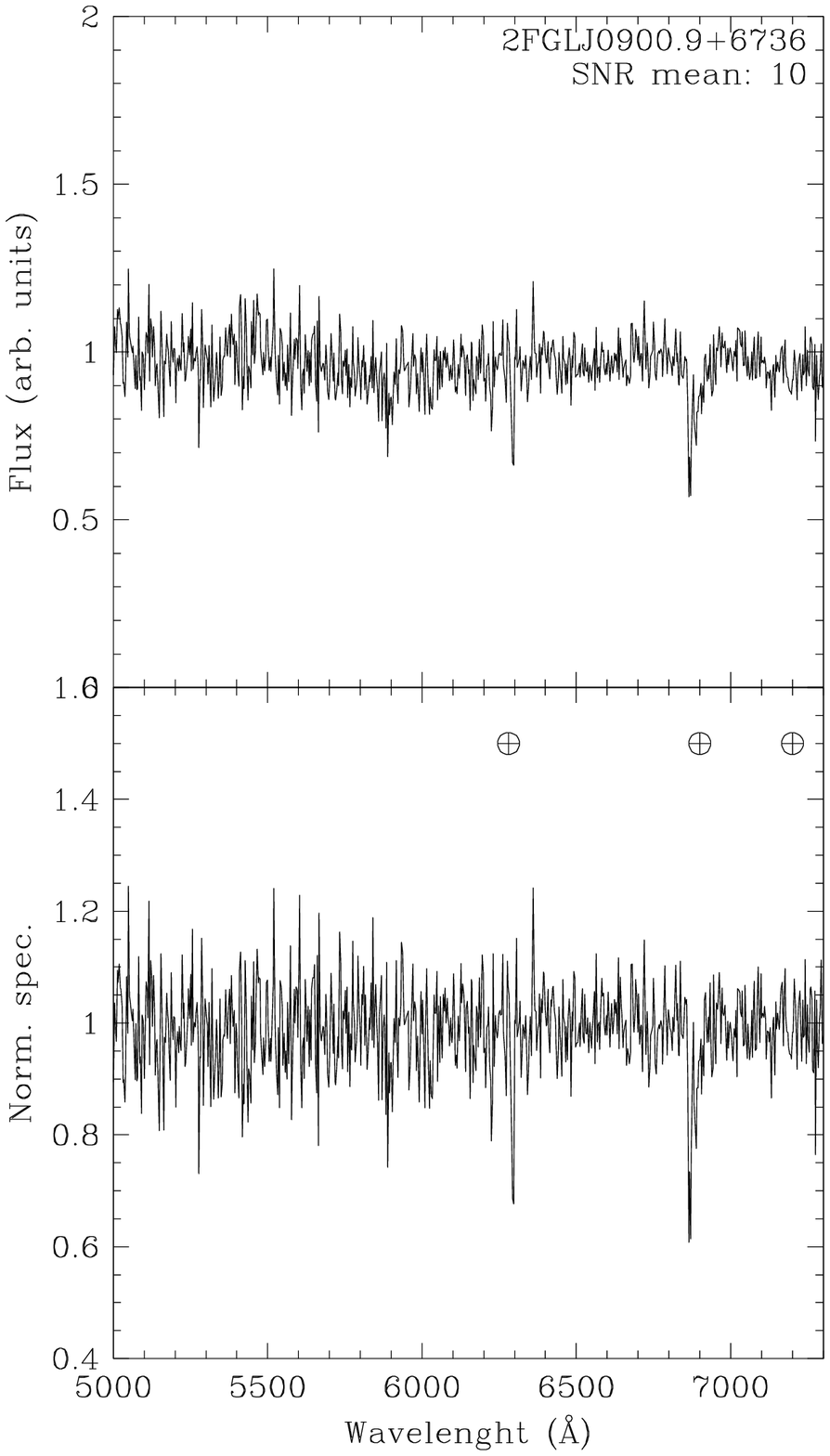}
\includegraphics[height=5.6cm,width=5.6cm,angle=0]{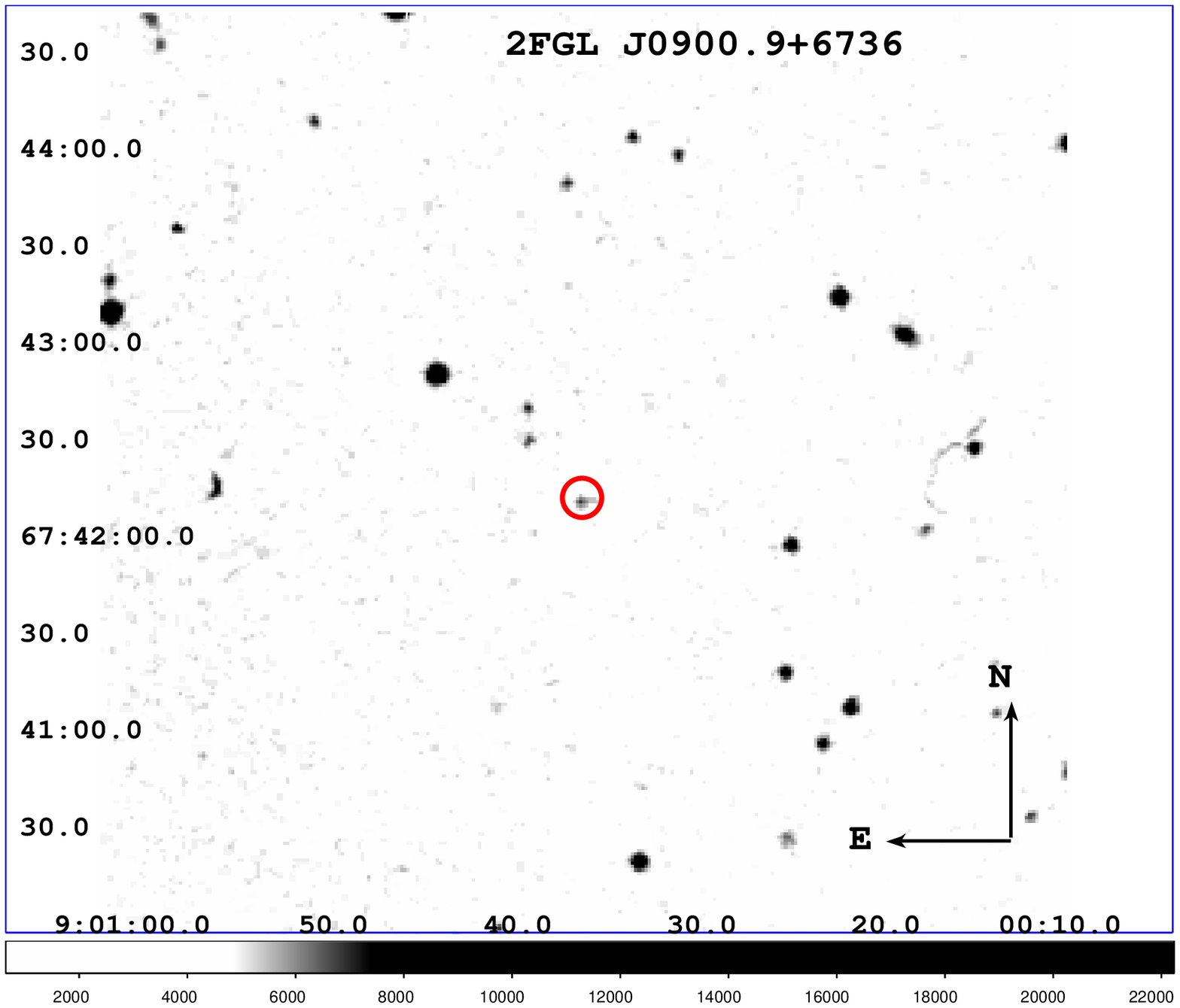}
\end{center}
\caption{Upper panel: optical spectra of WISE J090038.69+674223.3, potential counterpart of 
2FGL J0900.9+6736, classified as a BZB on the basis of its featureless continuum.
The average S/N is also indicated.
Middle panel: normalized spectrum is shown here.
Lower panel: 5\arcmin\,x\,5\arcmin\ finding chart from the Digital Sky Survey (red filter). 
The potential counterpart of  2FGL J0900.9+6736
 is indicated by the red circle.}
\label{fig:J0900.9}
\end{figure}

\begin{figure}[]
\begin{center}
\includegraphics[height=12.2cm,width=12.2cm,angle=0]{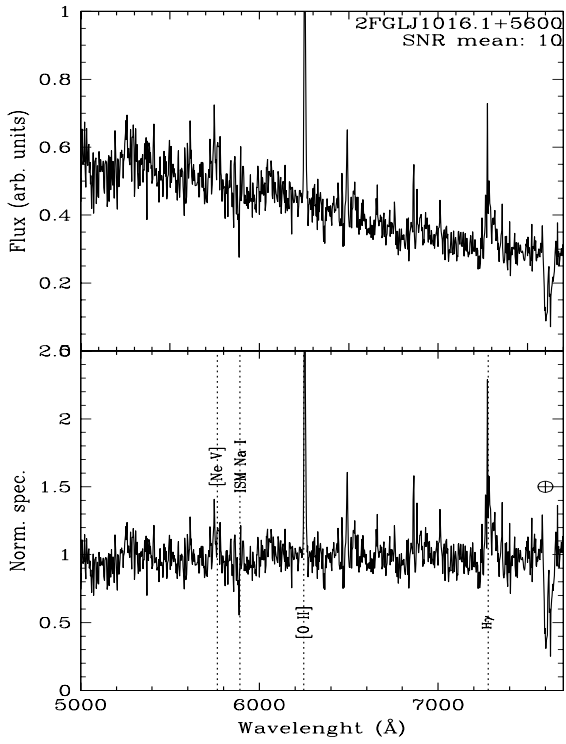}
\includegraphics[height=5.6cm,width=5.6cm,angle=0]{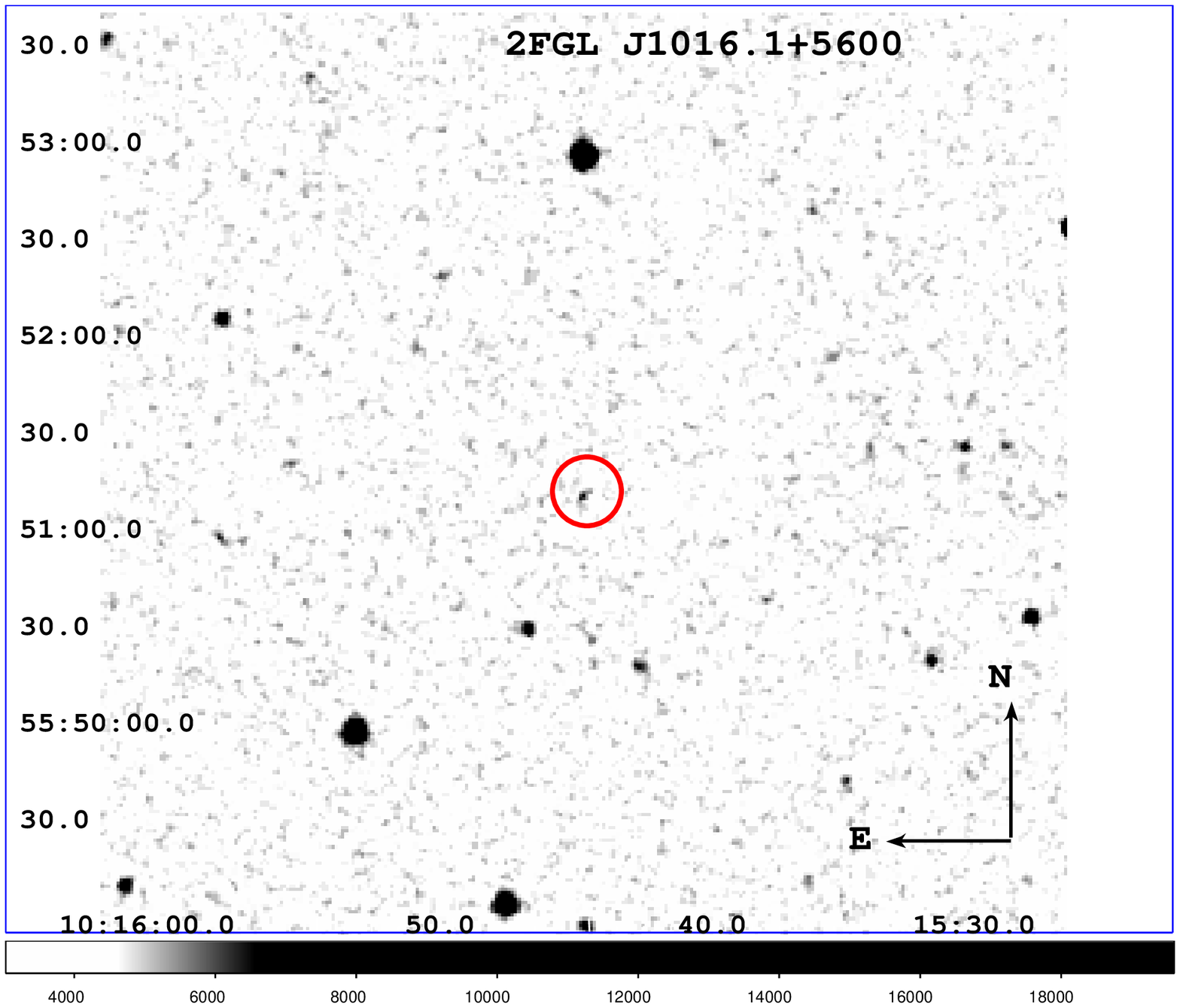}
\end{center}
\caption{Upper panel: optical spectra of WISE J101544.43+555100.6, potential counterpart of
2FGL J1016.1+5600, classified as a QSO at $z=$0.677 on the basis of the emission lines marked in the plot.
The average S/N is also indicated.
Middle panel: normalized spectrum.
Lower panel: 5\arcmin\,x\,5\arcmin\ finding chart from the Digital Sky Survey (red filter). 
The potential counterpart of  2FGL J1016.1+5600
 is indicated by the red circle.}
\label{fig:J1016.1}
\end{figure}

\begin{figure}[]
\begin{center}
\includegraphics[height=12.2cm,width=12.2cm,angle=0]{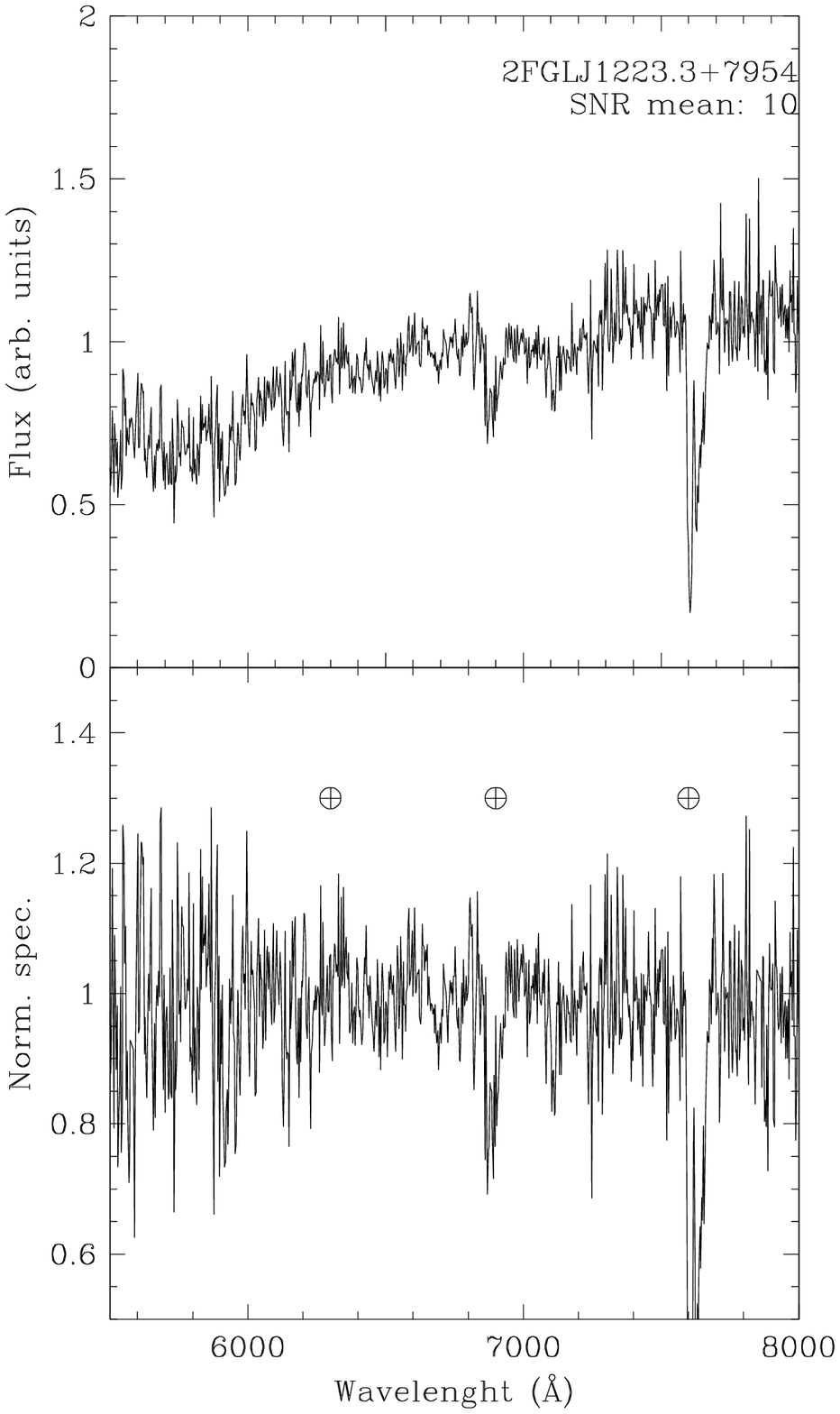}
\includegraphics[height=5.6cm,width=5.6cm,angle=0]{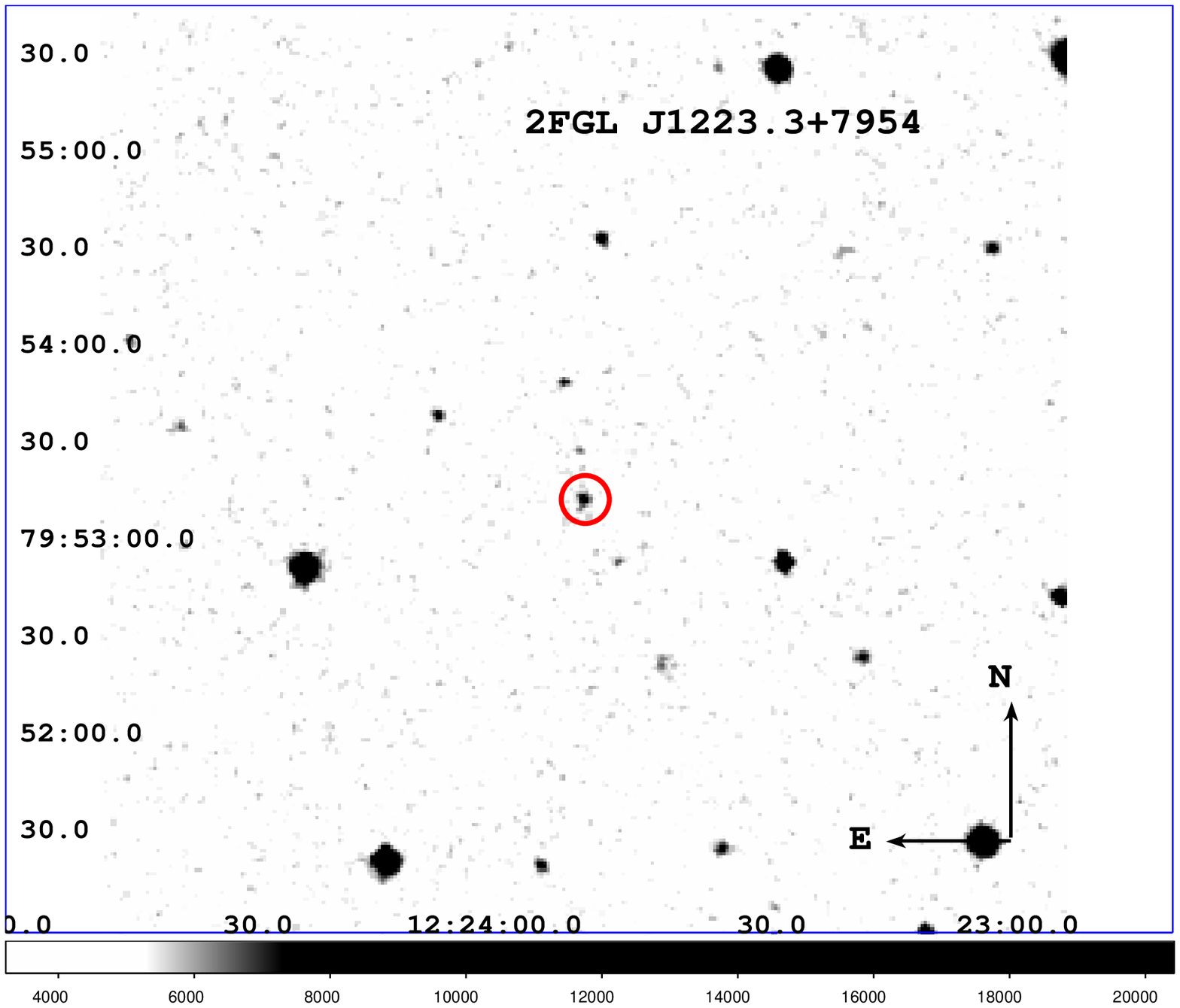}
\end{center}
\caption{Upper panel: optical spectra of WISE J122358.06+795328.2, potential counterpart of 
2FGL J1223.3+7954, classified as a BZB on the basis of its featureless continuum.
The average S/N is also indicated.
The depression blueward at $\sim$6000$\textrm{\AA}$ might be produced by the Balmer decrement. 
However, the rather poor S/N  has prevented us from detecting
the CaII H+K lines associated with the flux depression.
Middle panel: normalized spectrum.
Lower panel: 5\arcmin\,x\,5\arcmin\ finding chart from the Digital Sky Survey (red filter). 
The potential counterpart of  2FGL J1223.3+7954
 is indicated by the red circle.}
\label{fig:J1223.3}
\end{figure}

\begin{figure}[]
\begin{center}
\includegraphics[height=12.2cm,width=12.2cm,angle=0]{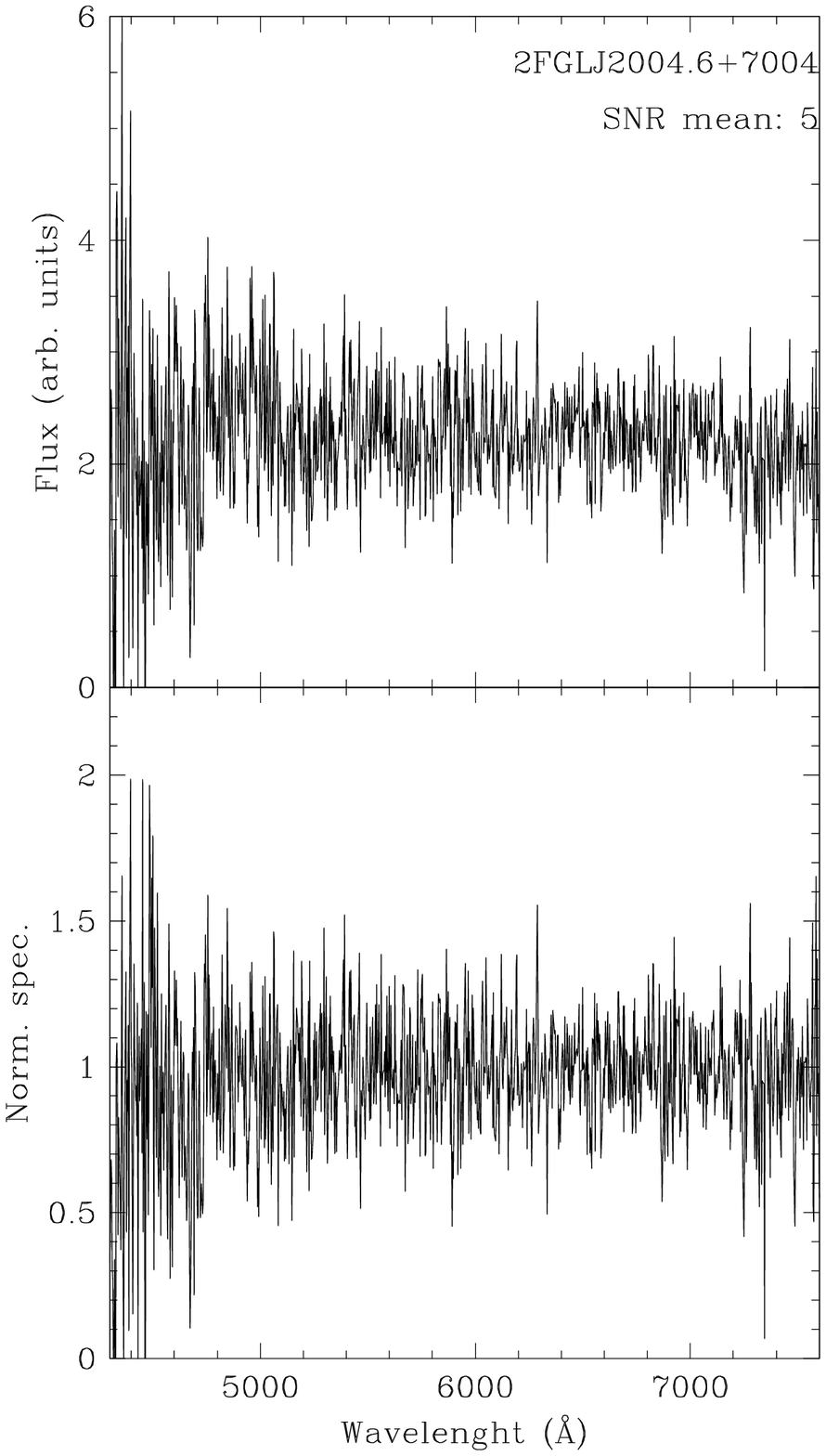}
\includegraphics[height=5.6cm,width=5.6cm,angle=0]{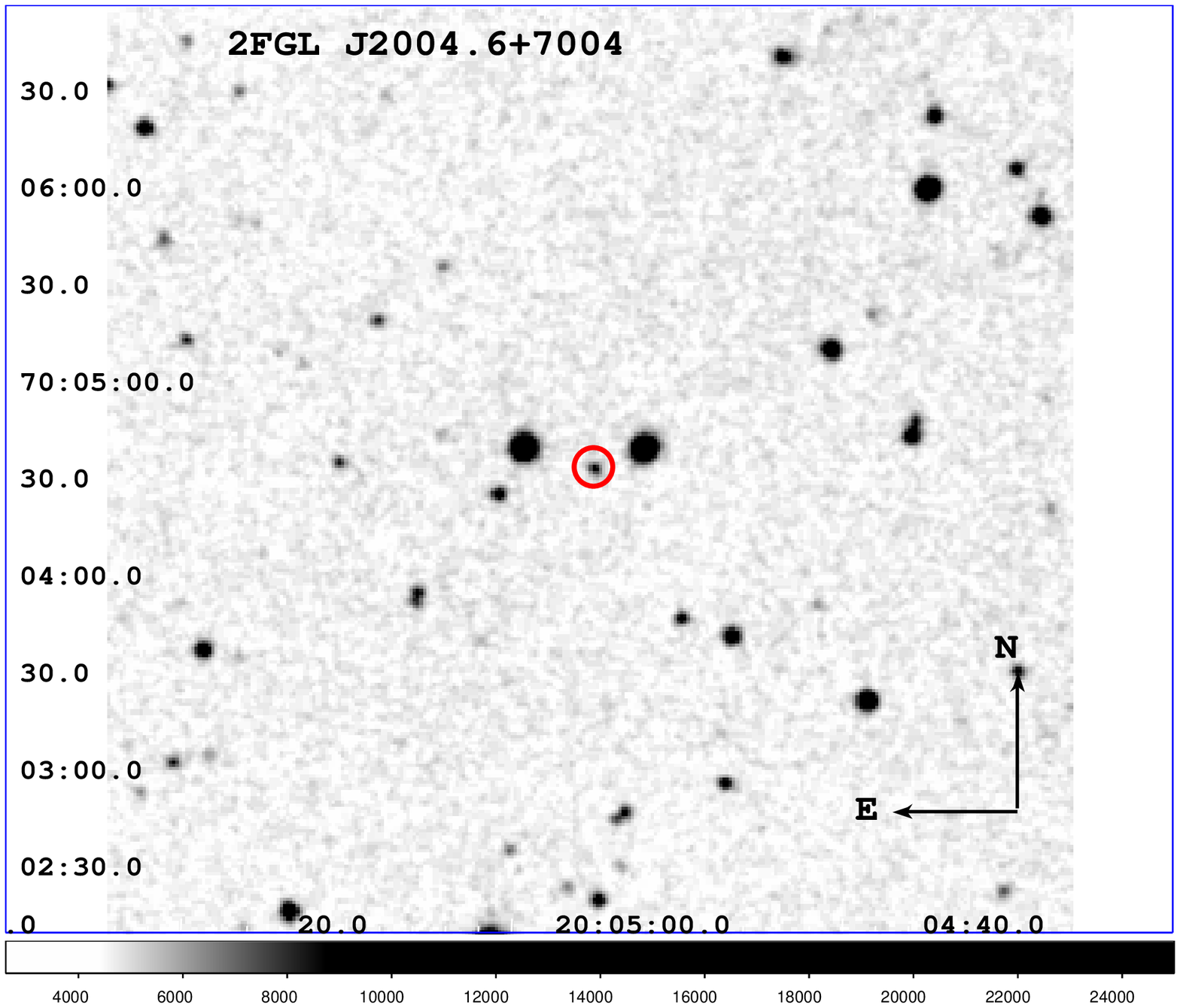}
\end{center}
\caption{Upper panel: optical spectra of WISE J200505.97+700439.5, potential counterpart of
2FGL J2004.6+7004, classified as a BZB on the basis of its featureless continuum.
The average S/N is also indicated.
Middle panel: normalized spectrum.
Lower panel: 5\arcmin\,x\,5\arcmin\ finding chart from the Digital Sky Survey (red filter). 
The potential counterpart of  2FGL J2004.6+7004
 is indicated by the red circle.}
\label{fig:J2004.6}
\end{figure}

\begin{figure}[]
\begin{center}
\includegraphics[height=12.2cm,width=12.2cm,angle=0]{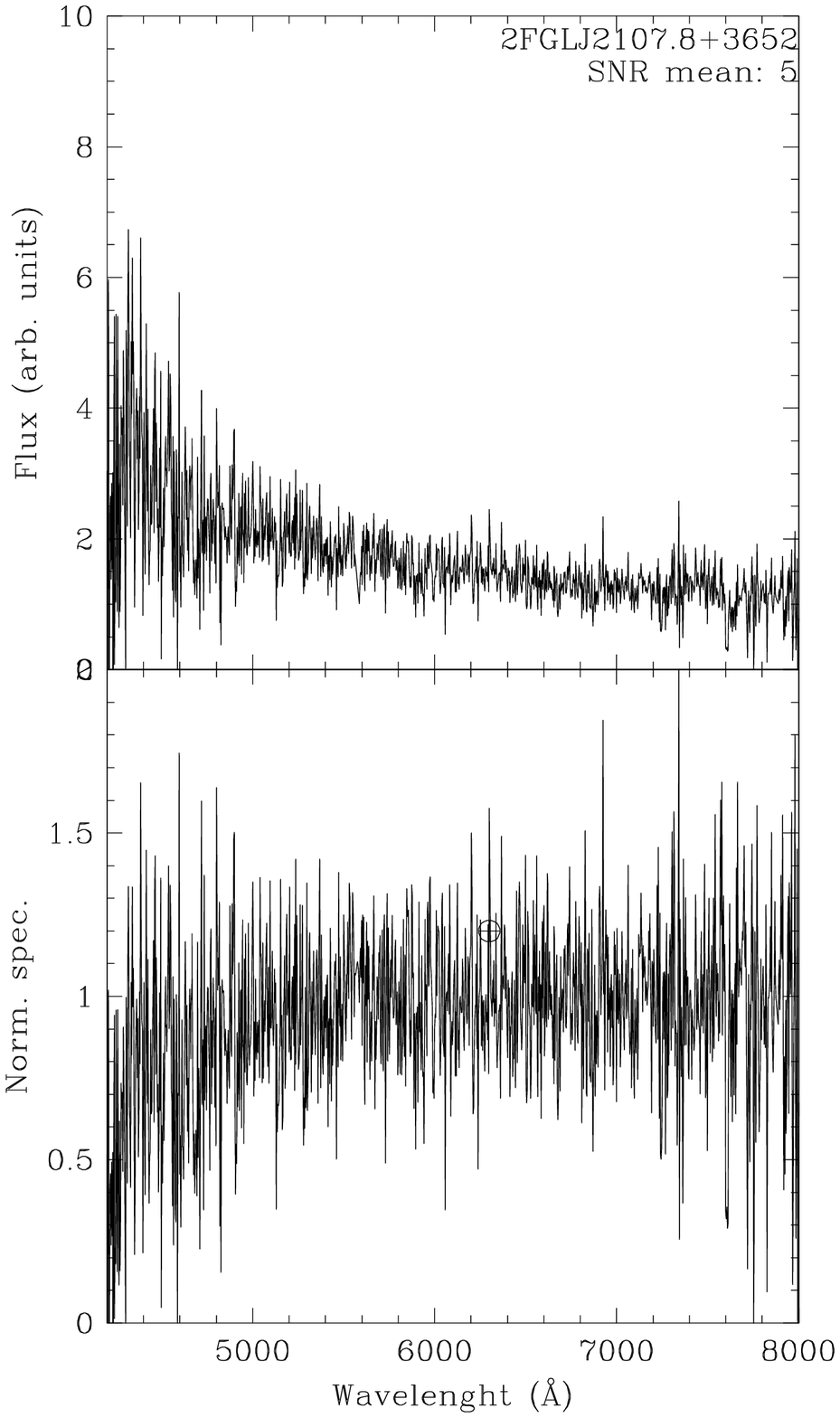}
\includegraphics[height=5.6cm,width=5.6cm,angle=0]{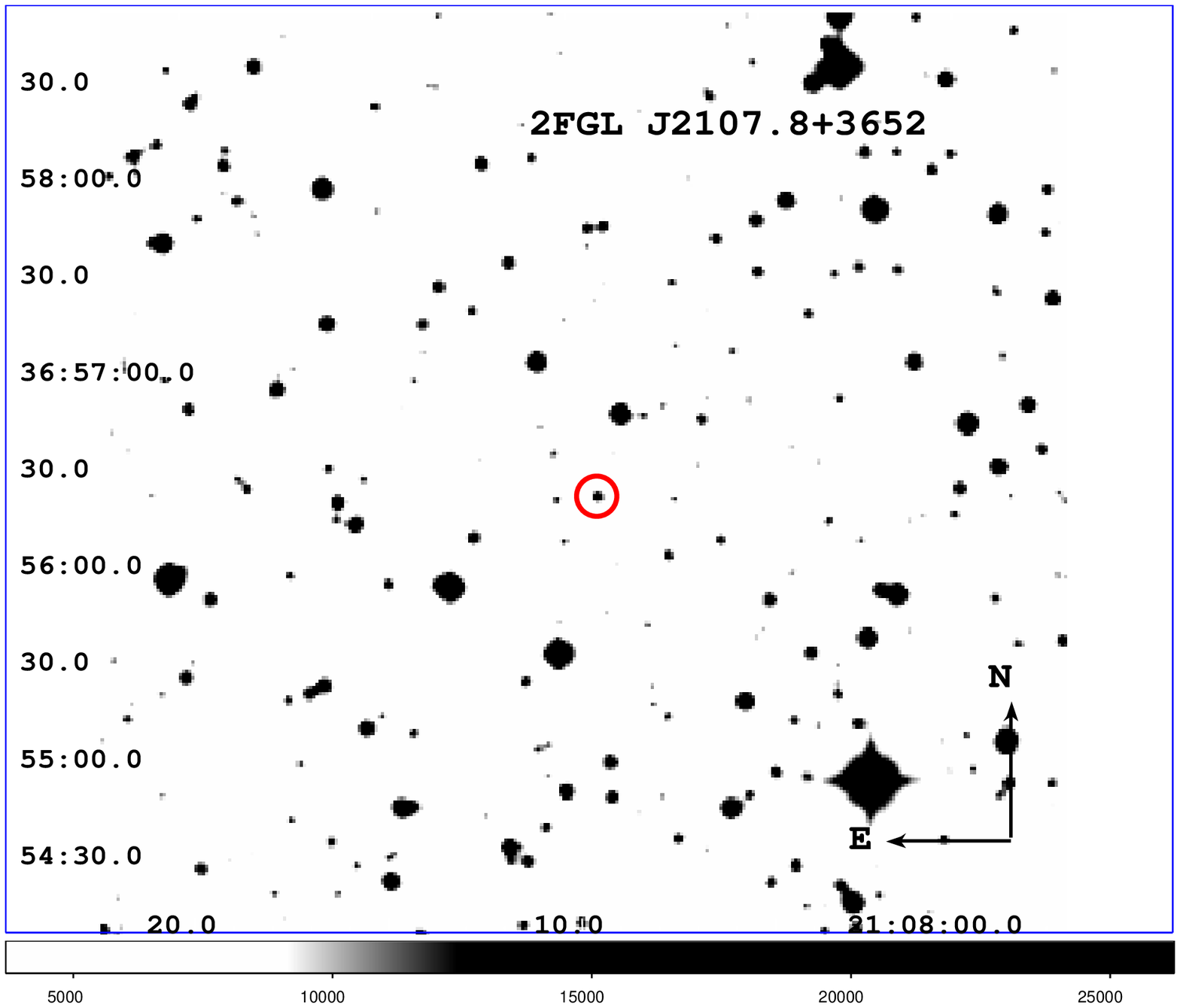}
\end{center}
\caption{Upper panel: optical spectra of WISE J210805.46+365526.5, potential counterpart of
2FGL J2107.8+3652, classified as a BZB on the basis of its featureless continuum.
The average S/N is also indicated.
Middle panel: normalized spectrum.
Lower panel: 5\arcmin\,x\,5\arcmin\ finding chart from the Digital Sky Survey (red filter). 
The potential counterpart of  2FGL J2107.8+3652 is indicated by the red circle.}
\label{fig:J2107.8}
\end{figure}

\begin{figure}[]
\begin{center}
\includegraphics[height=12.2cm,width=12.2cm,angle=0]{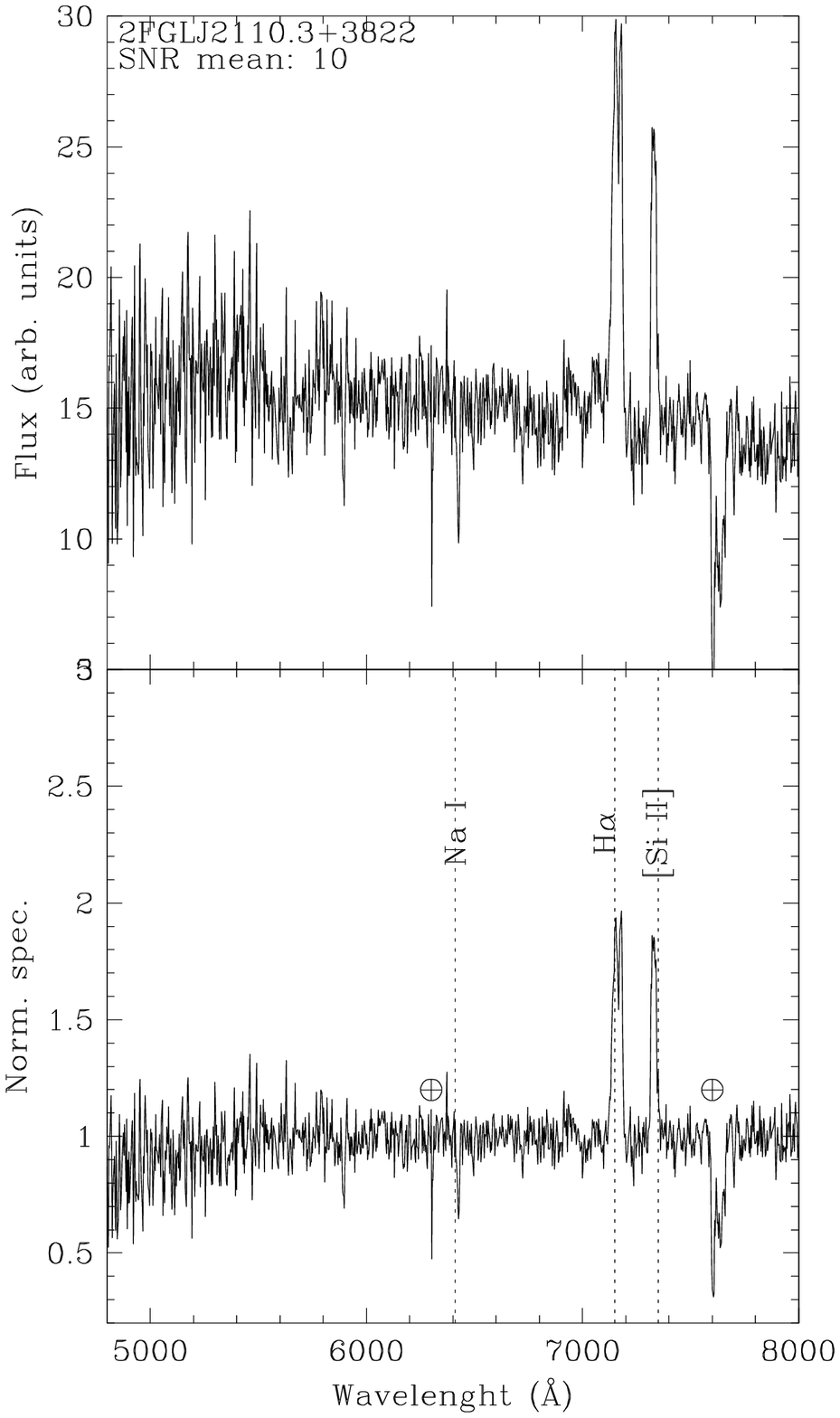}
\includegraphics[height=5.6cm,width=5.6cm,angle=0]{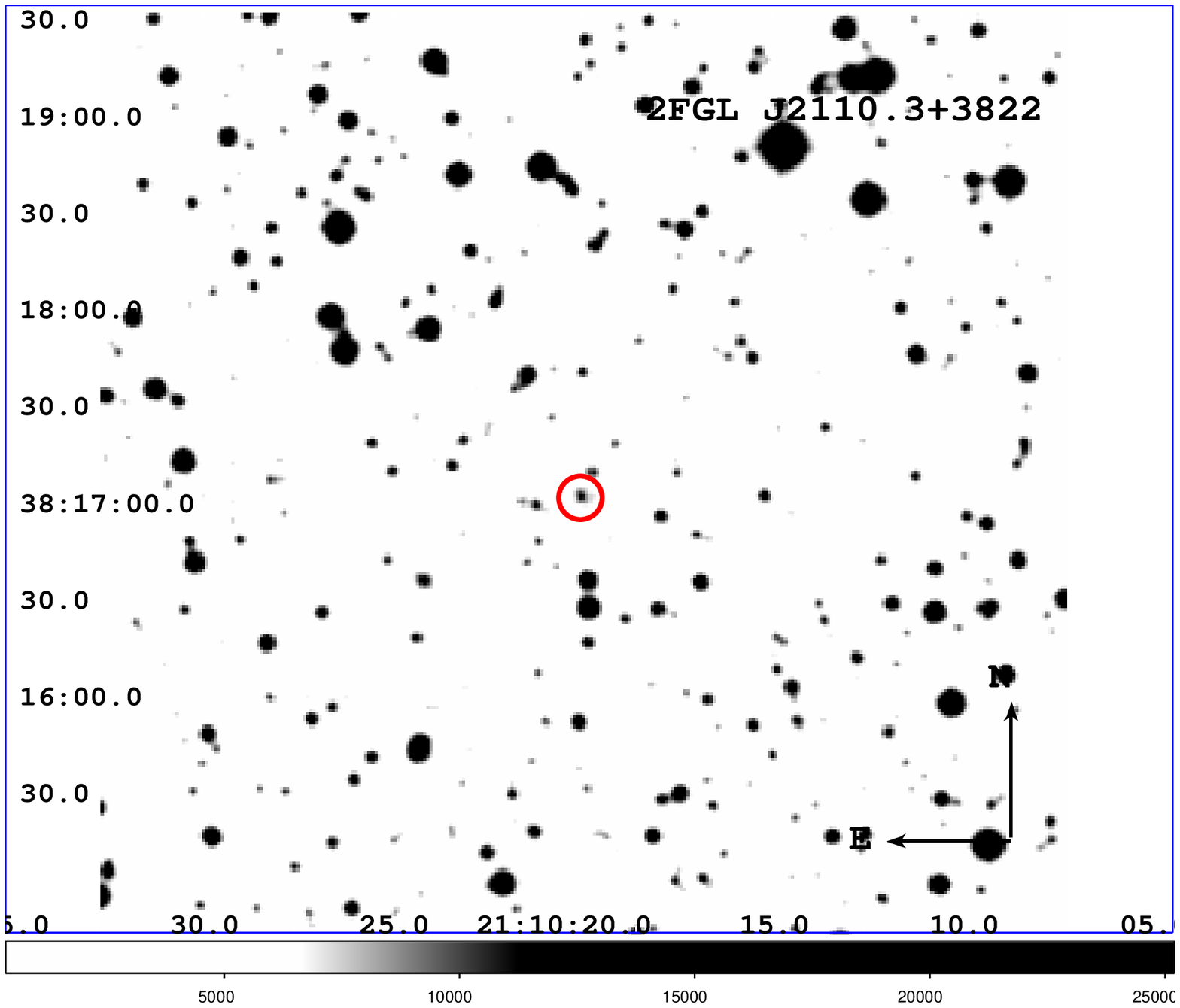}
\end{center}
\caption{Upper panel: optical spectra of WISE J211020.19+381659.2, potential counterpart of
2FGL J2110.3+3822, classified as a QSO at $z=$0.46 on the basis of the emission lines marked in the plot.
The average S/N is also indicated.
Middle panel: normalized spectrum.
Lower panel: 5\arcmin\,x\,5\arcmin\ finding chart from the Digital Sky Survey (red filter). 
The potential counterpart of  2FGL J2110.3+3822
 is indicated by the red circle.}
\label{fig:J2110.3}
\end{figure}

\begin{figure}[]
\begin{center}
\includegraphics[height=12.2cm,width=12.2cm,angle=0]{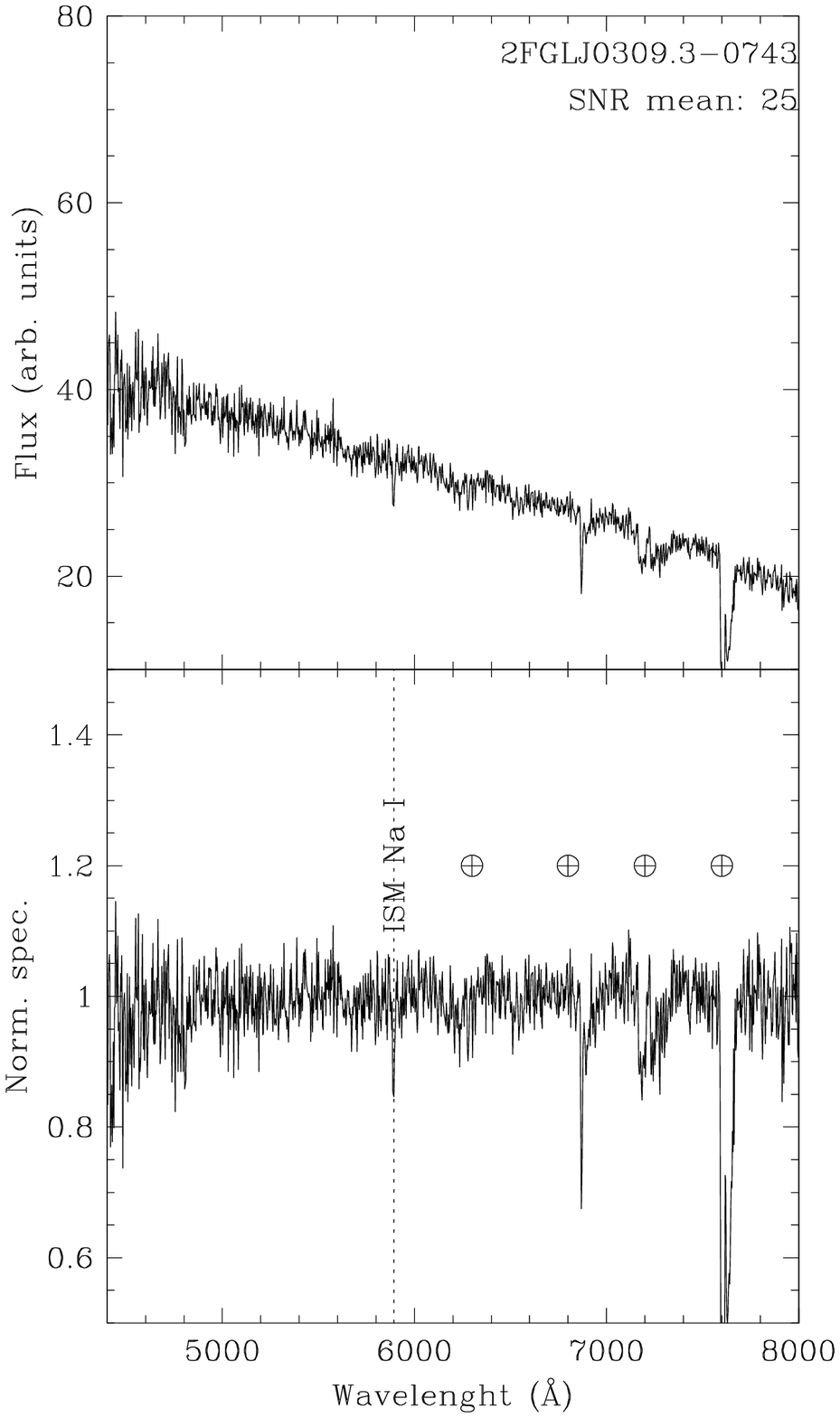}
\includegraphics[height=5.6cm,width=5.6cm,angle=0]{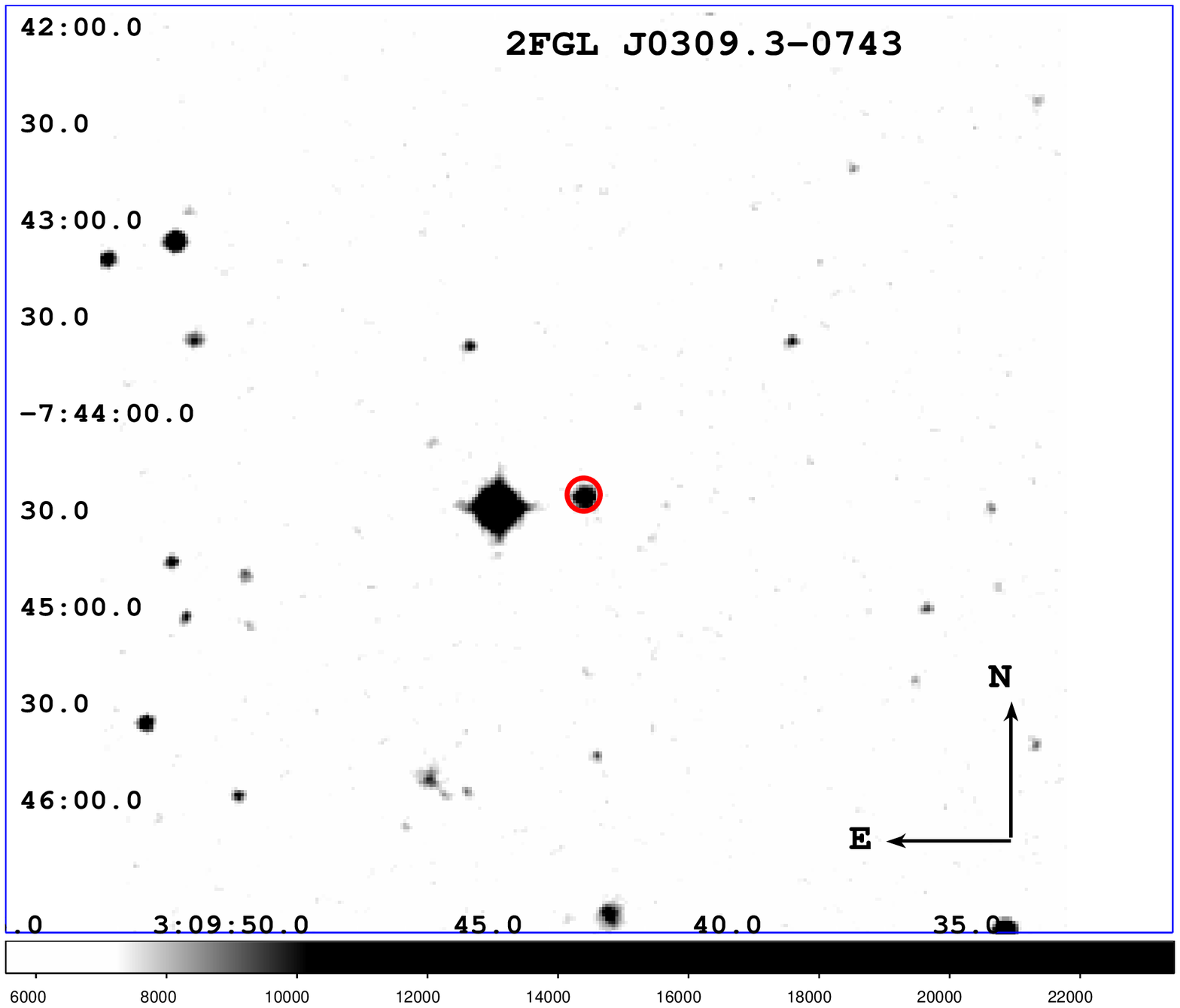}
\end{center}
\caption{Upper panel: optical spectra of WISE J030943.23-074427.5, counterpart associated with the AGU
2FGL J0309.3-0743, classified as a BZB on the basis of its featureless continuum.
The average S/N is also indicated.
Middle panel) The normalized spectrum.
Lower panel) The 5\arcmin\,x\,5\arcmin\ finding chart from the Digital Sky Survey (red filter). 
The potential counterpart of  2FGL J0309.3-0743
 is indicated by the red circle.}
\label{fig:J0309.3}
\end{figure}

\begin{figure}[]
\begin{center}
\includegraphics[height=12.2cm,width=12.2cm,angle=0]{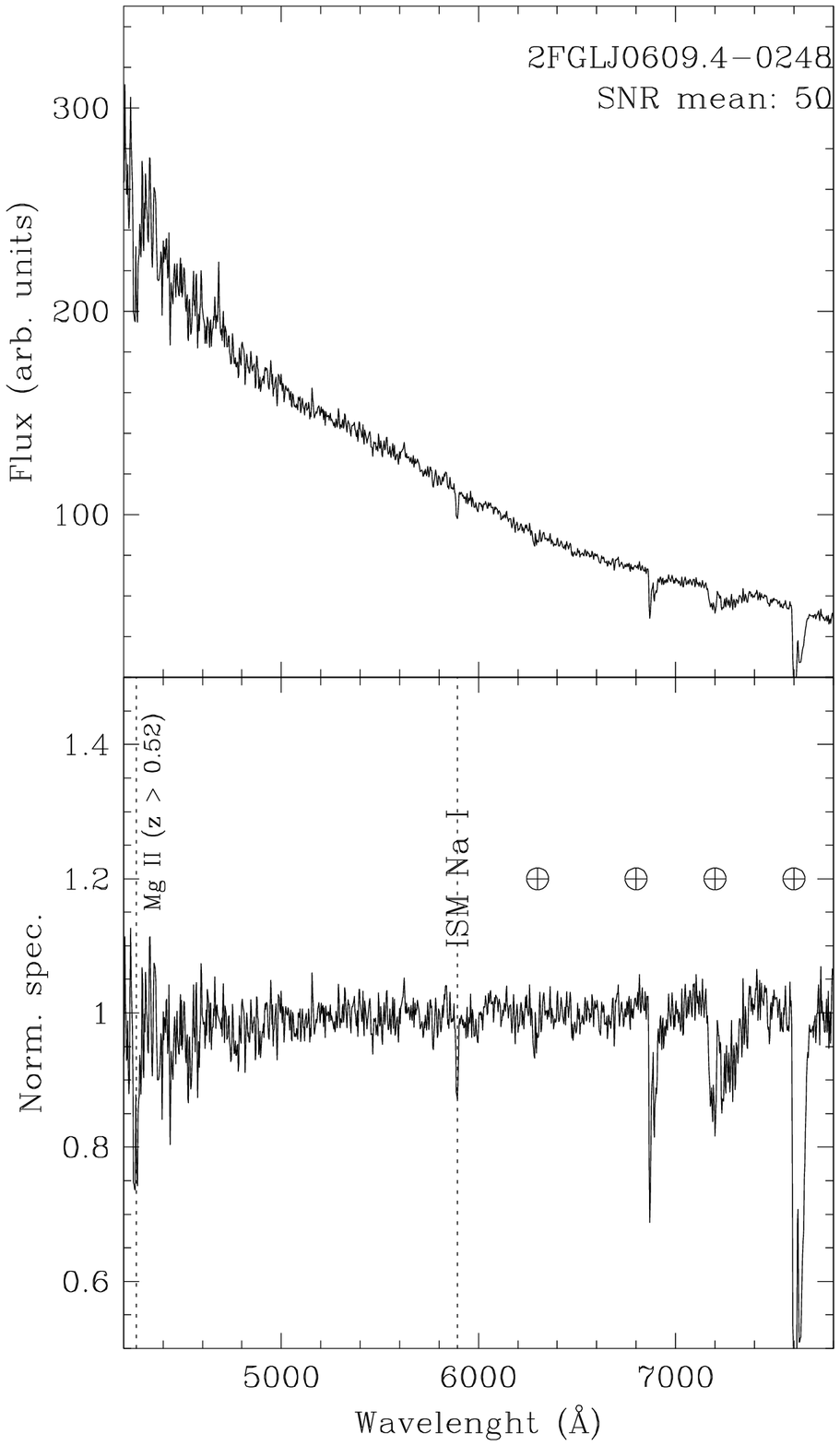}
\includegraphics[height=5.6cm,width=5.6cm,angle=0]{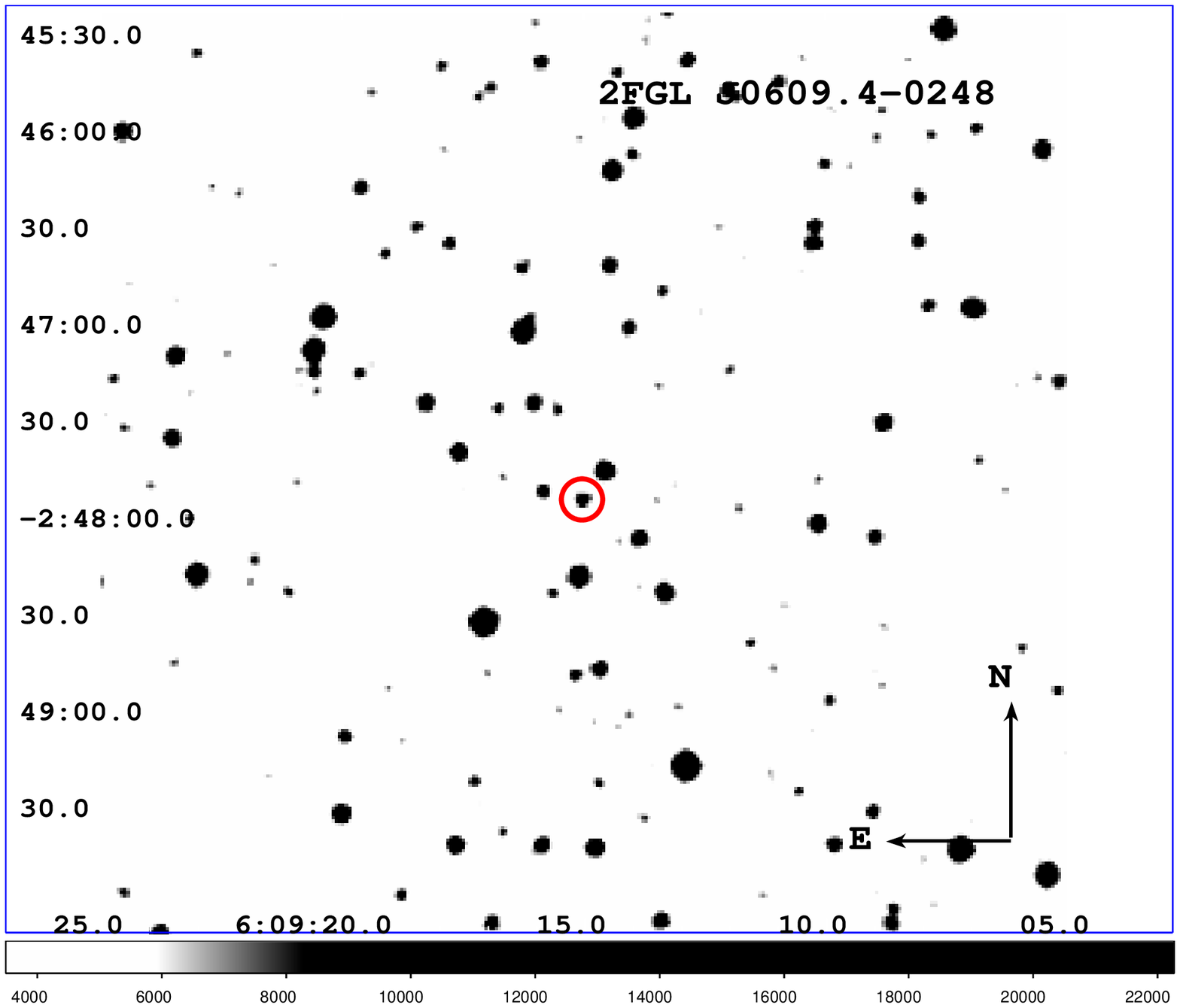}
\end{center}
\caption{Upper panel: optical spectra of WISE J060915.06-024754.5, counterpart associated with the AGU
2FGL J0609.4-0248, classified as a BZB on the basis of its featureless continuum.
Our results on this source agree with those of Shaw et al. (2013a).
The average S/N is also indicated.
Middle panel: normalized spectrum.
Lower panel: 5\arcmin\,x\,5\arcmin\ finding chart from the Digital Sky Survey (red filter). 
The potential counterpart of  2FGL J0609.4-0248
 is indicated by the red circle.}
\label{fig:J0609.4}
\end{figure}

\begin{figure}[]
\begin{center}
\includegraphics[height=12.2cm,width=12.2cm,angle=0]{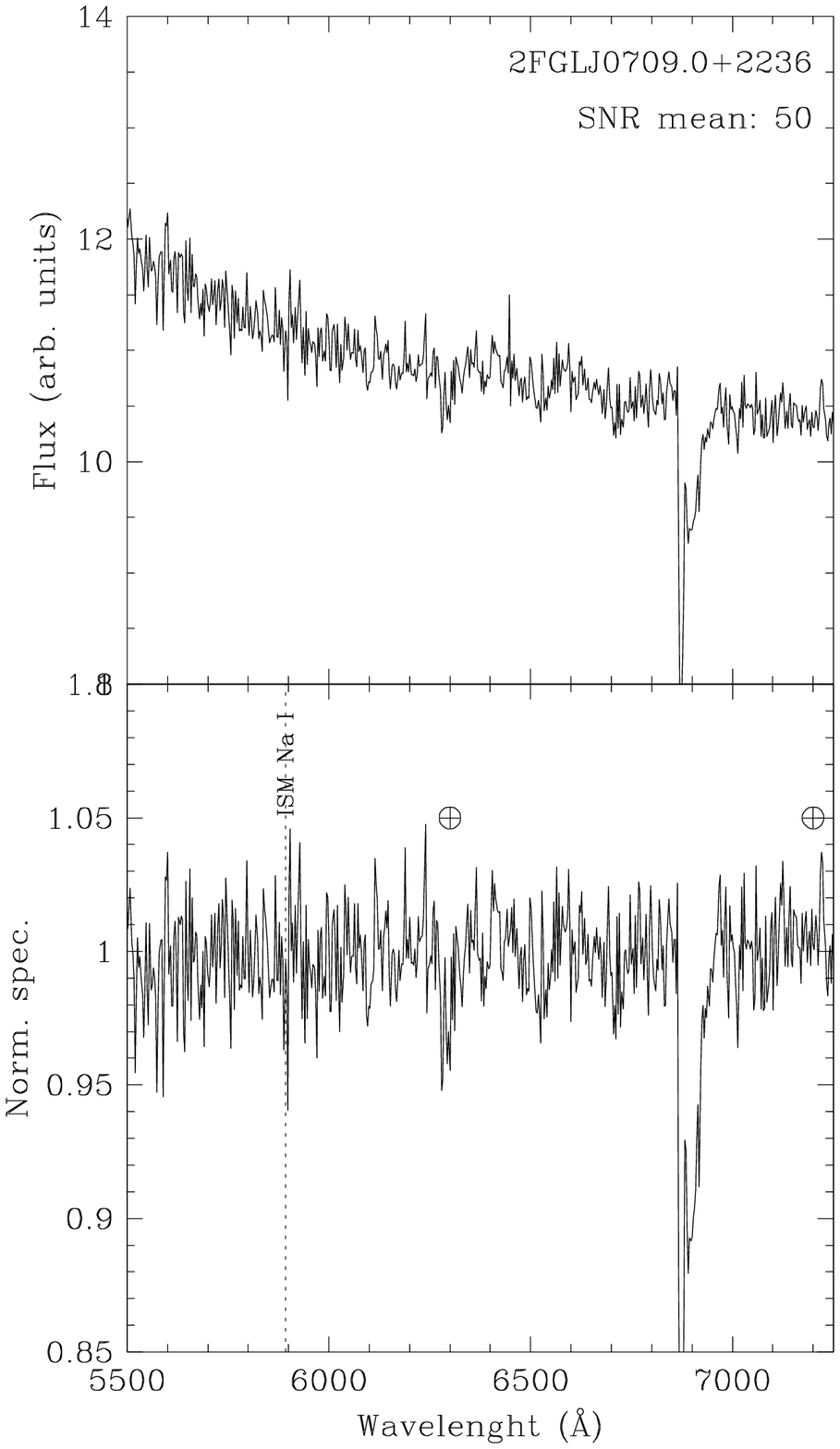}
\includegraphics[height=5.6cm,width=5.6cm,angle=0]{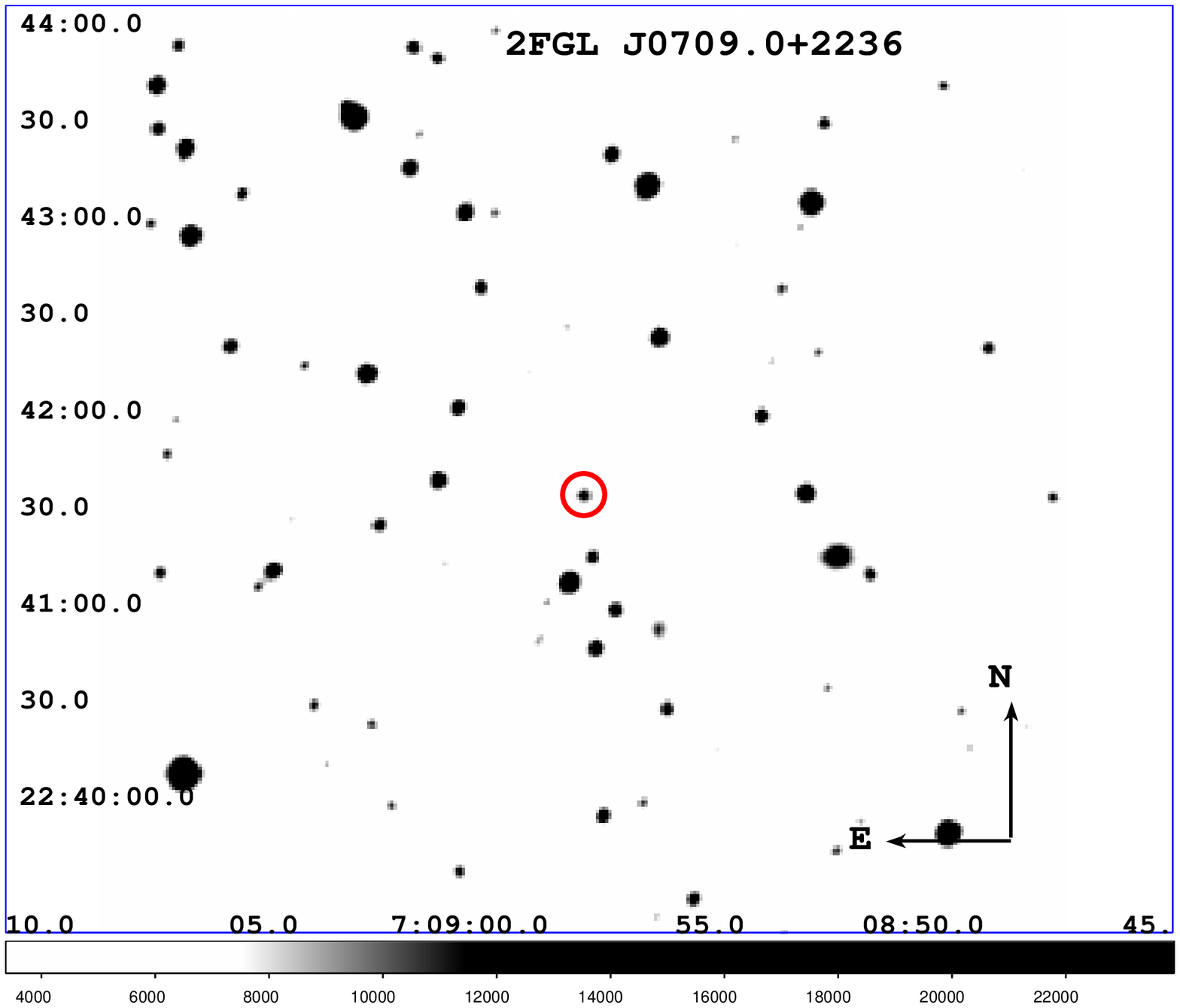}
\end{center}
\caption{Upper panel: optical spectra of WISE J070858.28+224135.4, counterpart associated with the AGU
2FGL J0709.0+2236 classified as a BZB on the basis of its featureless continuum.
The average S/N is also indicated.
Middle panel: normalized spectrum.
Lower panel: 5\arcmin\,x\,5\arcmin\ finding chart from the Digital Sky Survey (red filter). 
The potential counterpart of  2FGL J0709.0+2236 is indicated by the red circle.}
\label{fig:J0709.0}
\end{figure}

\begin{figure}[]
\begin{center}
\includegraphics[height=12.2cm,width=12.2cm,angle=0]{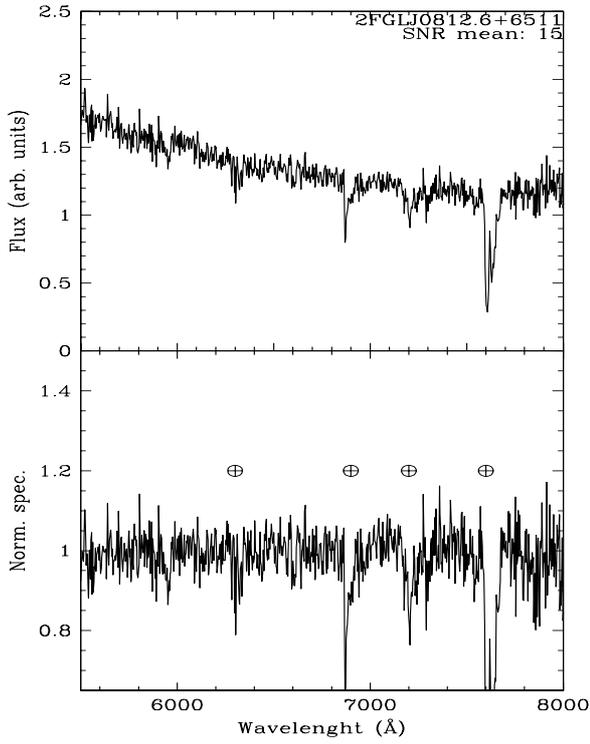}
\includegraphics[height=5.6cm,width=5.6cm,angle=0]{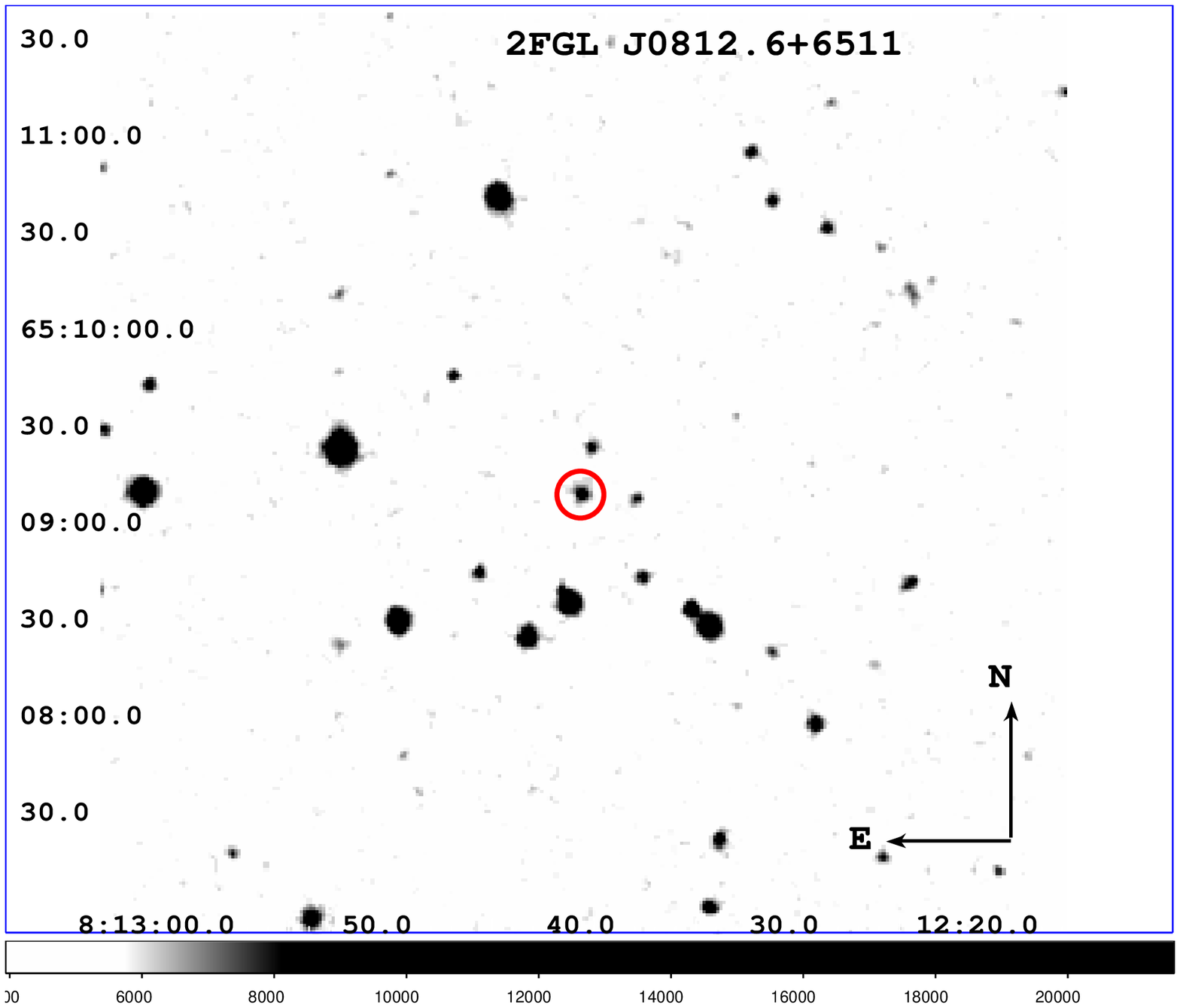}
\end{center}
\caption{Upper panel: optical spectra of WISE J081240.84+650911.1, counterpart associated with the AGU
2FGL J0812.6+6511, classified as a BZB on the basis of its featureless continuum.
The average S/N is also indicated.
Middle panel: normalized spectrum.
Lower panel: 5\arcmin\,x\,5\arcmin\ finding chart from the Digital Sky Survey (red filter). 
The potential counterpart of  2FGL J0812.6+6511
 is indicated by the red circle.}
\label{fig:J0812.6}
\end{figure}

\begin{figure}[]
\begin{center}
\includegraphics[height=12.2cm,width=12.2cm,angle=0]{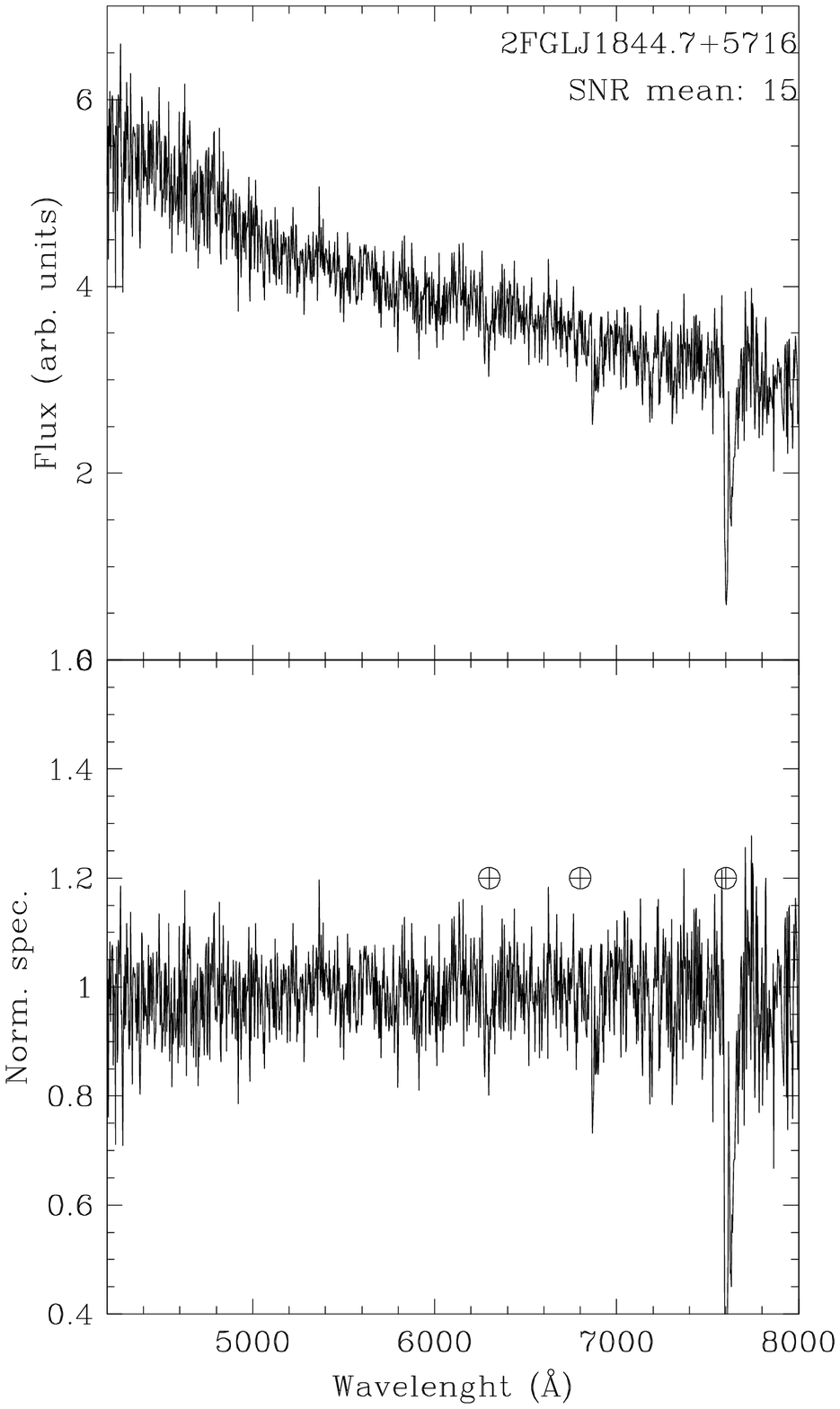}
\includegraphics[height=5.6cm,width=5.6cm,angle=0]{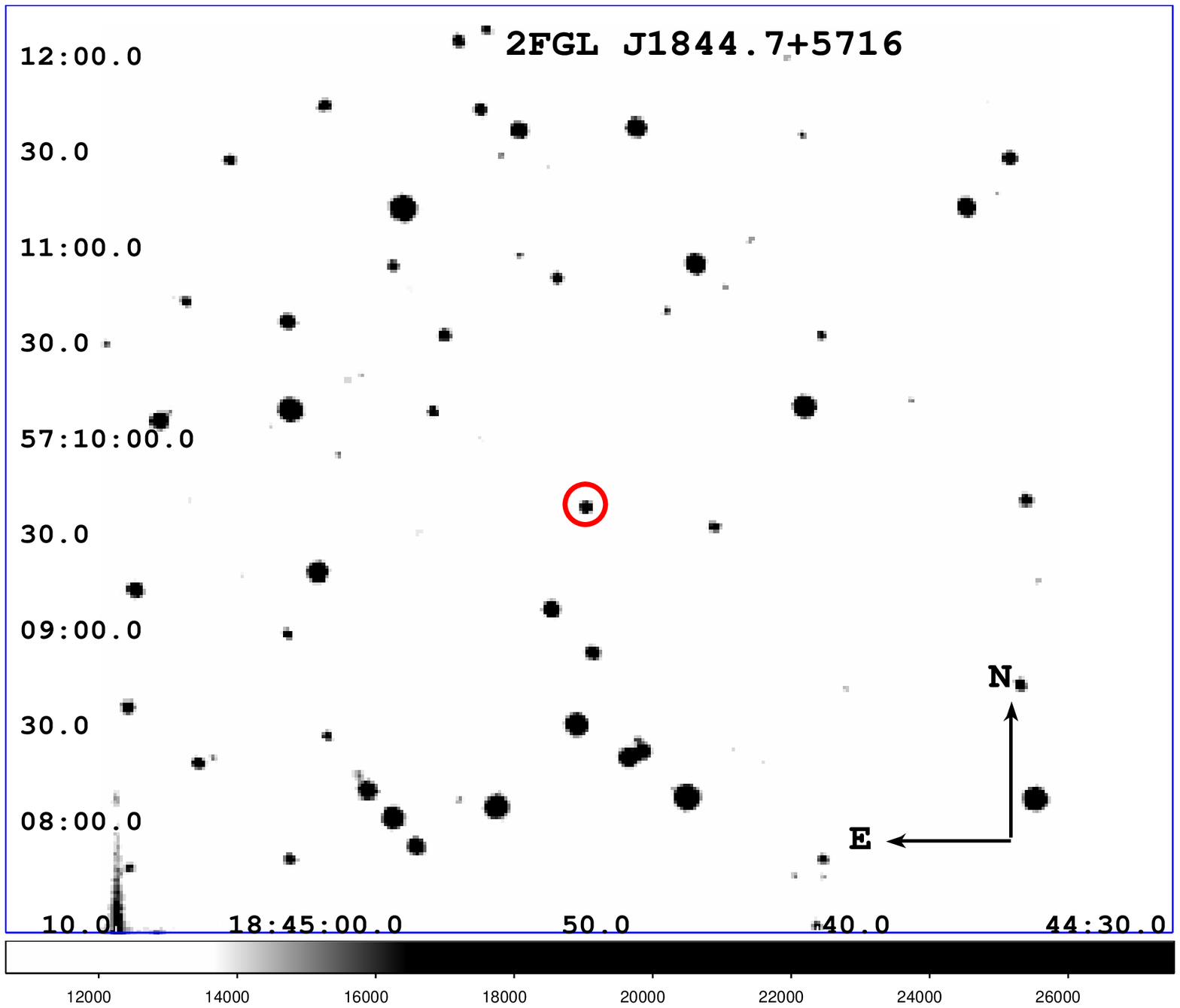}
\end{center}
\caption{Upper panel: optical spectra of WISE J184450.96+570938.6, counterpart associated with the AGU
2FGL J1844.7+5716, classified as a BZB on the basis of its featureless continuum.
The average S/Nis also indicated.
Middle panel: normalized spectrum.
Lower panel: 5\arcmin\,x\,5\arcmin\ finding chart from the Digital Sky Survey (red filter). 
The potential counterpart of  2FGL J1844.7+5716 is indicated by the red circle.}
\label{fig:J1844.7}
\end{figure}

\begin{figure}[]
\begin{center}
\includegraphics[height=12.2cm,width=12.2cm,angle=0]{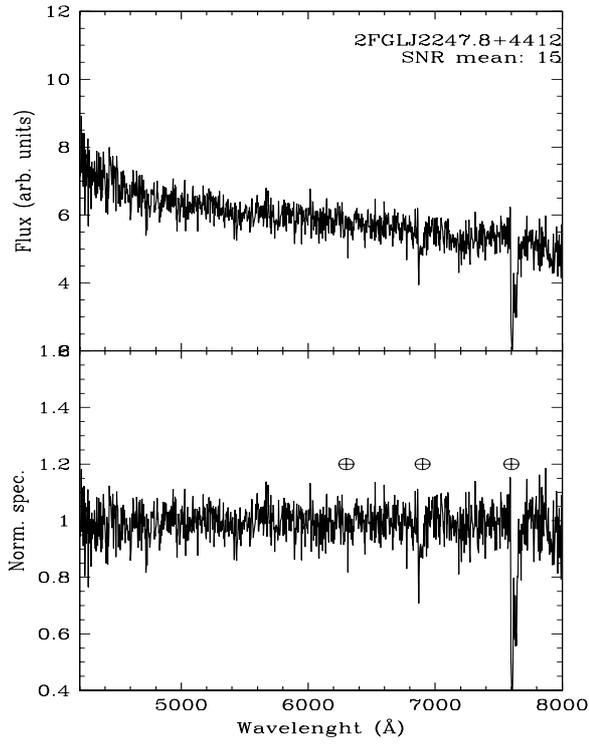}
\includegraphics[height=5.6cm,width=5.6cm,angle=0]{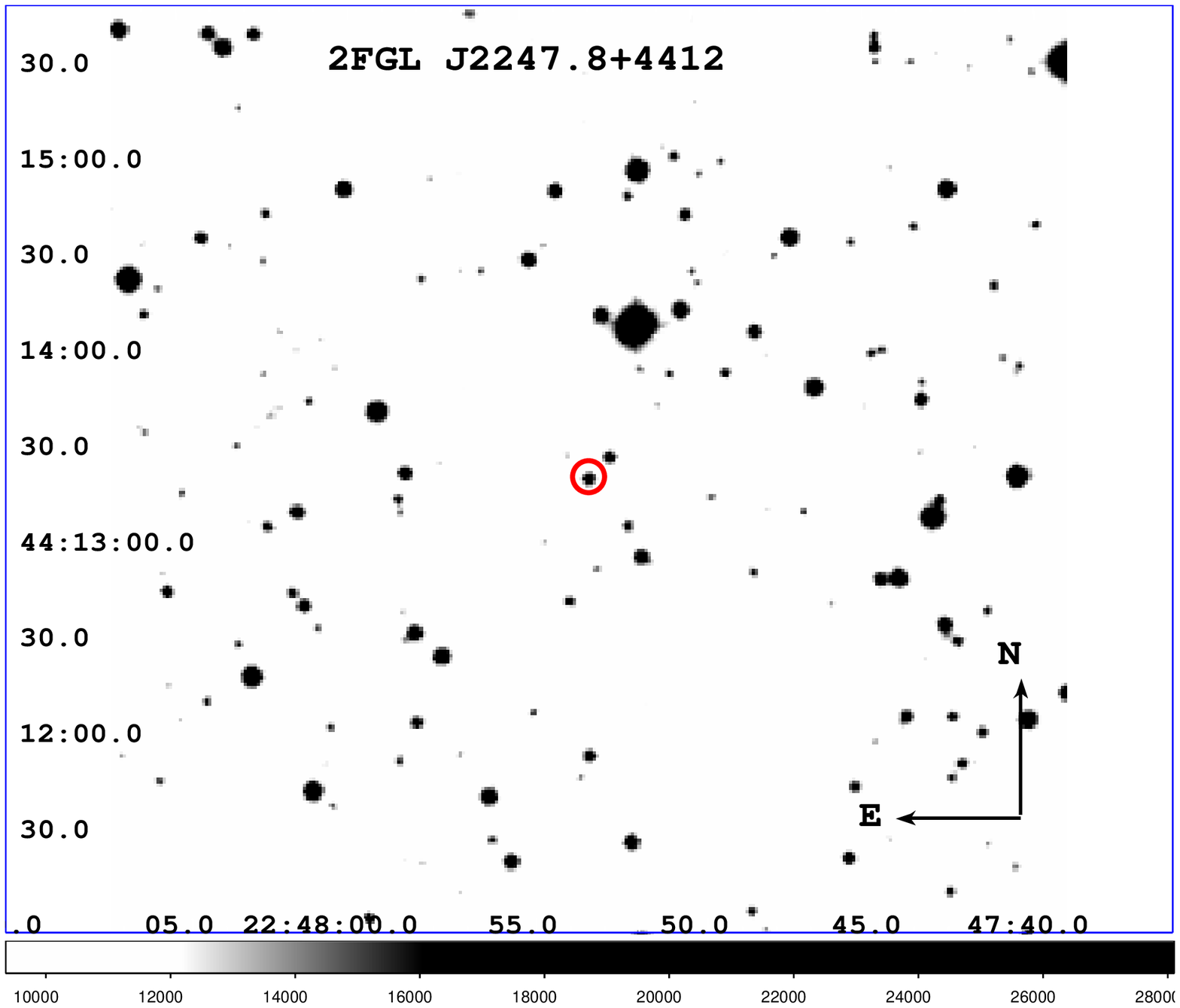}
\end{center}
\caption{Upper panel: optical spectra of the WISE J224753.22+441315.5, counterpart associated with the AGU
2FGL J2247.8+4412, classified as a BZB on the basis of its featureless continuum.
The average S/N is also indicated.
Middle panel: normalized spectrum.
Lower panel: 5\arcmin\,x\,5\arcmin\ finding chart from the Digital Sky Survey (red filter). 
The potential counterpart of  2FGL J2247.8+4412 is indicated by the red circle.}
\label{fig:J2247.8}
\end{figure}

\begin{figure}[]
\begin{center}
\includegraphics[height=12.2cm,width=12.2cm,angle=0]{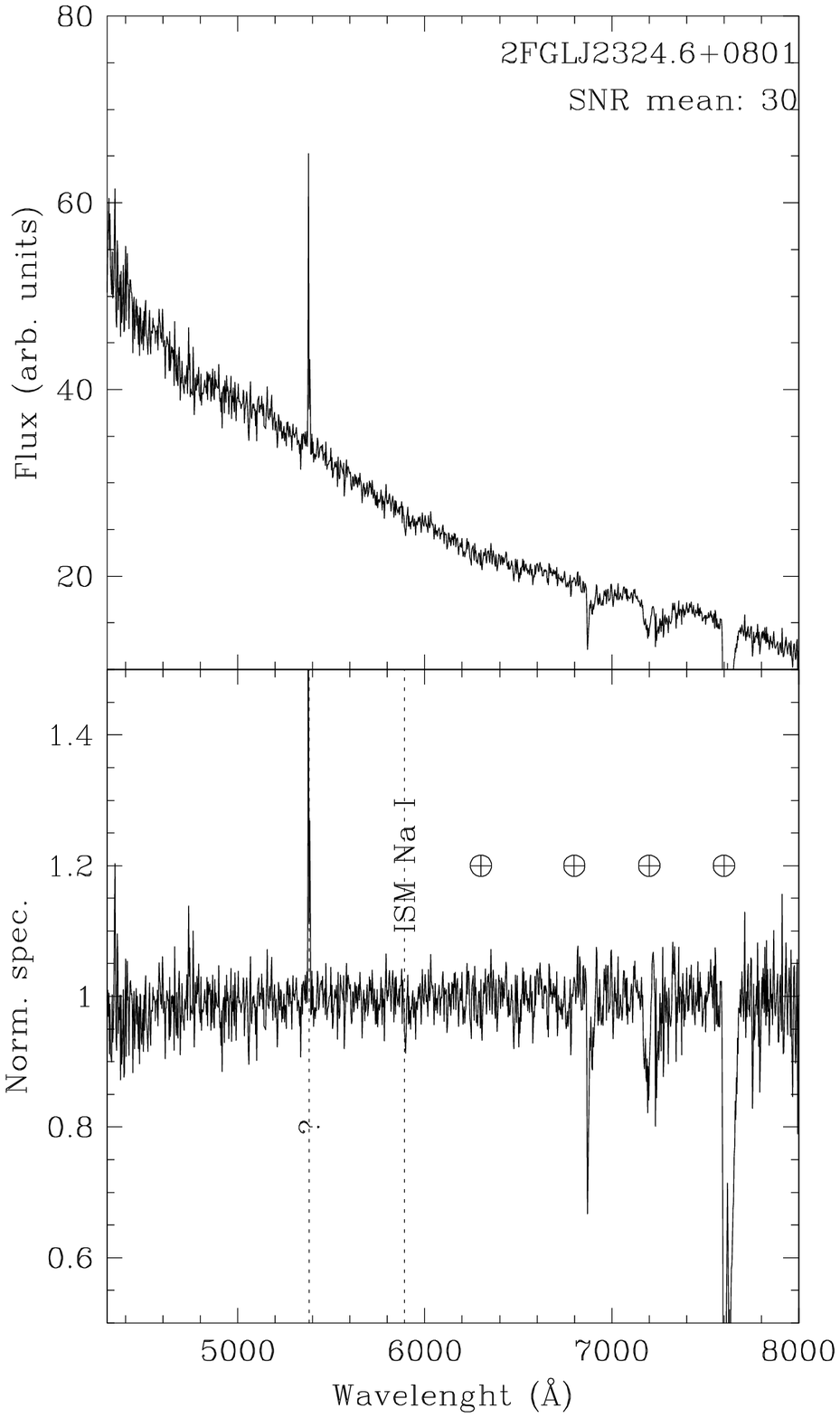}
\includegraphics[height=5.6cm,width=5.6cm,angle=0]{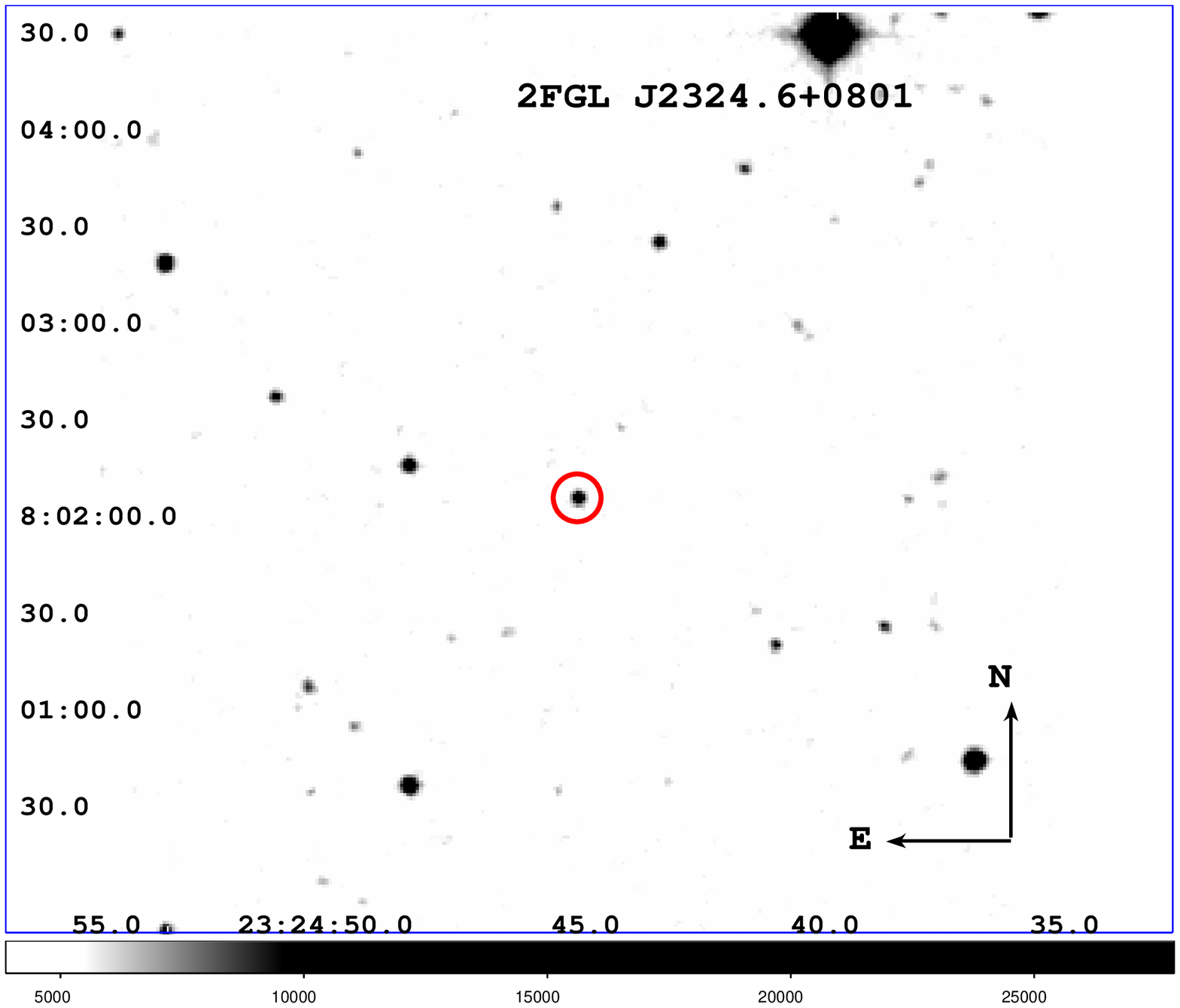}
\end{center}
\caption{Upper panel: optical spectra of WISE J232445.32+080206.1, counterpart associated with the AGU
2FGL J2324.6+0801, classified as a BZB on the basis of its featureless continuum.
The average S/N is also indicated.
In the spectrum of this source we also detected an unknown absorption feature at $\sim$5400$\textrm{\AA}$
as marked in the figure above.
Middle panel: normalized spectrum.
Lower panel: 5\arcmin\,x\,5\arcmin\ finding chart from the Digital Sky Survey (red filter). 
The potential counterpart of  2FGL J2324.6+0801 is indicated by the red circle.}
\label{fig:J2324.6}
\end{figure}

\begin{figure}[]
\begin{center}
\includegraphics[height=12.2cm,width=12.2cm,angle=0]{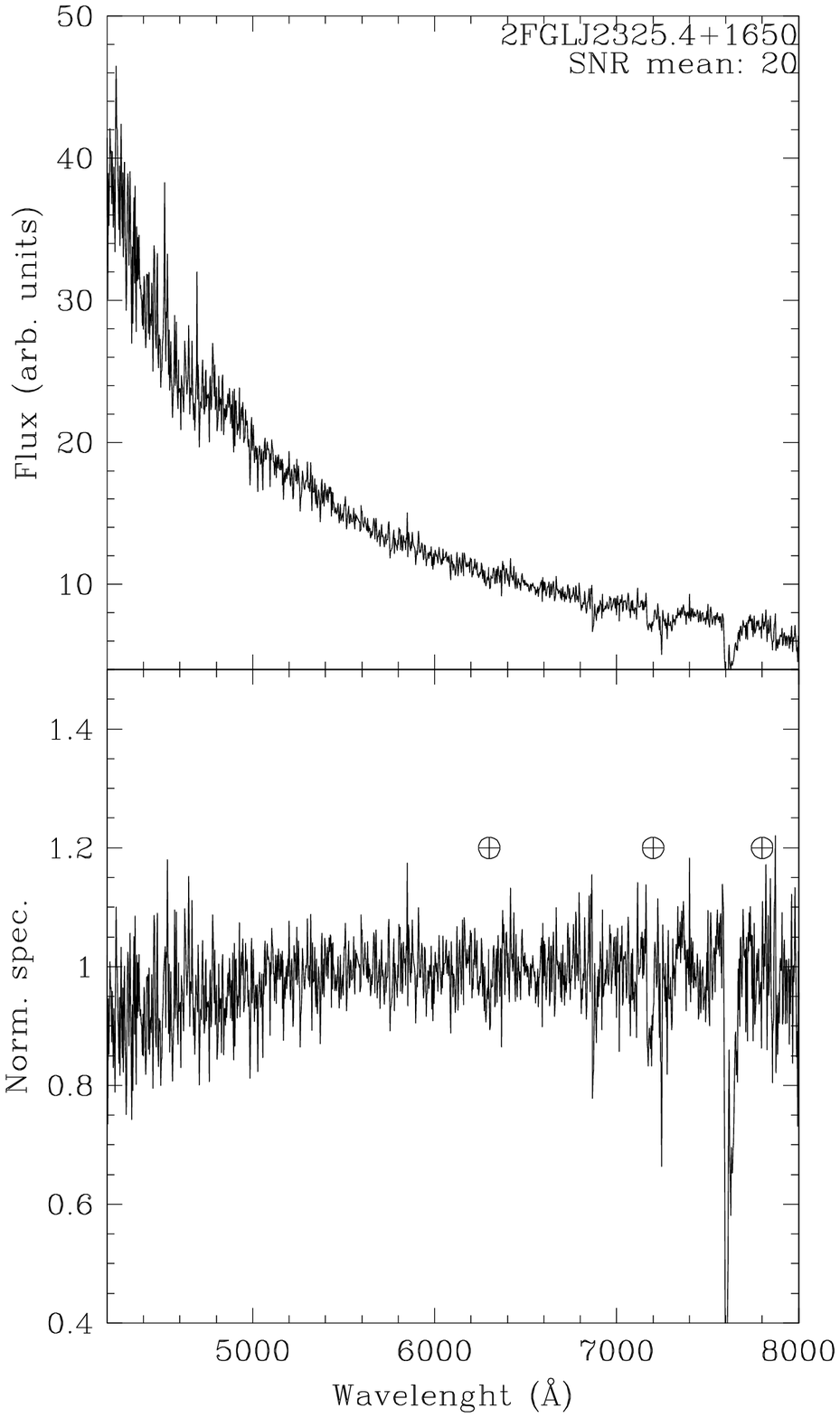}
\includegraphics[height=5.6cm,width=5.6cm,angle=0]{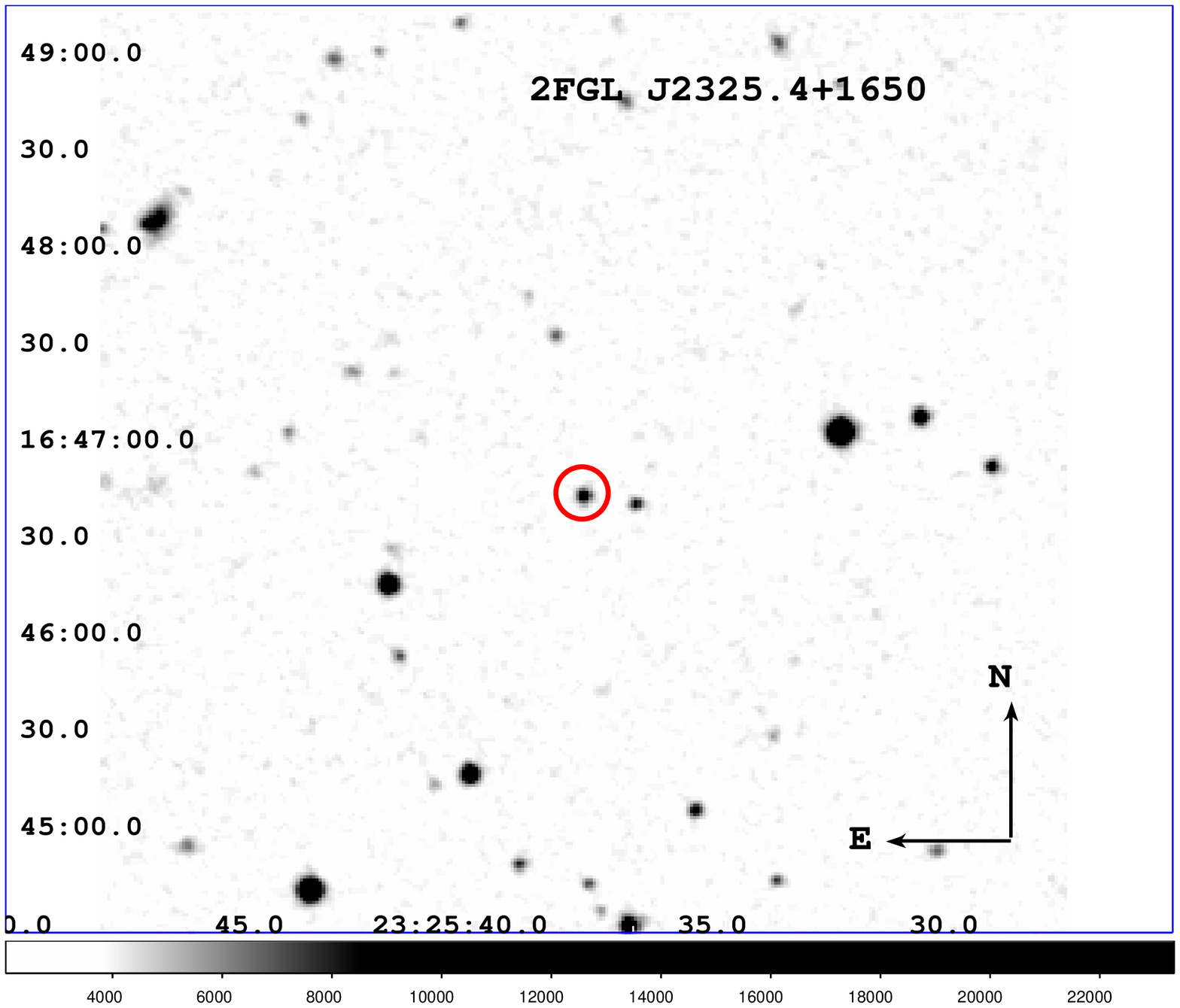}
\end{center}
\caption{Upper panel: optical spectra of WISE J232538.11+164642.7, counterpart associated with the AGU
2FGL J2325.4+1650, classified as a BZB on the basis of its featureless continuum.
The average S/N is also indicated.
Middle panel: normalized spectrum.
Lower panel: 5\arcmin\,x\,5\arcmin\ finding chart from the Digital Sky Survey (red filter). 
The potential counterpart of  2FGL J2325.4+1650 is indicated by the red circle.}
\label{fig:J2325.4}
\end{figure}

\begin{figure}[]
\begin{center}
\includegraphics[height=12.2cm,width=12.2cm,angle=0]{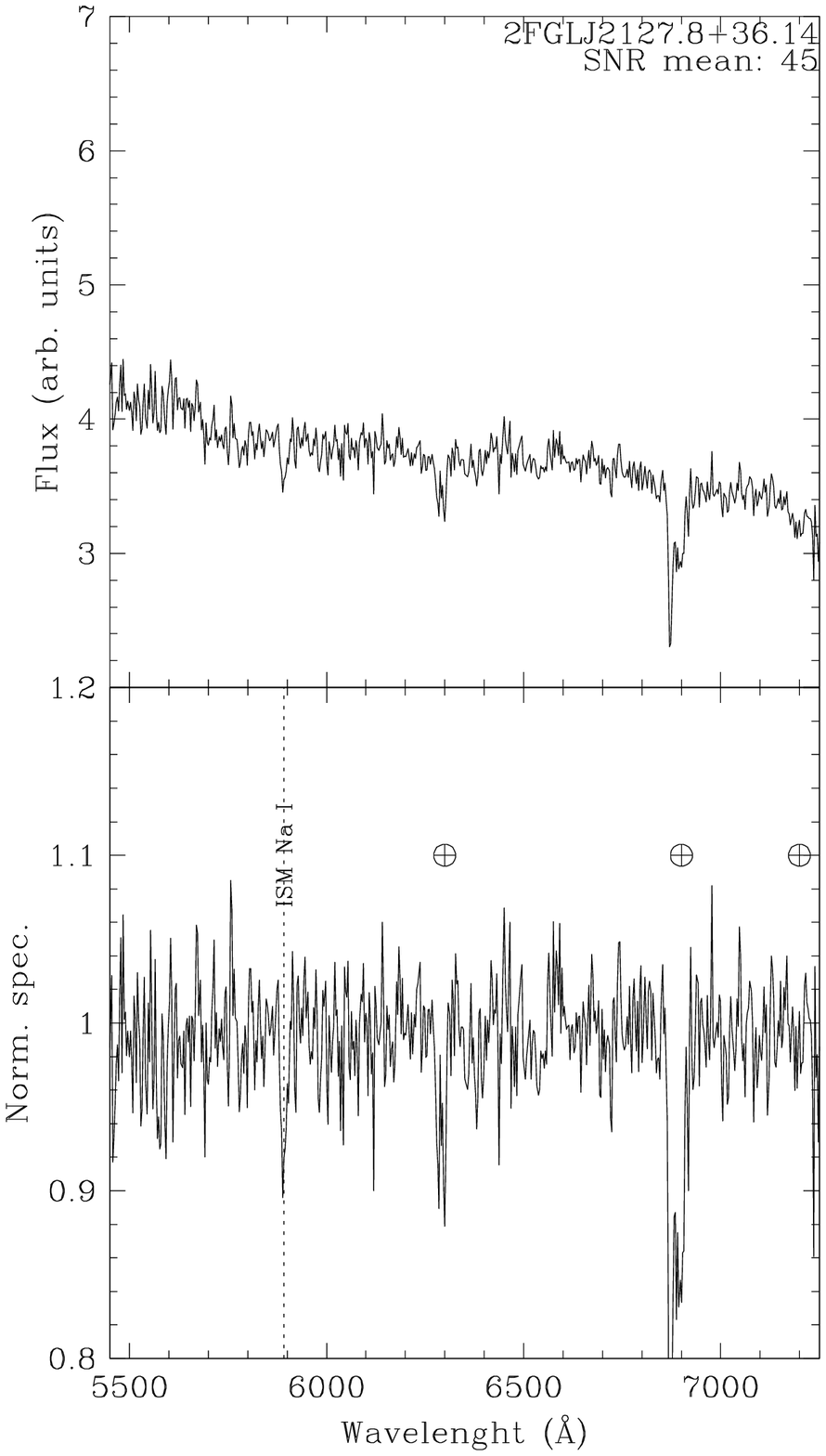}
\includegraphics[height=5.6cm,width=5.6cm,angle=0]{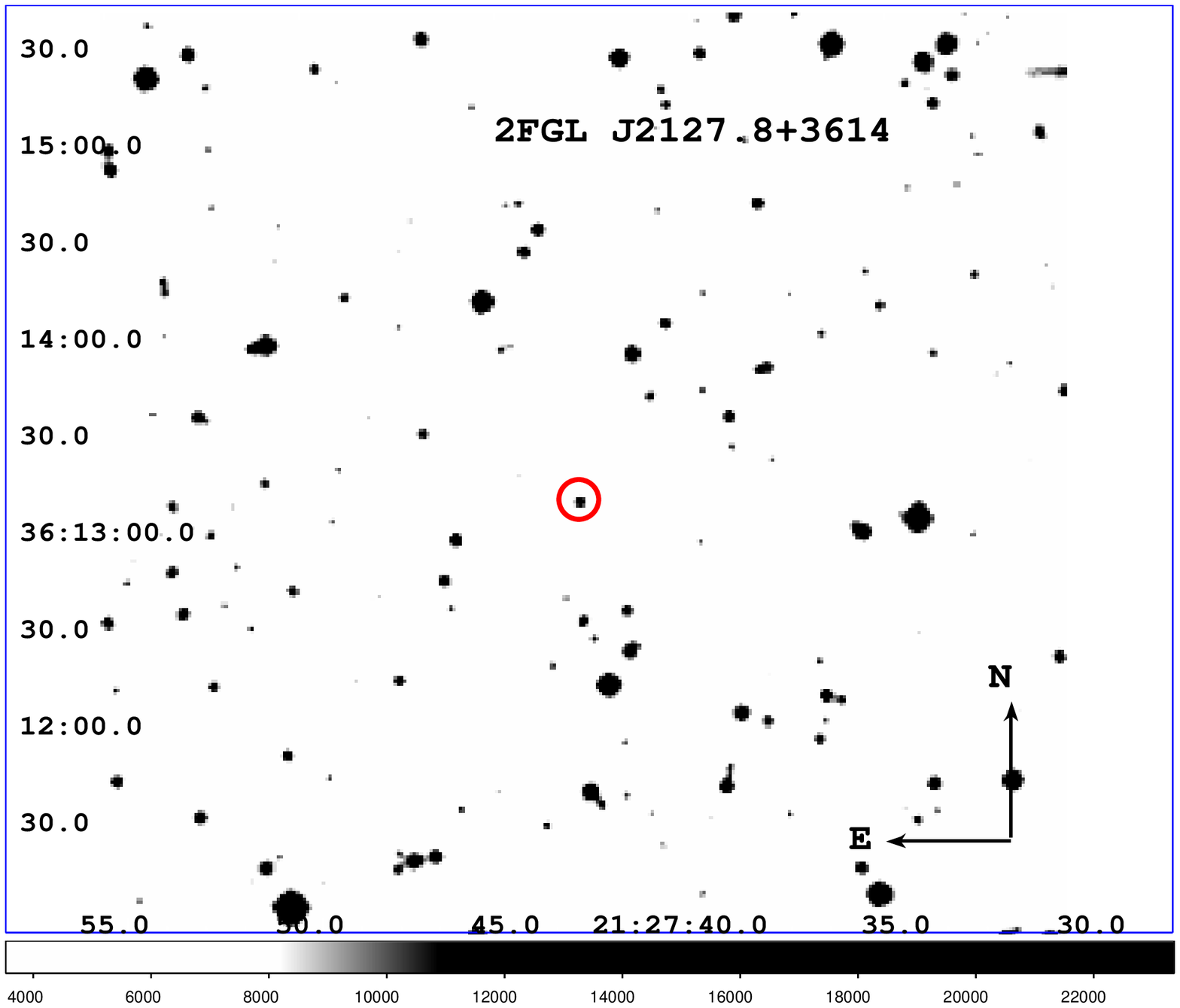}
\end{center}
\caption{Upper panel: optical spectra of WISE J212743.03+361305.7, counterpart associated with the BZB
2FGL J2127.8+3614 classified by Ackermann et al. (2011a).
Our observation clearly shows a featureless continuum, which confirms its classification.
The average S/N also indicated.
Middle panel: normalized spectrum.
Lower panel: 5\arcmin\,x\,5\arcmin\ finding chart from the Digital Sky Survey (red filter). 
The potential counterpart of  2FGL J2127.8+3614 is indicated by the red circle.}
\label{fig:J2127.8}
\end{figure}

\begin{figure}[]
\begin{center}
\includegraphics[height=12.2cm,width=12.2cm,angle=0]{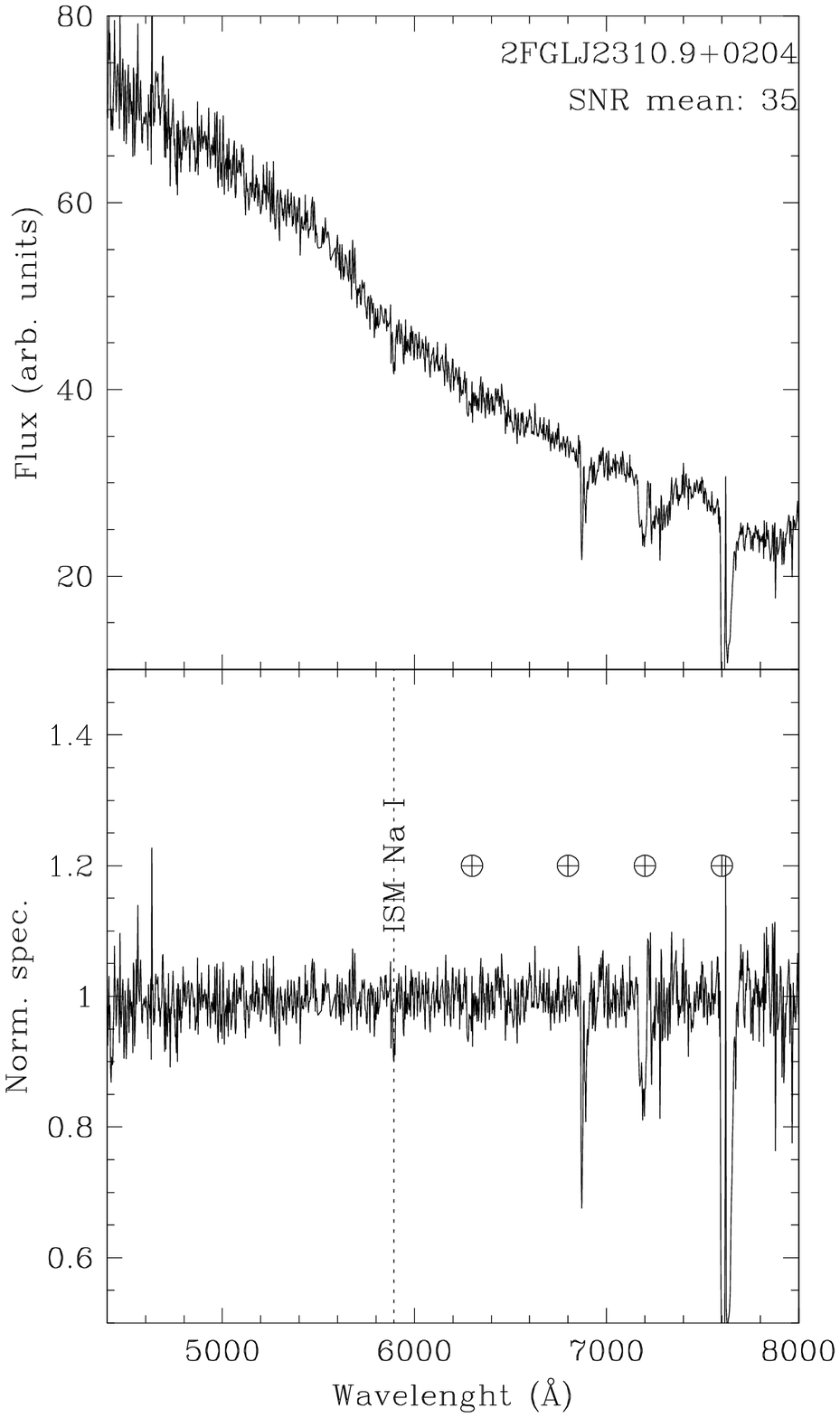}
\includegraphics[height=5.6cm,width=5.6cm,angle=0]{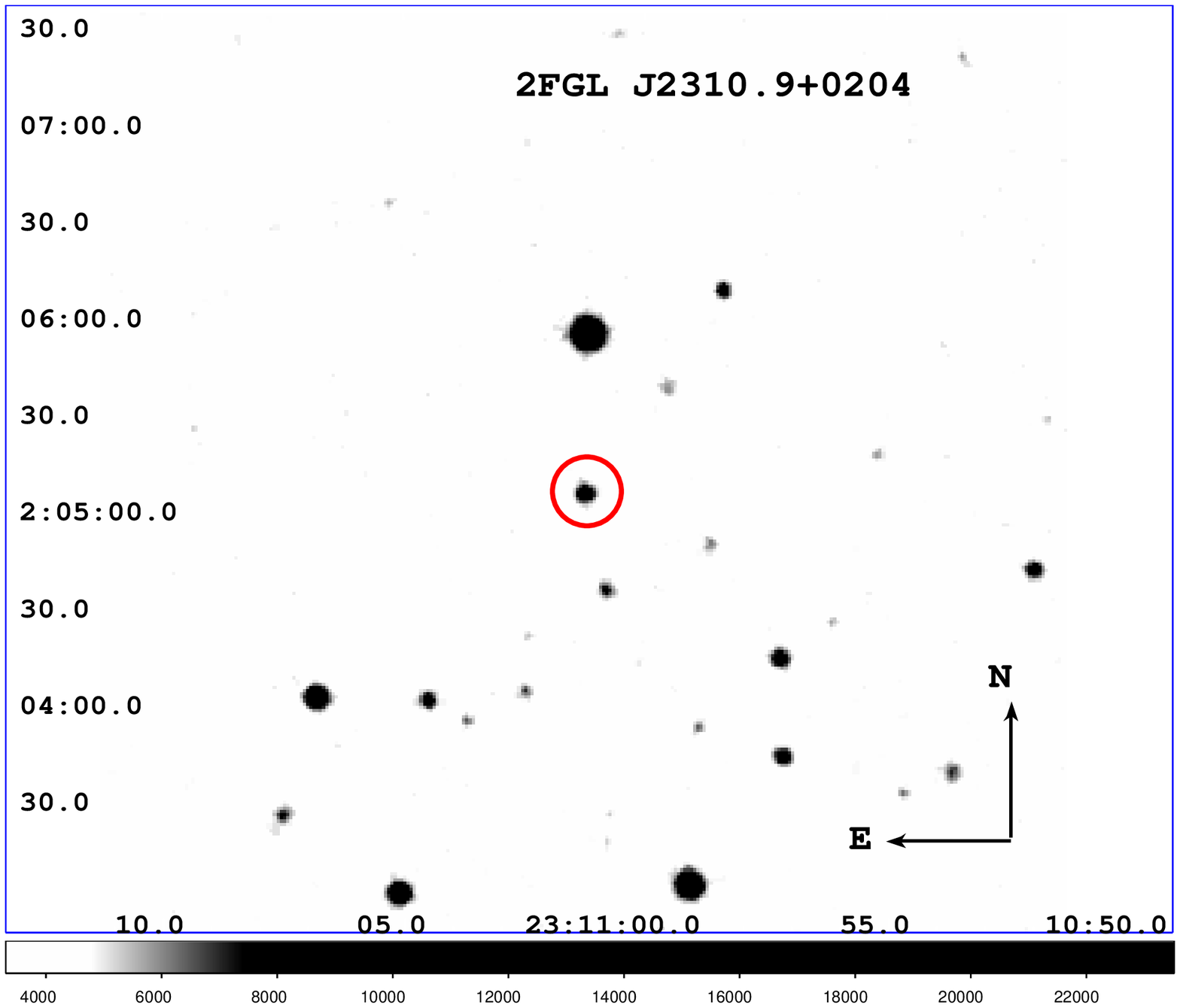}
\end{center}
\caption{Upper panel: optical spectra of WISE J231101.29+020505.3, counterpart associated with the BZB
2FGL J2310.9+0204 classified by Ackermann et al. (2011a).
Our observation clearly shows a featureless continuum, which confirms its classification,
in agreement with the recent observations of Shaw et al. (2013a).
The average S/N is also indicated.
Middle panel: normalized spectrum.
Lower panel: 5\arcmin\,x\,5\arcmin\ finding chart from the Digital Sky Survey (red filter). 
The potential counterpart of  2FGL J2310.9+0204 is indicated by the red circle.}
\label{fig:J2310.9}
\end{figure}

\begin{figure}[]
\begin{center}
\includegraphics[height=12.2cm,width=12.2cm,angle=0]{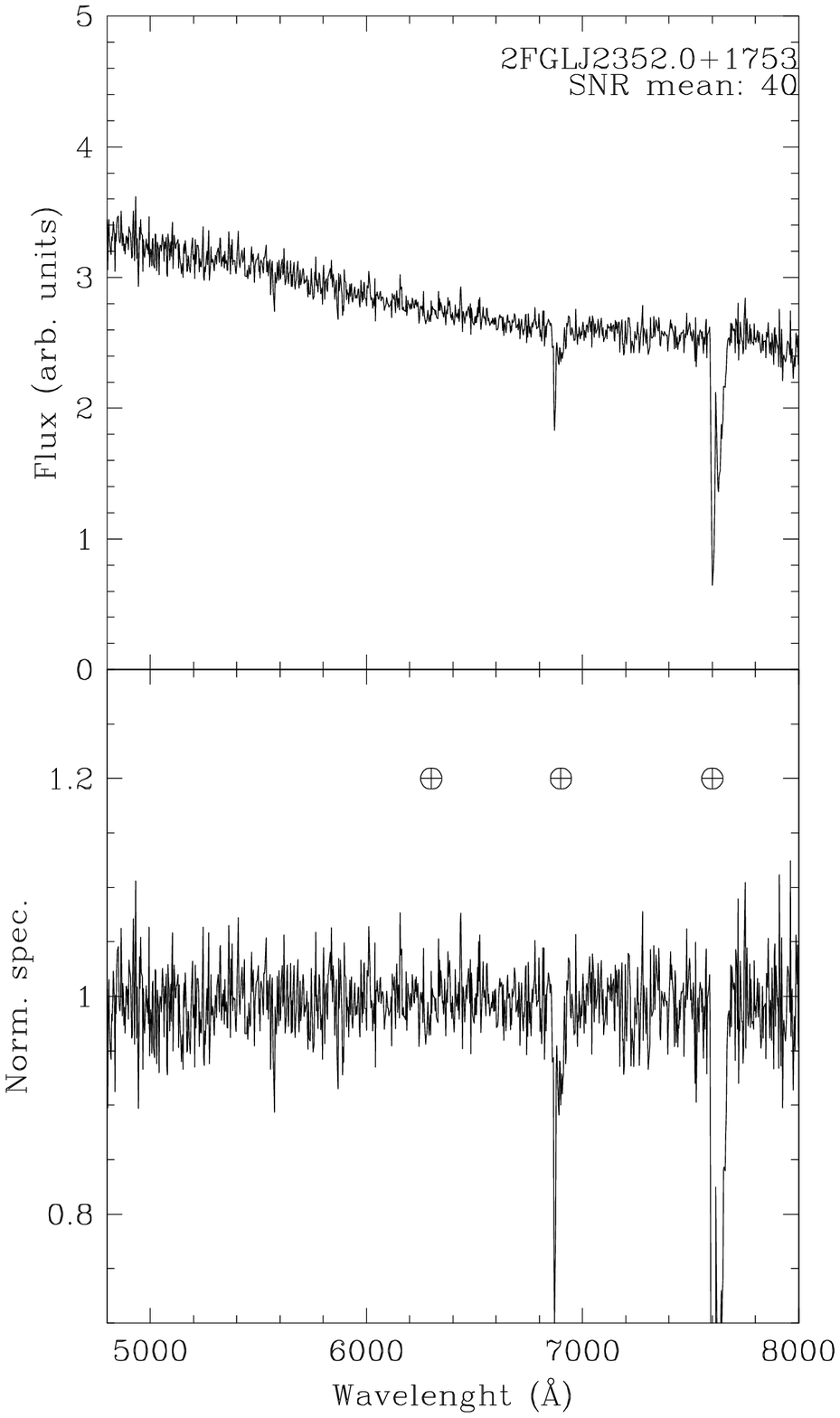}
\includegraphics[height=5.6cm,width=5.6cm,angle=0]{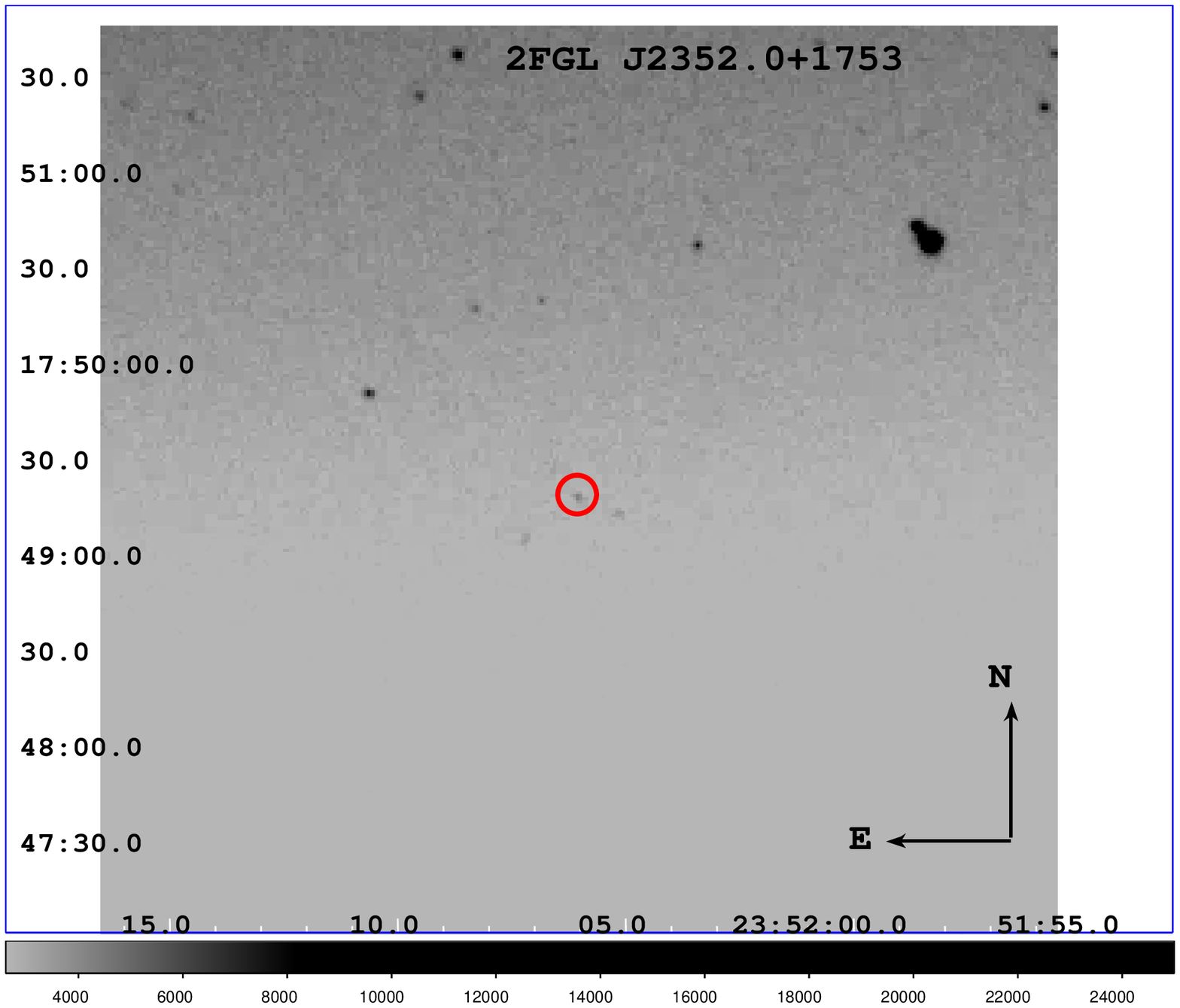}
\end{center}
\caption{Upper panel: optical spectra of WISE J235205.84+174913.7, counterpart associated with the BZB
2FGL J2352.0+1753 classified by Ackermann et al. (2011a).
Our observation clearly shows a featureless continuum, which confirms its classification,
in agreement with the recent observations of Shaw et al. (2013a).
The average S/N is also indicated.
Middle panel: normalized spectrum.
Lower panel: 5\arcmin\,x\,5\arcmin\ finding chart from the Digital Sky Survey (red filter). 
The potential counterpart of  2FGL J2352.0+1753 is indicated by the red circle.}
\label{fig:J2352.0}
\end{figure}

\begin{figure}[]
\begin{center}
\includegraphics[height=12.2cm,width=12.2cm,angle=0]{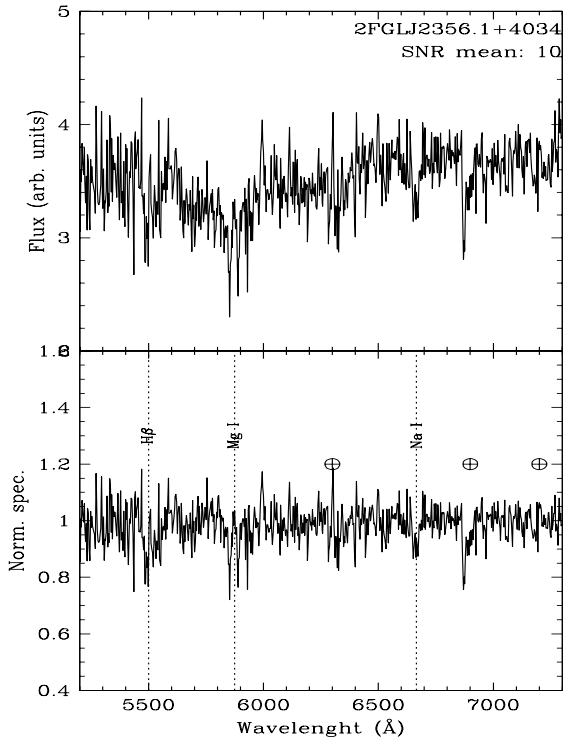}
\includegraphics[height=5.6cm,width=5.6cm,angle=0]{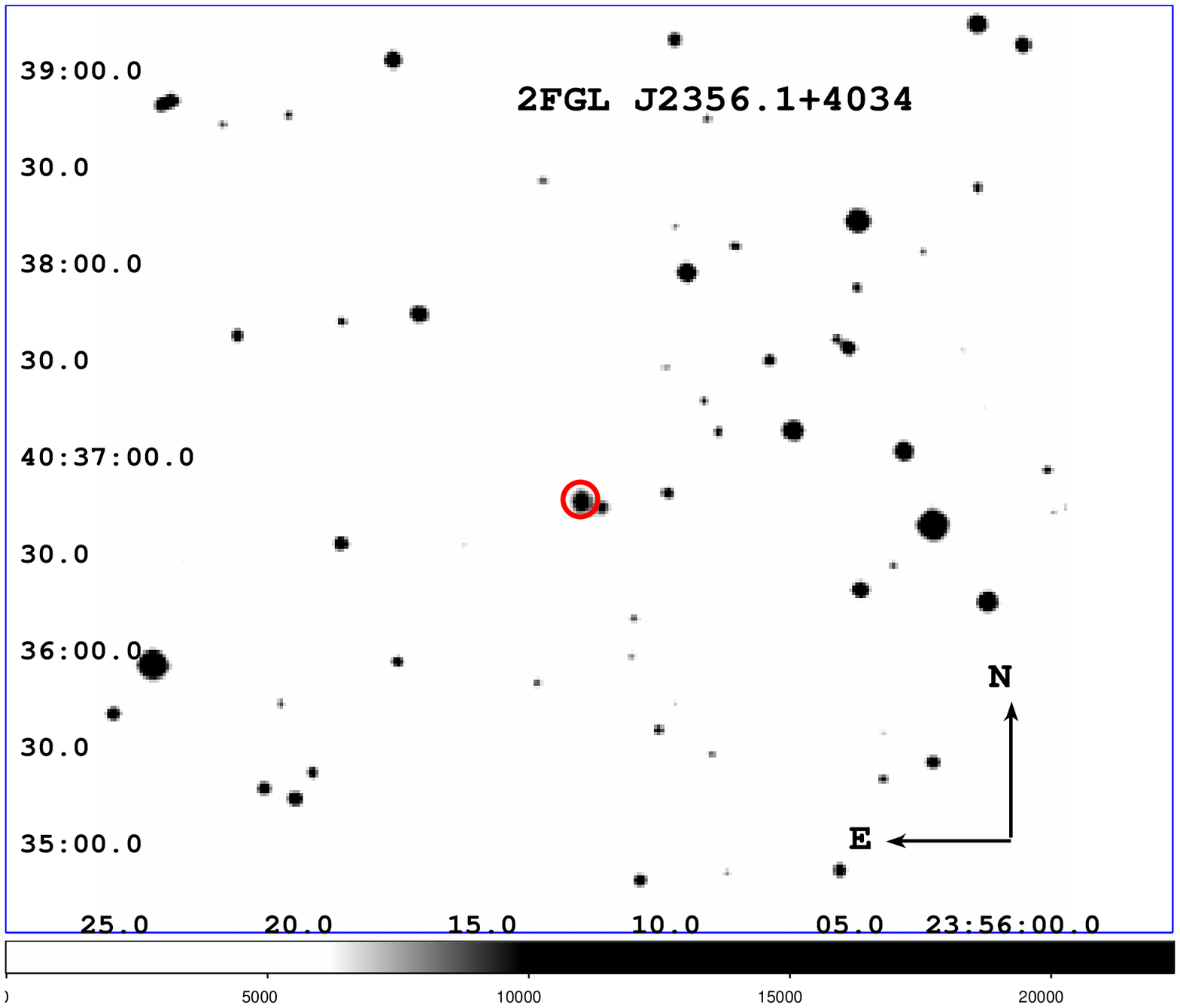}
\end{center}
\caption{Upper panel: optical spectra of WISE J235612.70+403646.8, counterpart associated with the BZB
2FGL J2356.1+4034 classified by Ackermann et al. (2011a).
In our observation its spectrum appears as that of a normal elliptical galaxy lying at redshift 0.131
instead of 0.331, as previously reported in Ackermann et al. (2011a).
The average S/N is also indicated.
Middle panel: normalized spectrum.
Lower panel: 5\arcmin\,x\,5\arcmin\ finding chart from the Digital Sky Survey (red filter). 
The potential counterpart of  2FGL J2356.1+4034 is indicated by the red circle.}
\label{fig:J2356.1}
\end{figure}

\begin{figure}[]
\begin{center}
\includegraphics[height=12.2cm,width=12.2cm,angle=0]{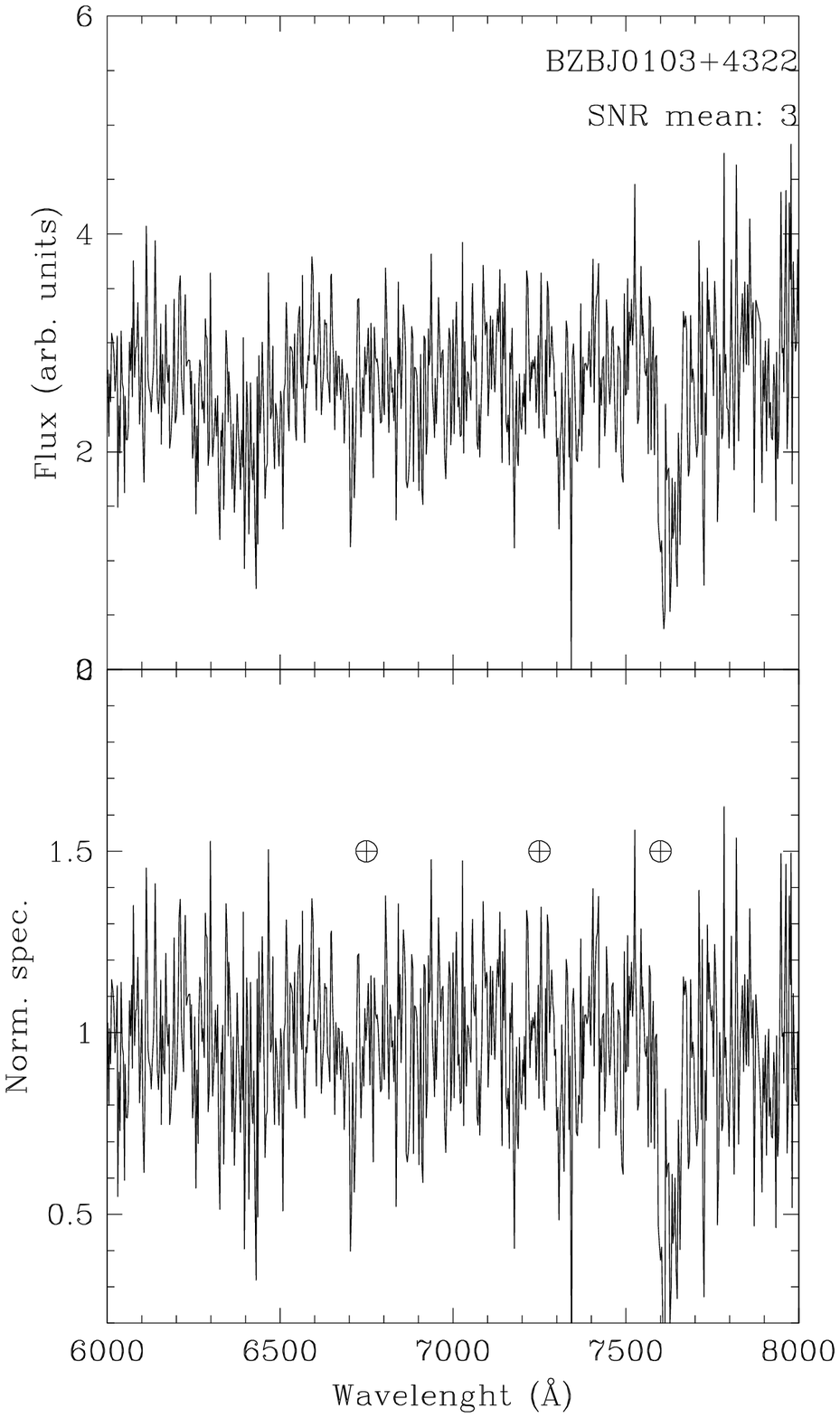}
\includegraphics[height=5.6cm,width=5.6cm,angle=0]{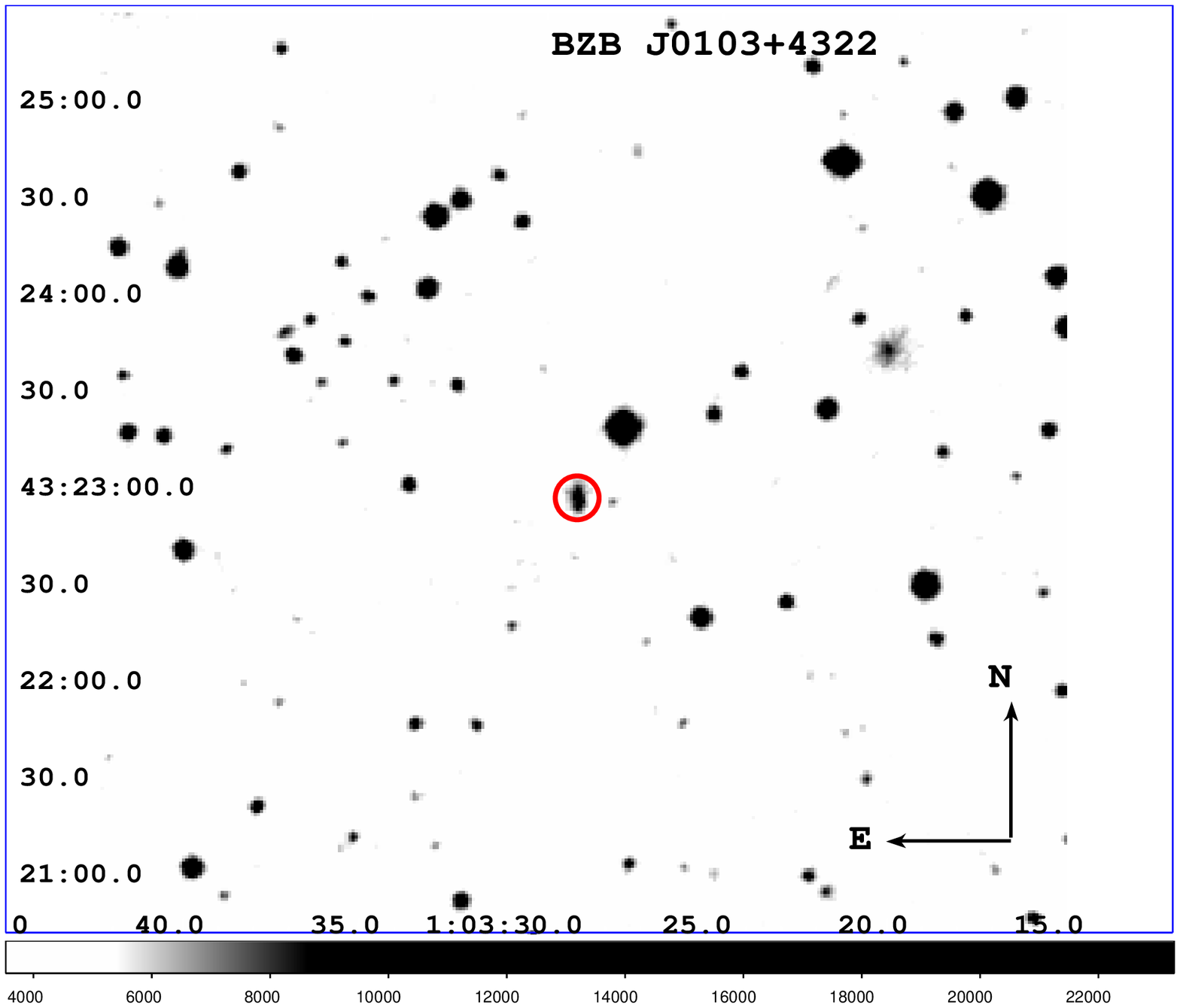}
\end{center}
\caption{Upper panel: optical spectra of the BL Lac candidate
BZB J0103+4322 listed in the \bzcat\ v4.1.
Our observation clearly shows a featureless continuum and allows us to verify its classification.
The average S/N is also indicated.
Middle panel: normalized spectrum.
Lower panel: 5\arcmin\,x\,5\arcmin\ finding chart from the Digital Sky Survey (red filter). 
The source is indicated by the red circle.}
\label{fig:J0103}
\end{figure}

\begin{figure}[]
\begin{center}
\includegraphics[height=12.2cm,width=12.2cm,angle=0]{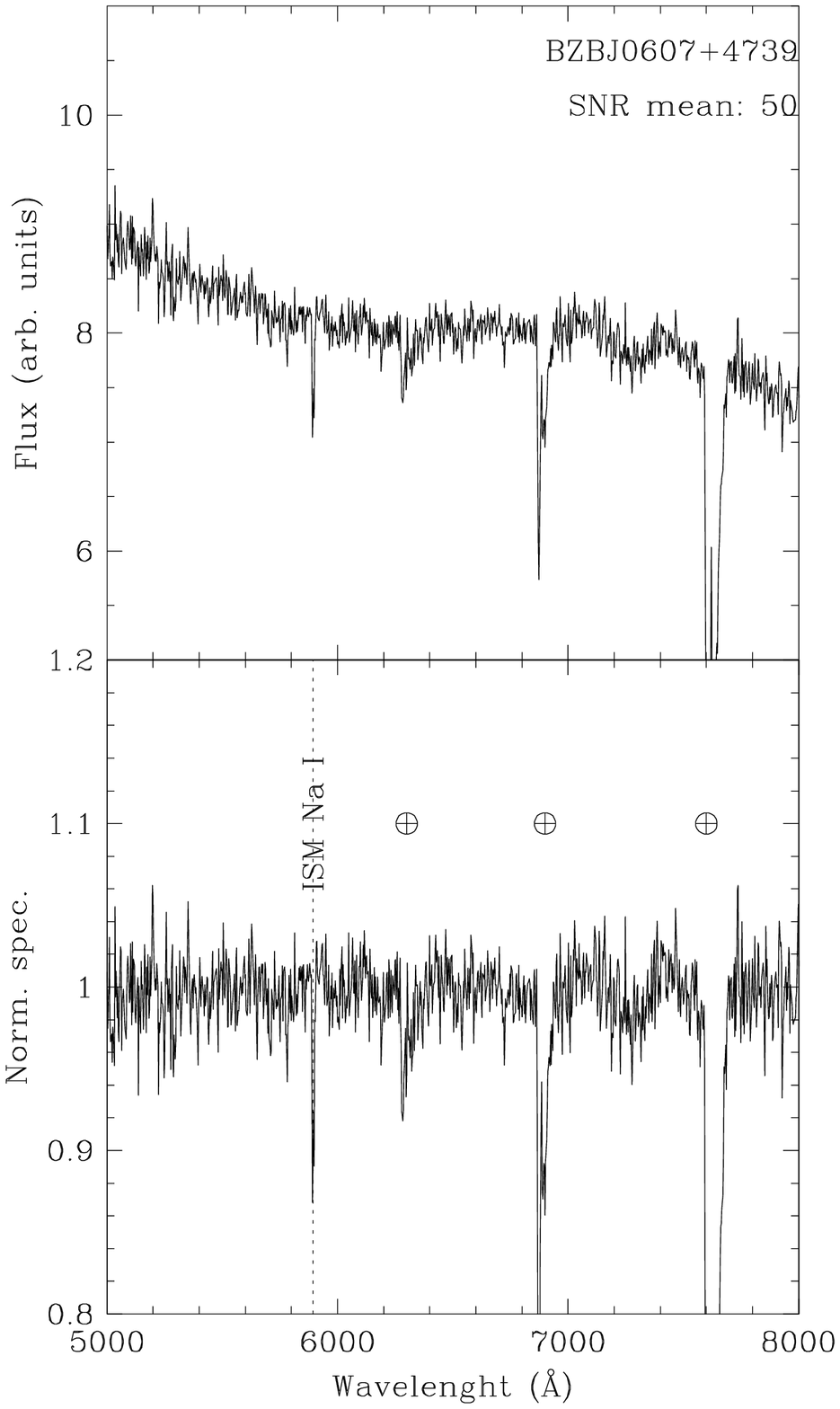}
\includegraphics[height=5.6cm,width=5.6cm,angle=0]{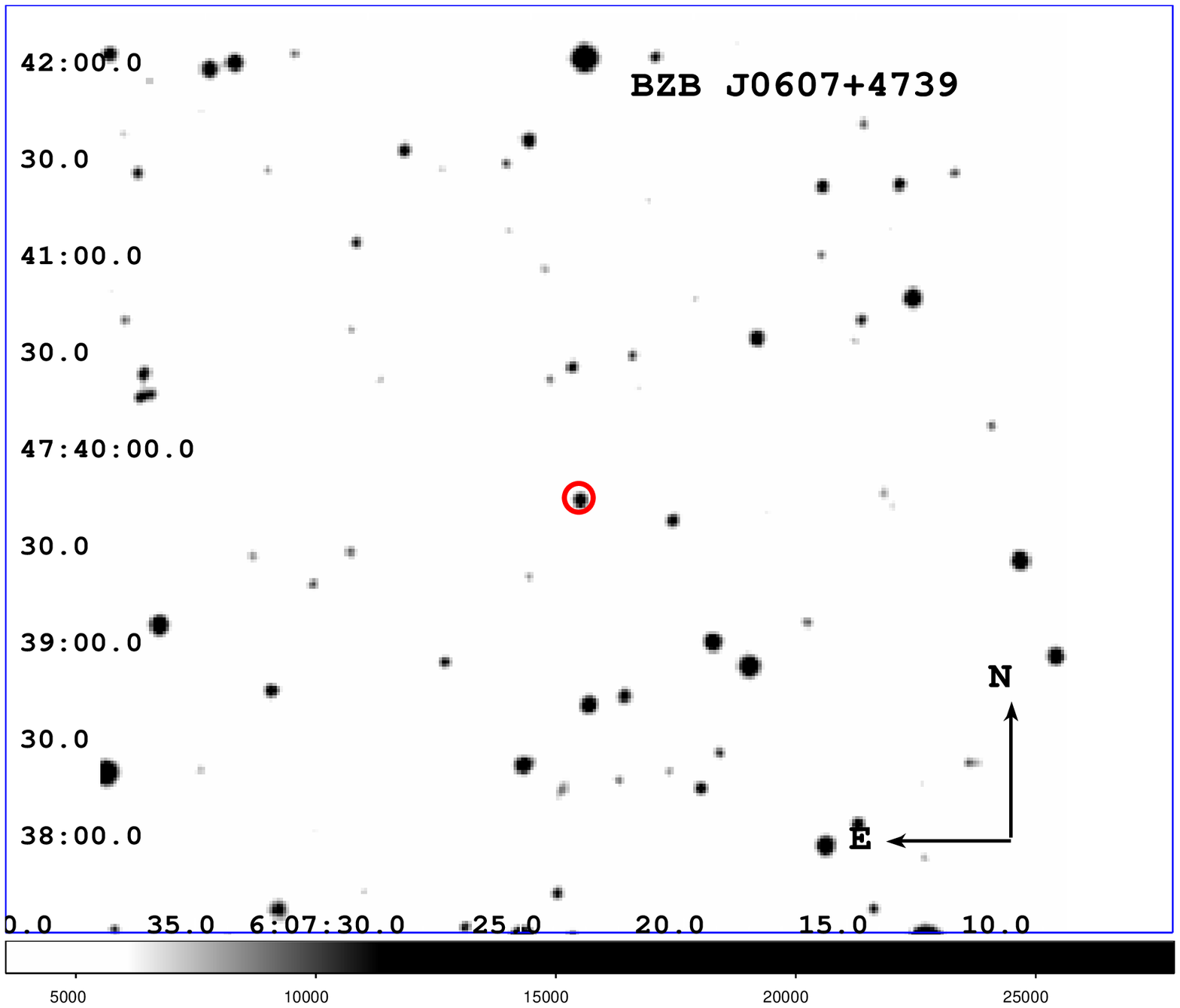}
\end{center}
\caption{Upper panel: optical spectra of the \fer\ BL Lac candidate
BZB J0607+4739 listed in the \bzcat\ v4.1.
Our observation clearly shows a featureless continuum and allows us to verify its classification.
The average S/N is also indicated.
Middle panel: normalized spectrum.
Lower panel: 5\arcmin\,x\,5\arcmin\ finding chart from the Digital Sky Survey (red filter). 
The source  is indicated by the red circle.}
\label{fig:J0607}
\end{figure}

\begin{figure}[]
\begin{center}
\includegraphics[height=12.2cm,width=12.2cm,angle=0]{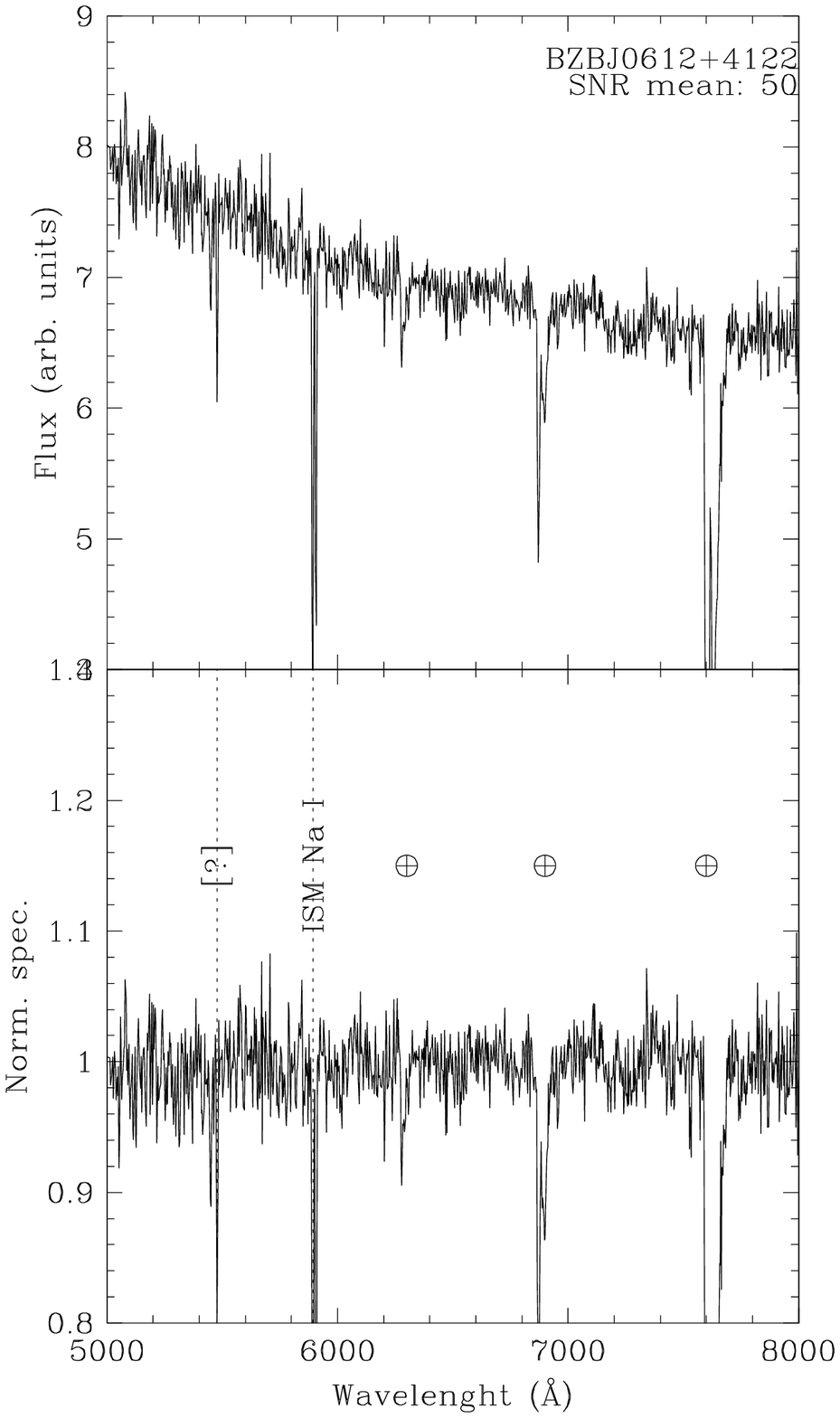}
\includegraphics[height=5.6cm,width=5.6cm,angle=0]{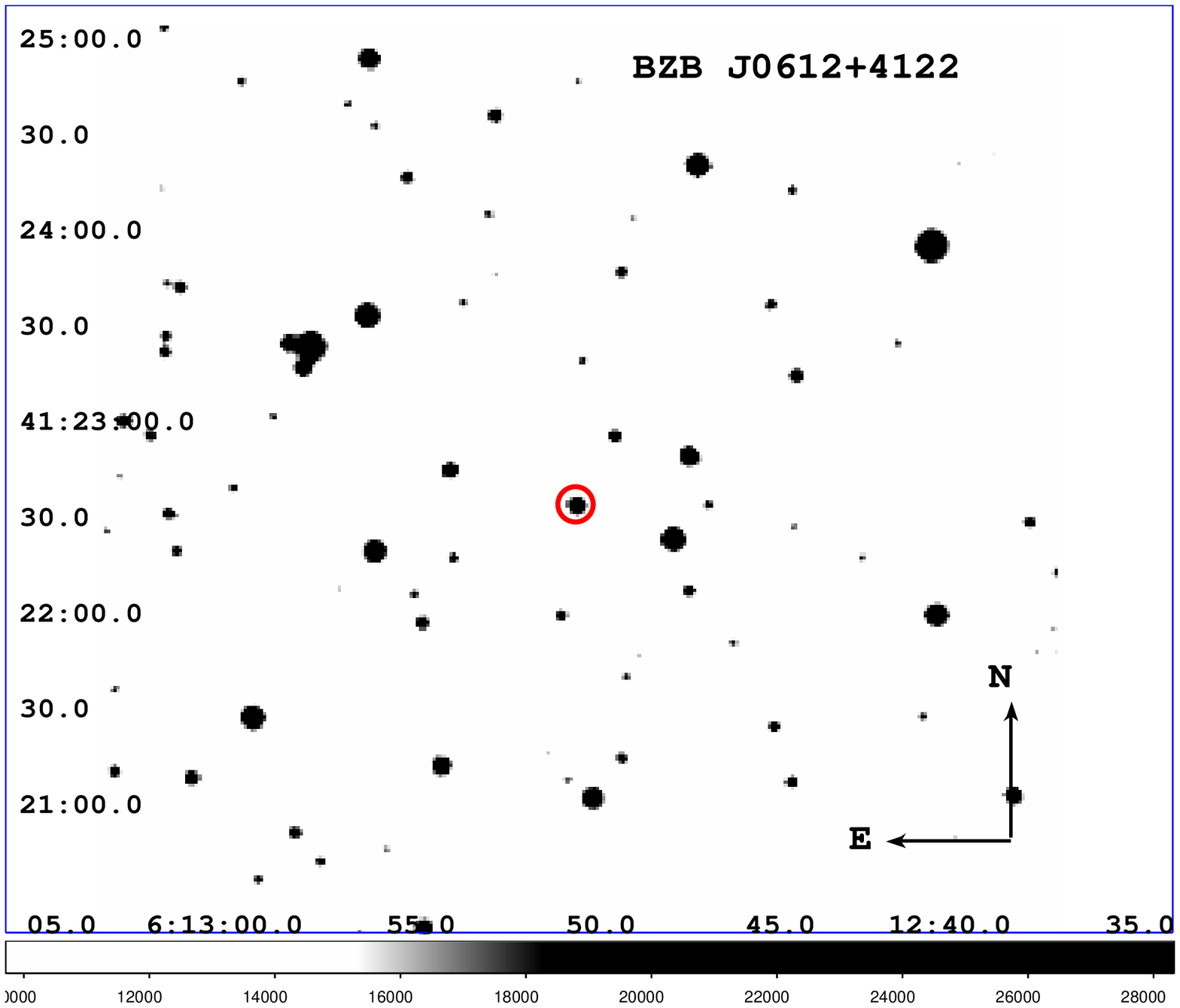}
\end{center}
\caption{Upper panel: optical spectra of the \fer\ BL Lac candidate
BZB J0612+4122 listed in the \bzcat\ v4.1.
Our observation clearly shows a featureless continuum and allows us to verify its classification.
The average S/N is also indicated.
Middle panel: normalized spectrum.
Lower panel: 5\arcmin\,x\,5\arcmin\ finding chart from the Digital Sky Survey (red filter). 
The source is indicated by the red circle.}
\label{fig:J0612}
\end{figure}

\begin{figure}[]
\begin{center}
\includegraphics[height=12.2cm,width=12.2cm,angle=0]{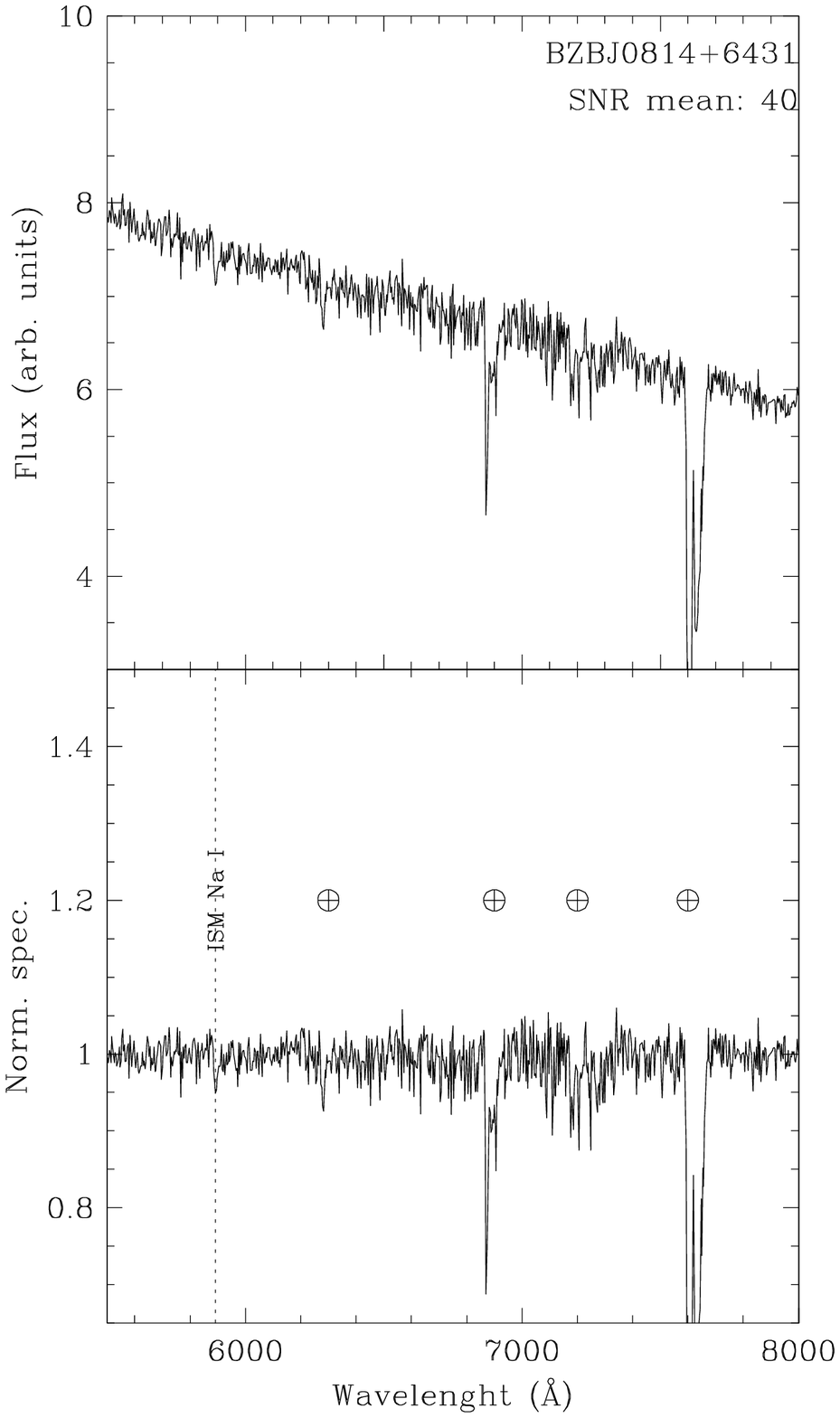}
\includegraphics[height=5.6cm,width=5.6cm,angle=0]{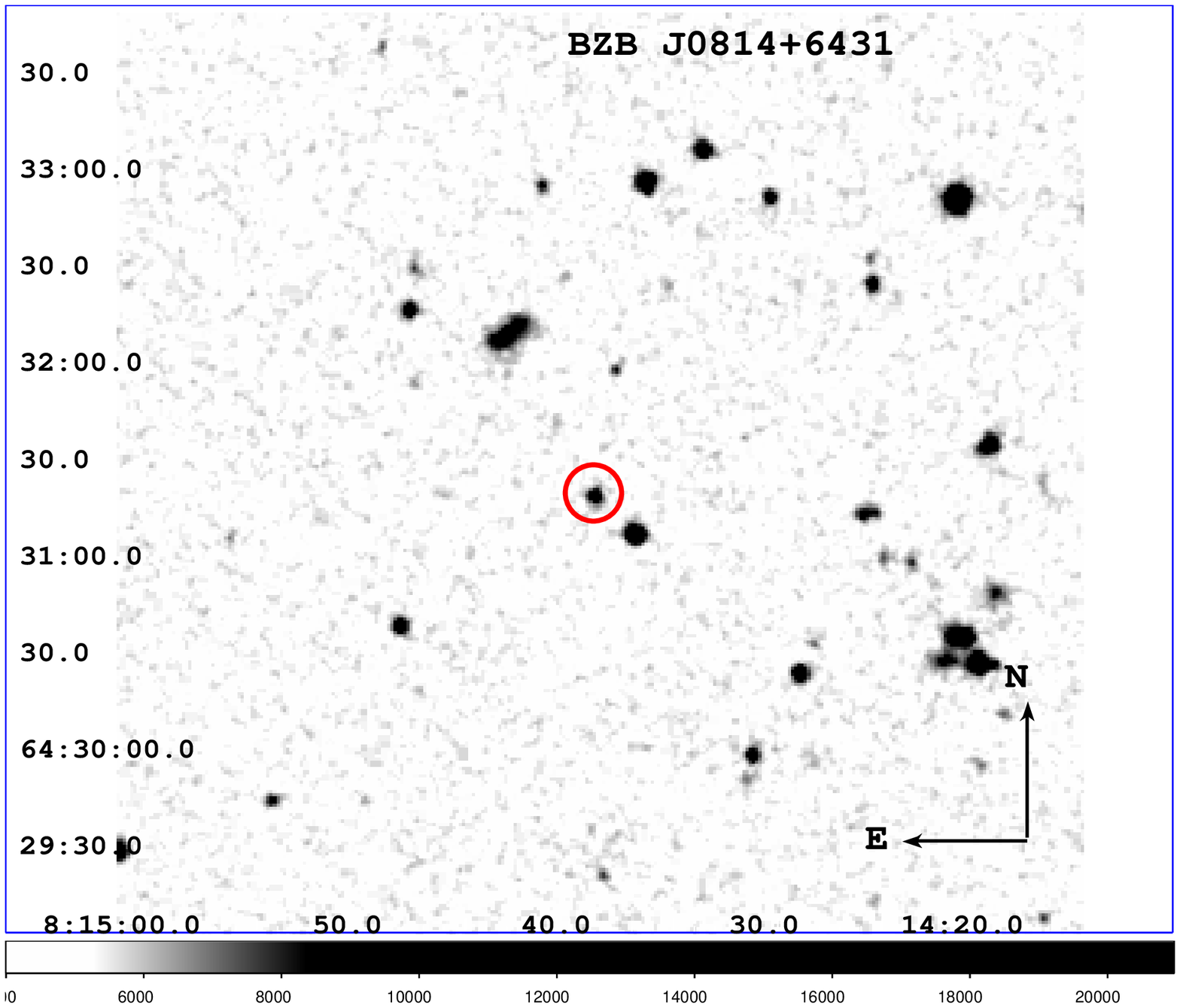}
\end{center}
\caption{Upper panel: optical spectra of the \fer\ BL Lac candidate
BZB J0814+6431 listed in the \bzcat\ v4.1.
Our observation clearly shows a featureless continuum and allows us to verify its classification.
The average S/N is also indicated.
Middle panel: normalized spectrum.
Lower panel: 5\arcmin\,x\,5\arcmin\ finding chart from the Digital Sky Survey (red filter). 
The source  is indicated by the red circle.}
\label{fig:J0814}
\end{figure}

\begin{figure}[]
\begin{center}
\includegraphics[height=12.2cm,width=12.2cm,angle=0]{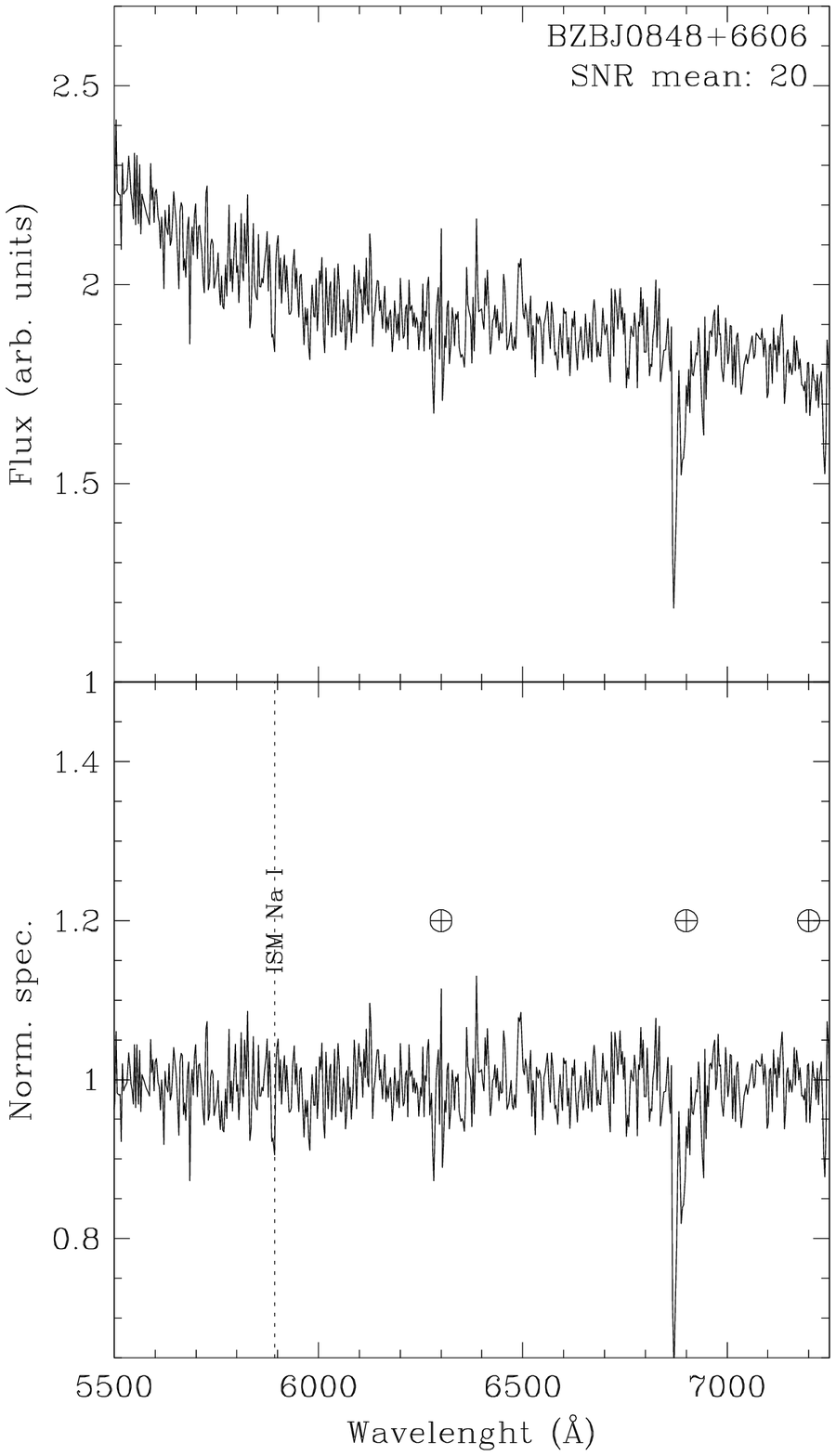}
\includegraphics[height=5.6cm,width=5.6cm,angle=0]{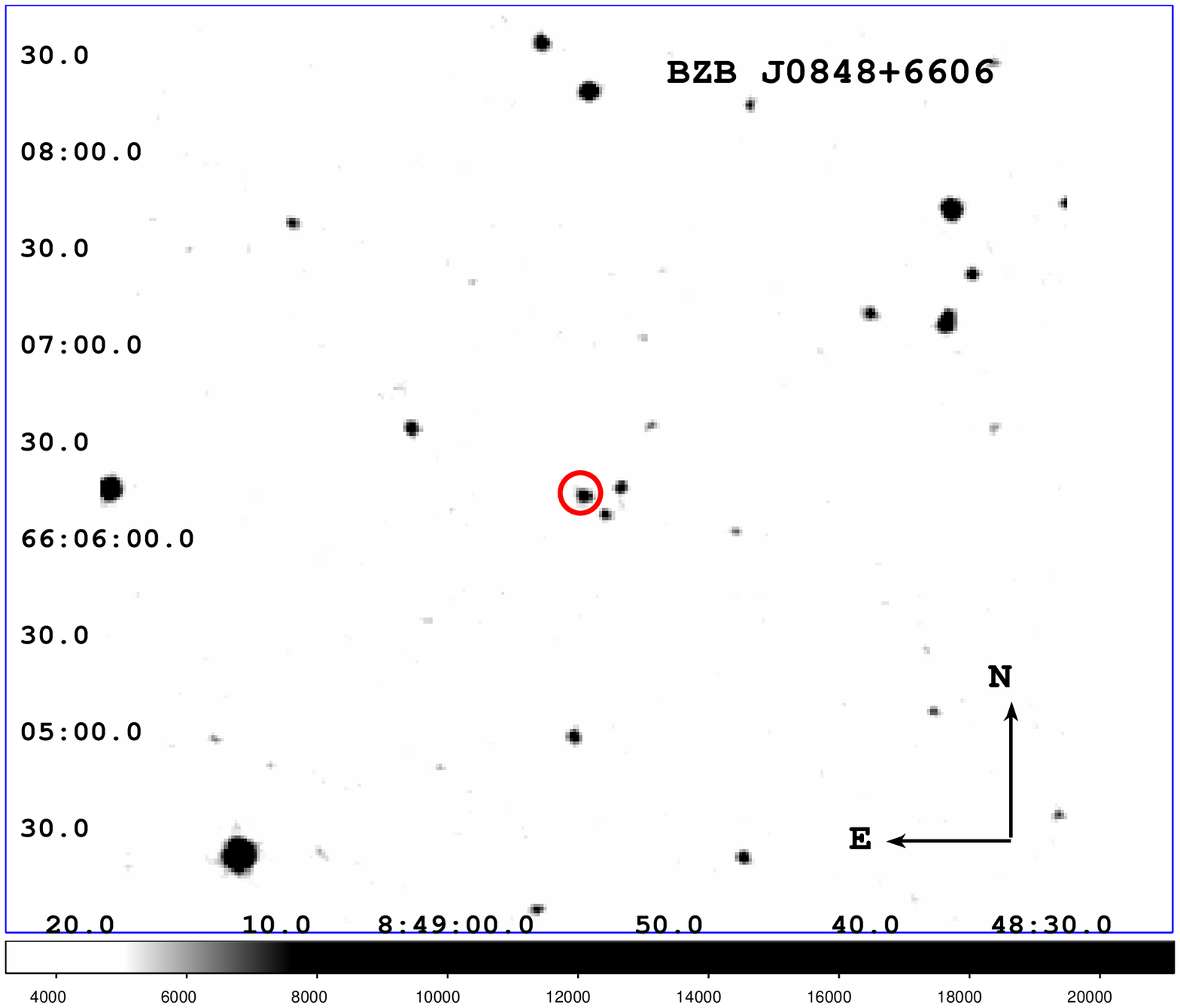}
\end{center}
\caption{Upper panel: optical spectra of the \fer\ BL Lac candidate
BZB J0848+6606 listed in the \bzcat\ v4.1.
Our observation clearly shows a featureless continuum and allows us to verify its classification.
The average S/N is also indicated.
Middle panel: normalized spectrum.
Lower panel: 5\arcmin\,x\,5\arcmin\ finding chart from the Digital Sky Survey (red filter). 
The source is indicated by the red circle.}
\label{fig:J0848}
\end{figure}

\begin{figure}[]
\begin{center}
\includegraphics[height=12.2cm,width=12.2cm,angle=0]{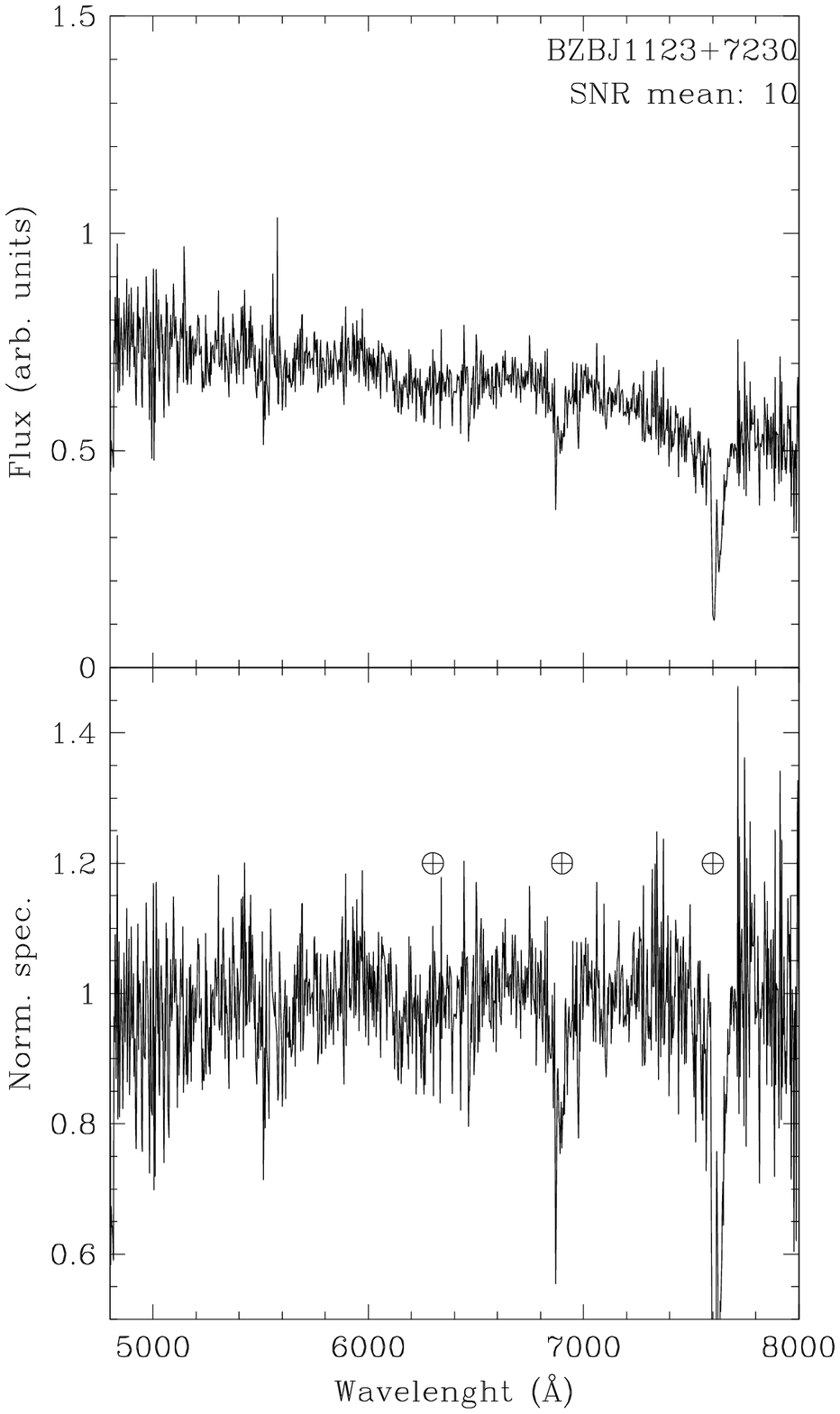}
\includegraphics[height=5.6cm,width=5.6cm,angle=0]{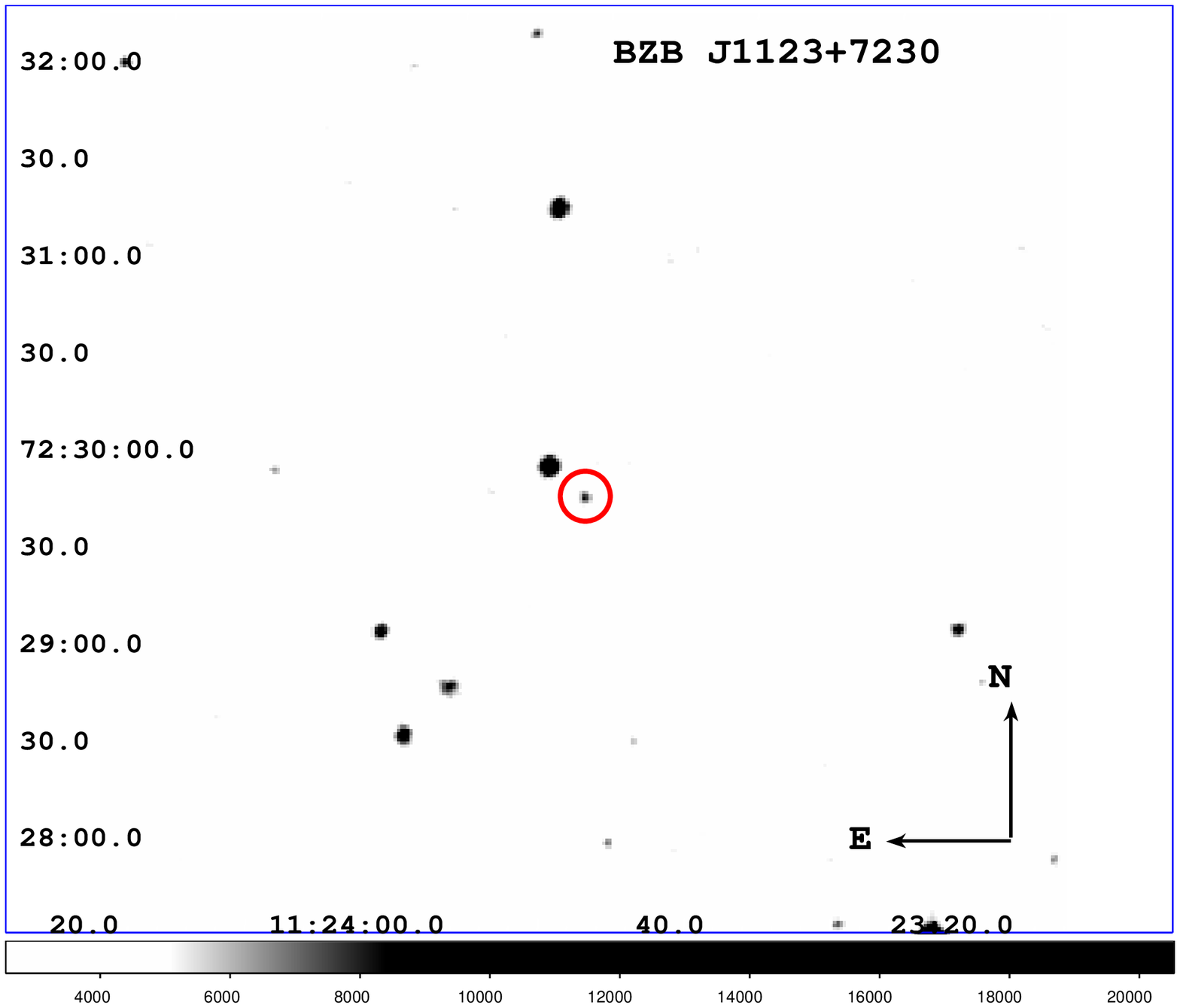}
\end{center}
\caption{Upper panel: optical spectra of the BL Lac candidate
BZB J1123+7230 listed in the \bzcat\ v4.1.
Our observation clearly shows a featureless continuum and allows us to verify its classification.
The average S/N is also indicated.
Middle panel: normalized spectrum.
Lower panel: 5\arcmin\,x\,5\arcmin\ finding chart from the Digital Sky Survey (red filter). 
The source is indicated by the red circle.}
\label{fig:J1123}
\end{figure}

\begin{figure}[]
\begin{center}
\includegraphics[height=12.2cm,width=12.2cm,angle=0]{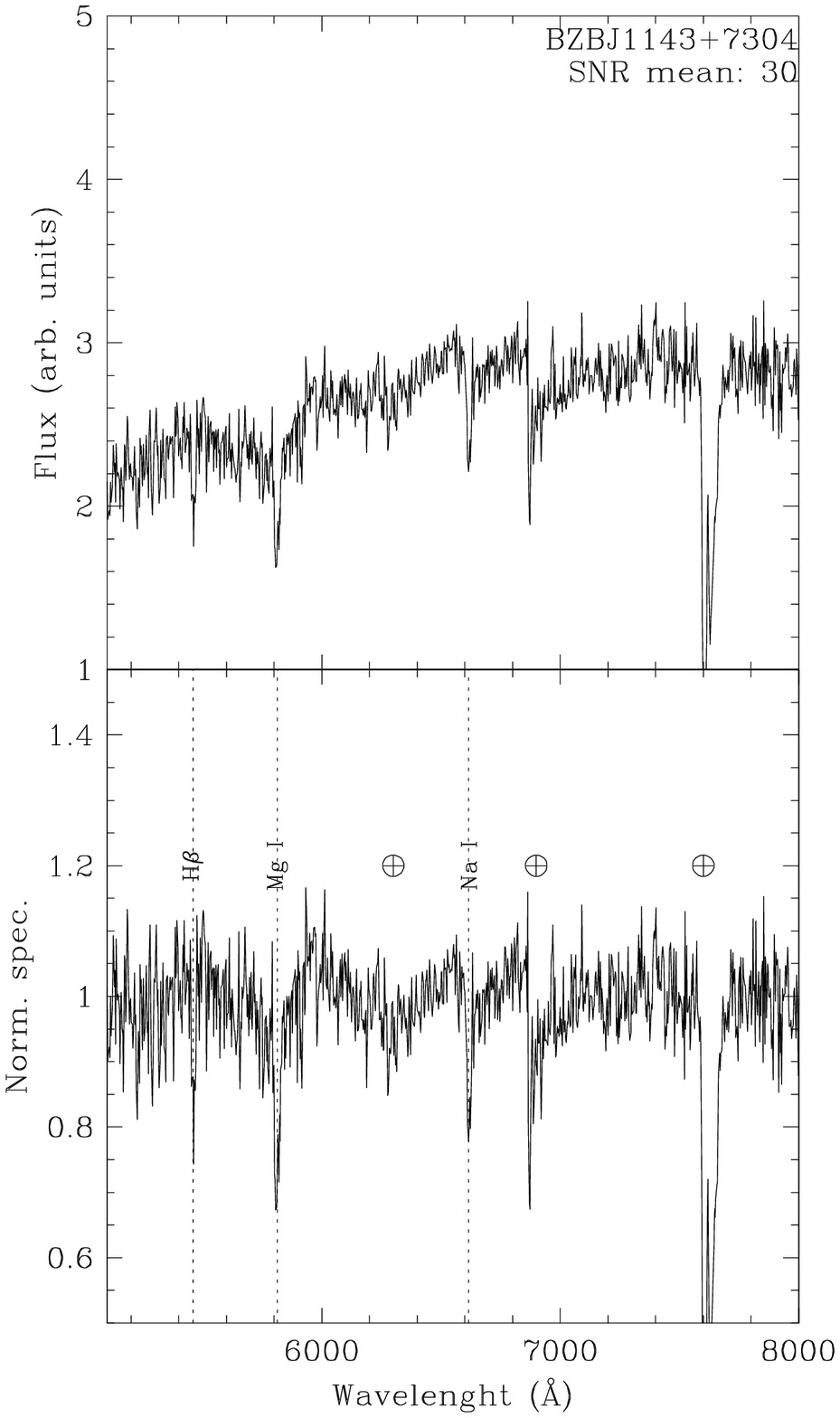}
\includegraphics[height=5.6cm,width=5.6cm,angle=0]{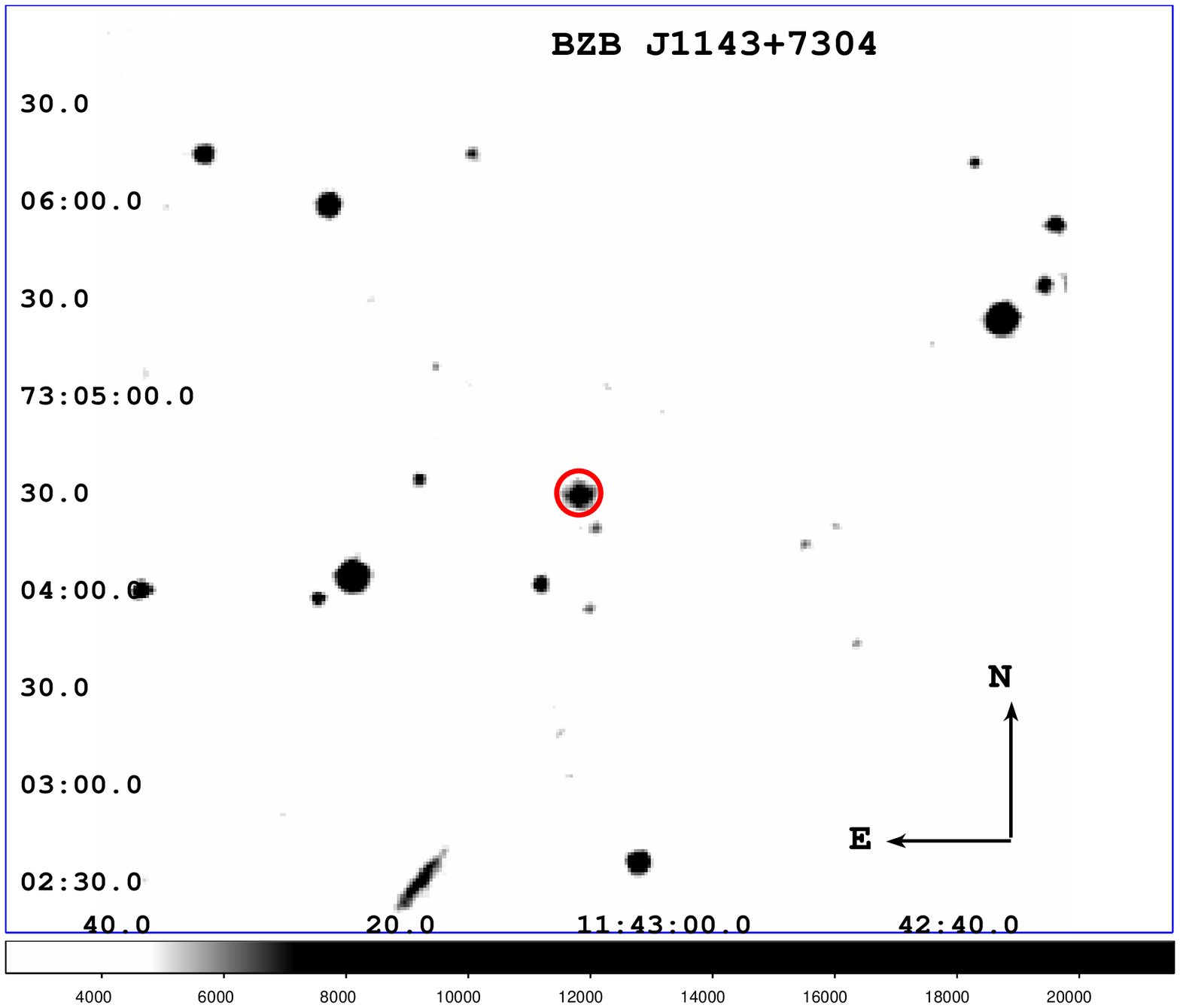}
\end{center}
\caption{Upper panel: optical spectra of the BL Lac candidate
BZB J1143+7304 listed in the \bzcat\ v4.1.
Our observation clearly shows a featureless continuum, however ,
the source spectrum appears to be that of a normal elliptical galaxy lying at redshift 0.123.
The average S/N is also indicated.
Middle panel: normalized spectrum.
Lower panel: 5\arcmin\,x\,5\arcmin\ finding chart from the Digital Sky Survey (red filter). 
The source  is indicated by the red circle.}
\label{fig:J1143}
\end{figure}

\begin{figure}[]
\begin{center}
\includegraphics[height=12.2cm,width=12.2cm,angle=0]{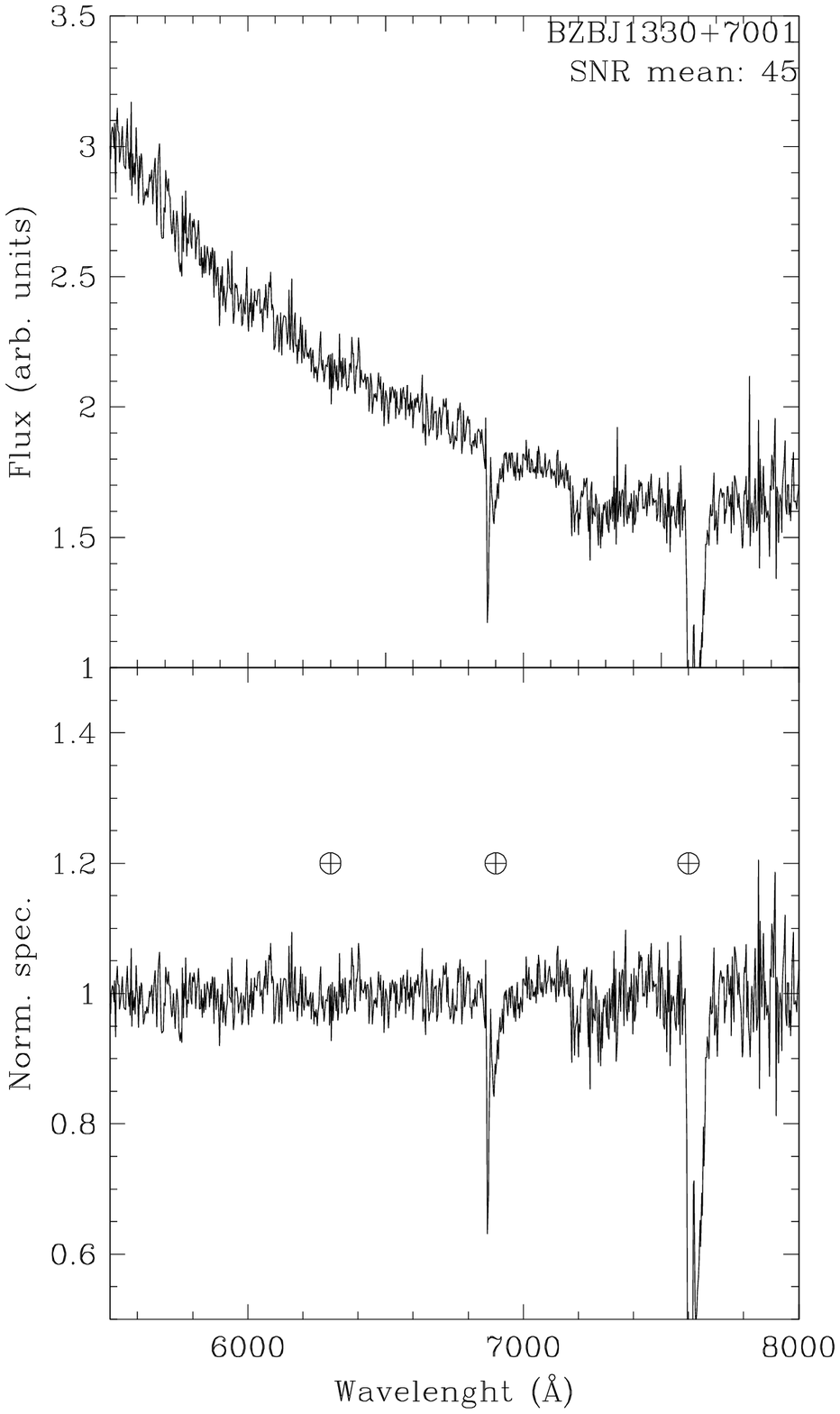}
\includegraphics[height=5.6cm,width=5.6cm,angle=0]{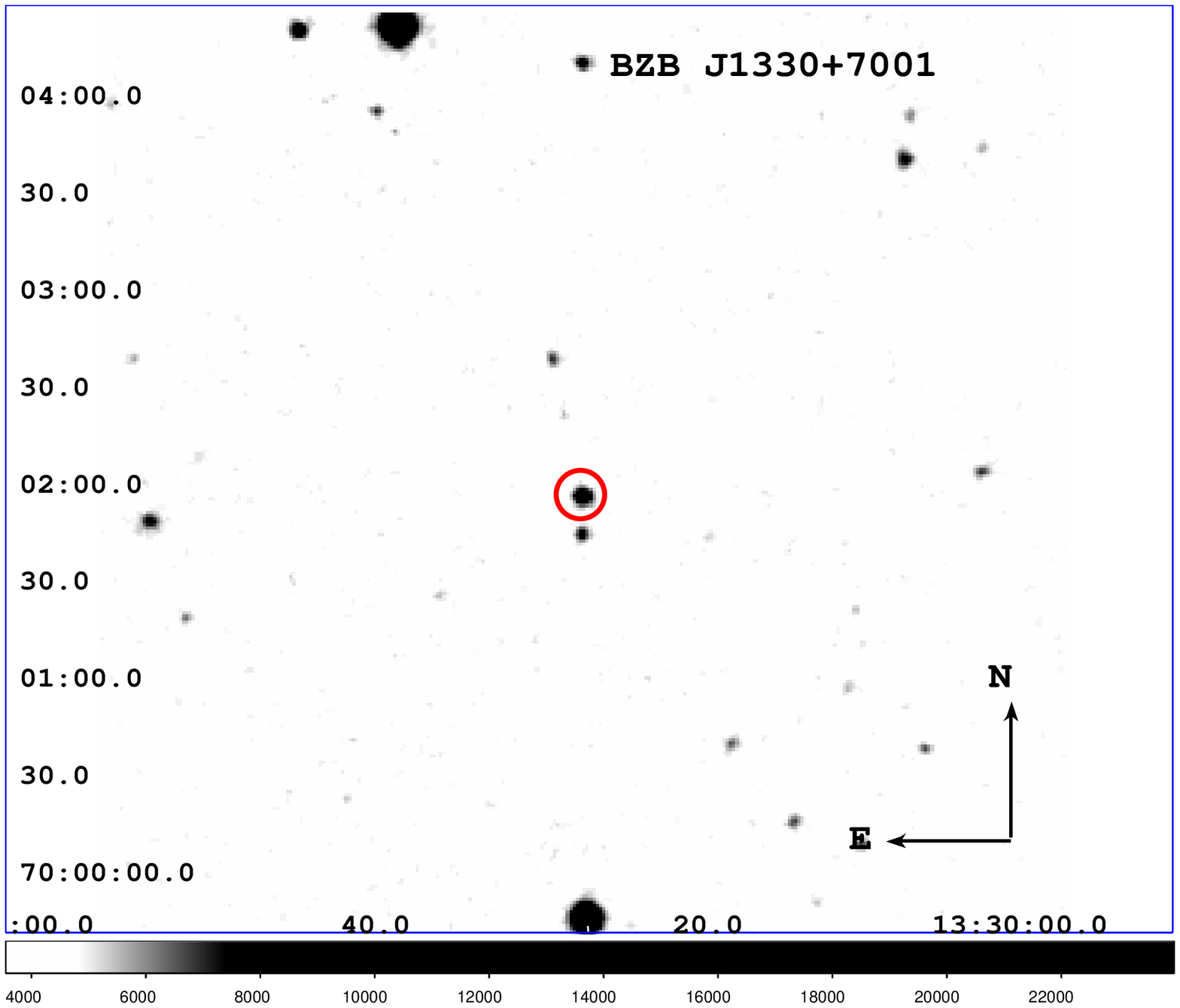}
\end{center}
\caption{Upper panel: optical spectra of the \fer\ BL Lac candidate
BZB J1330+7001 listed in the \bzcat\ v4.1.
Our observation clearly shows a featureless continuum and allows us to verify its classification.
The average S/N is also indicated.
Middle panel: normalized spectrum.
Lower panel: 5\arcmin\,x\,5\arcmin\ finding chart from the Digital Sky Survey (red filter). 
The source is indicated by the red circle.}
\label{fig:J1330}
\end{figure}

\begin{figure}[]
\begin{center}
\includegraphics[height=12.2cm,width=12.2cm,angle=0]{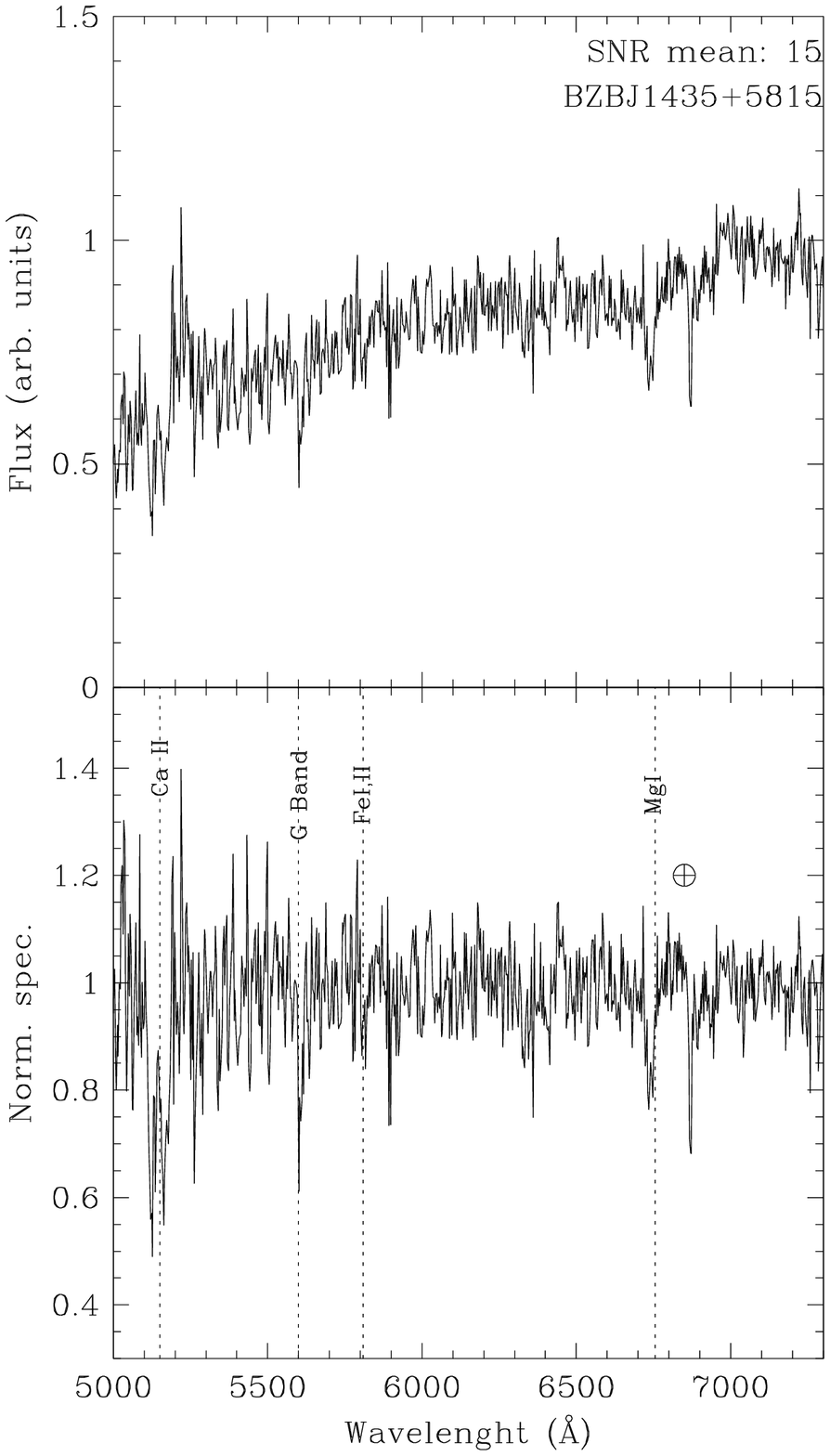}
\includegraphics[height=5.6cm,width=5.6cm,angle=0]{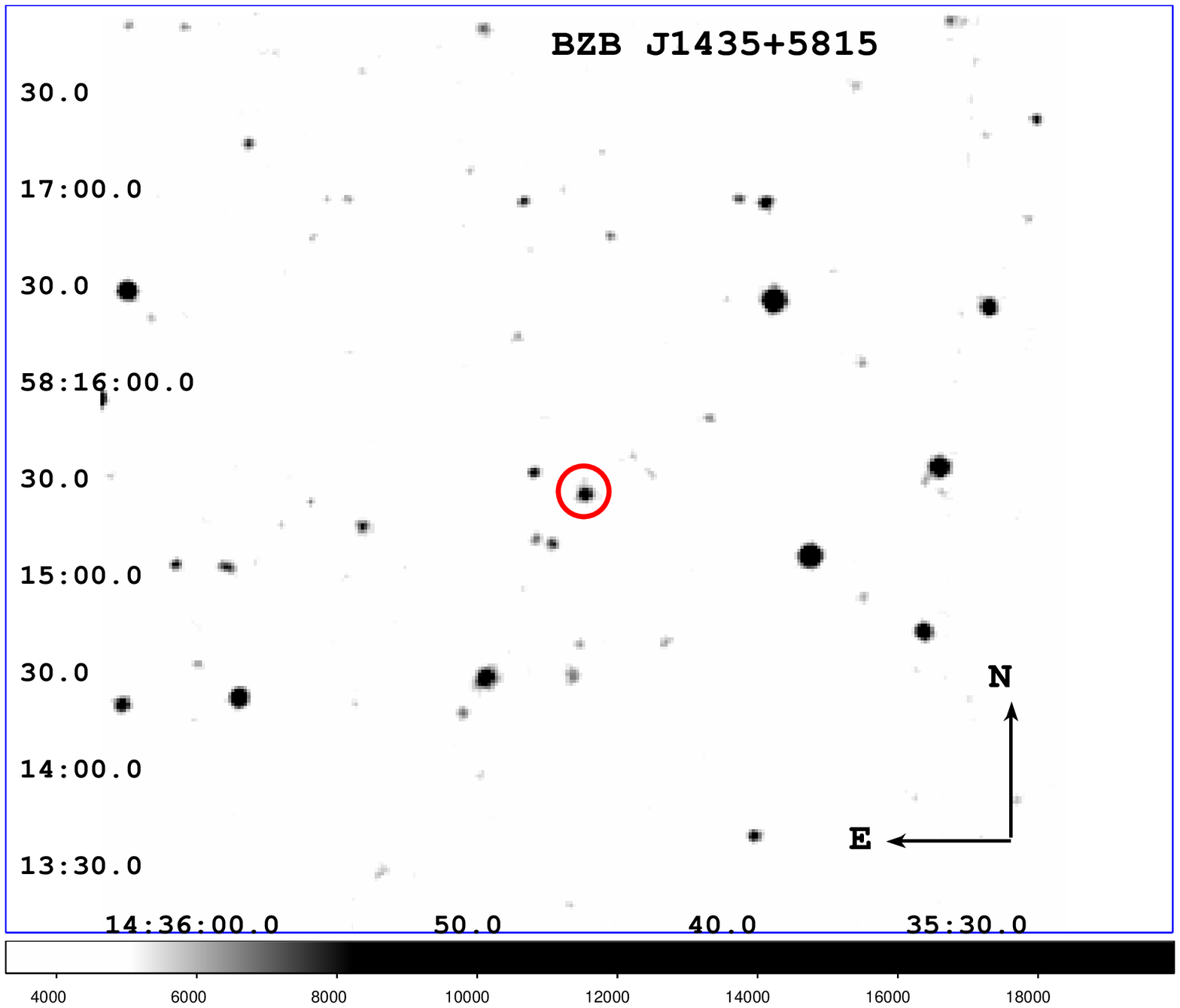}
\end{center}
\caption{Upper panel: optical spectra of the BL Lac candidate
BZB J1435+5815 listed in the \bzcat\ v4.1.
Our observation clearly shows a featureless continuum, however, 
the source spectrum appears to be that of a normal elliptical galaxy lying at redshift 0.299.
The average S/N is also indicated.
Middle panel: normalized spectrum.
Lower panel: 5\arcmin\,x\,5\arcmin\ finding chart from the Digital Sky Survey (red filter). 
The source  is indicated by the red circle.}
\label{fig:J1435}
\end{figure}

\begin{figure}[]
\begin{center}
\includegraphics[height=12.2cm,width=12.2cm,angle=0]{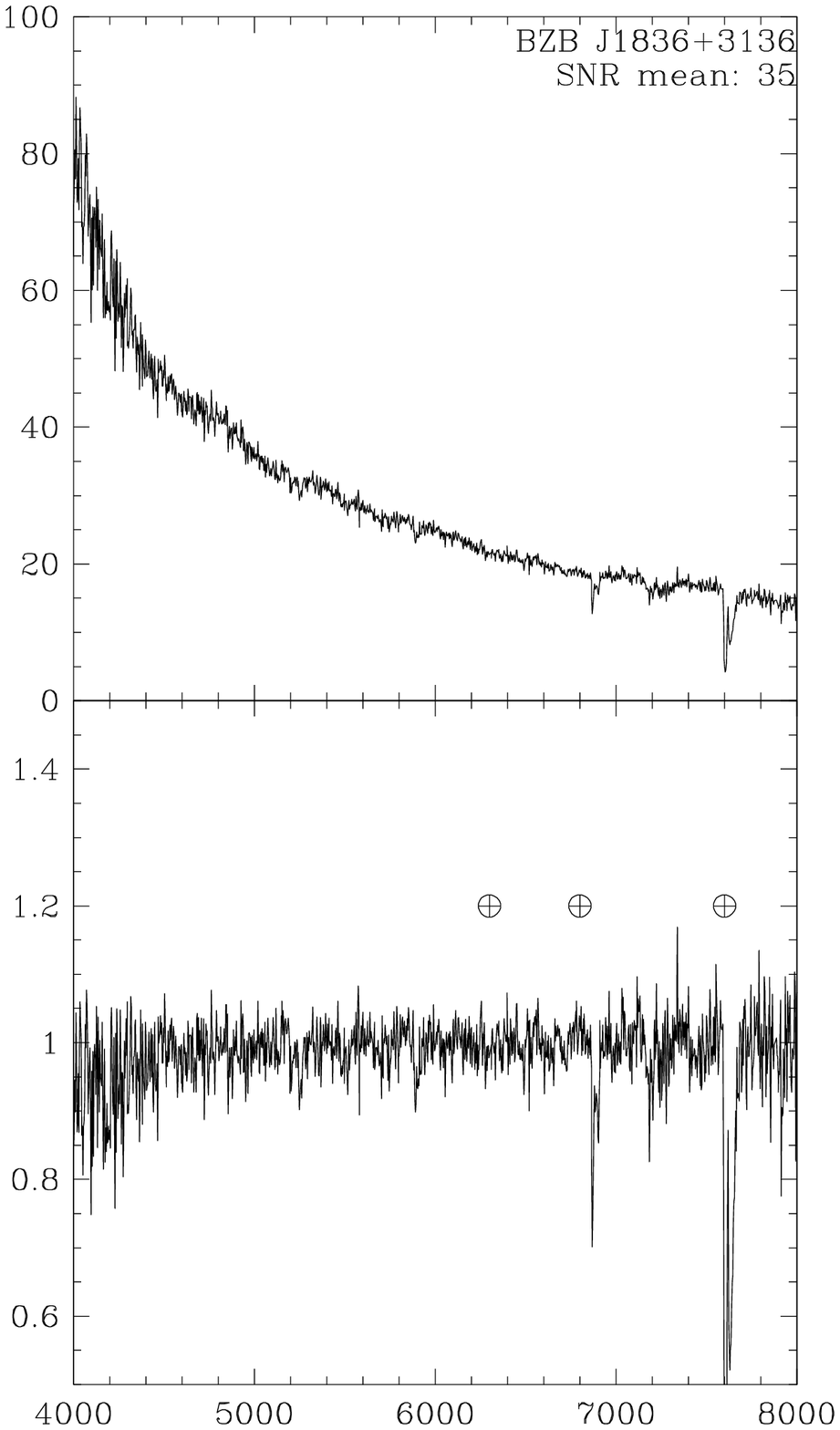}
\includegraphics[height=5.6cm,width=5.6cm,angle=0]{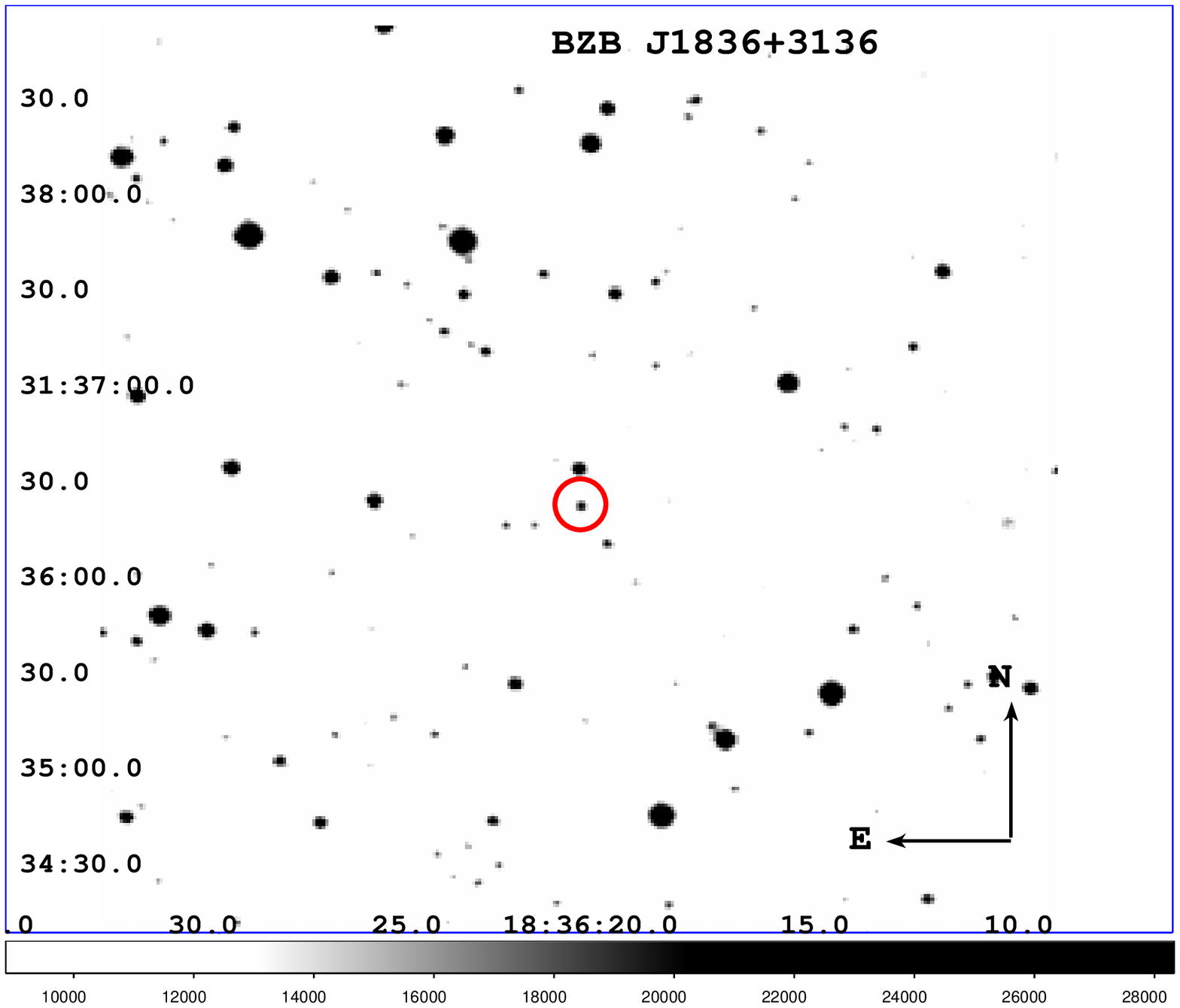}
\end{center}
\caption{Upper panel: optical spectra of the \fer\ BL Lac candidate
BZB J1836+3136 listed in the \bzcat\ v4.1.
Our observation clearly shows a featureless continuum and allows us to verify its classification.
The average S/N is also indicated.
Middle panel: normalized spectrum.
Lower panel: 5\arcmin\,x\,5\arcmin\ finding chart from the Digital Sky Survey (red filter). 
The source is indicated by the red circle.}
\label{fig:J1836}
\end{figure}

\begin{figure}[]
\begin{center}
\includegraphics[height=12.2cm,width=12.2cm,angle=0]{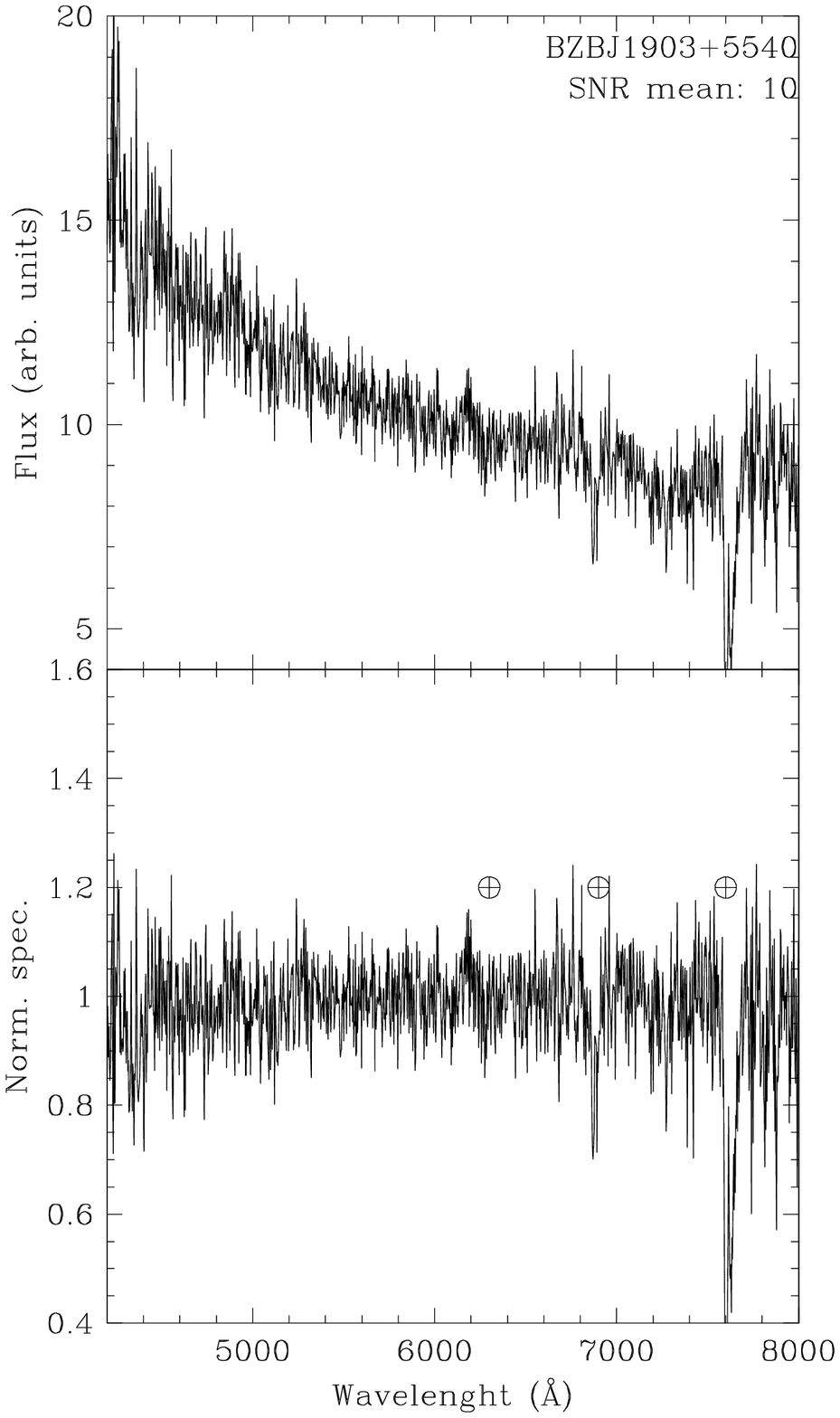}
\includegraphics[height=5.6cm,width=5.6cm,angle=0]{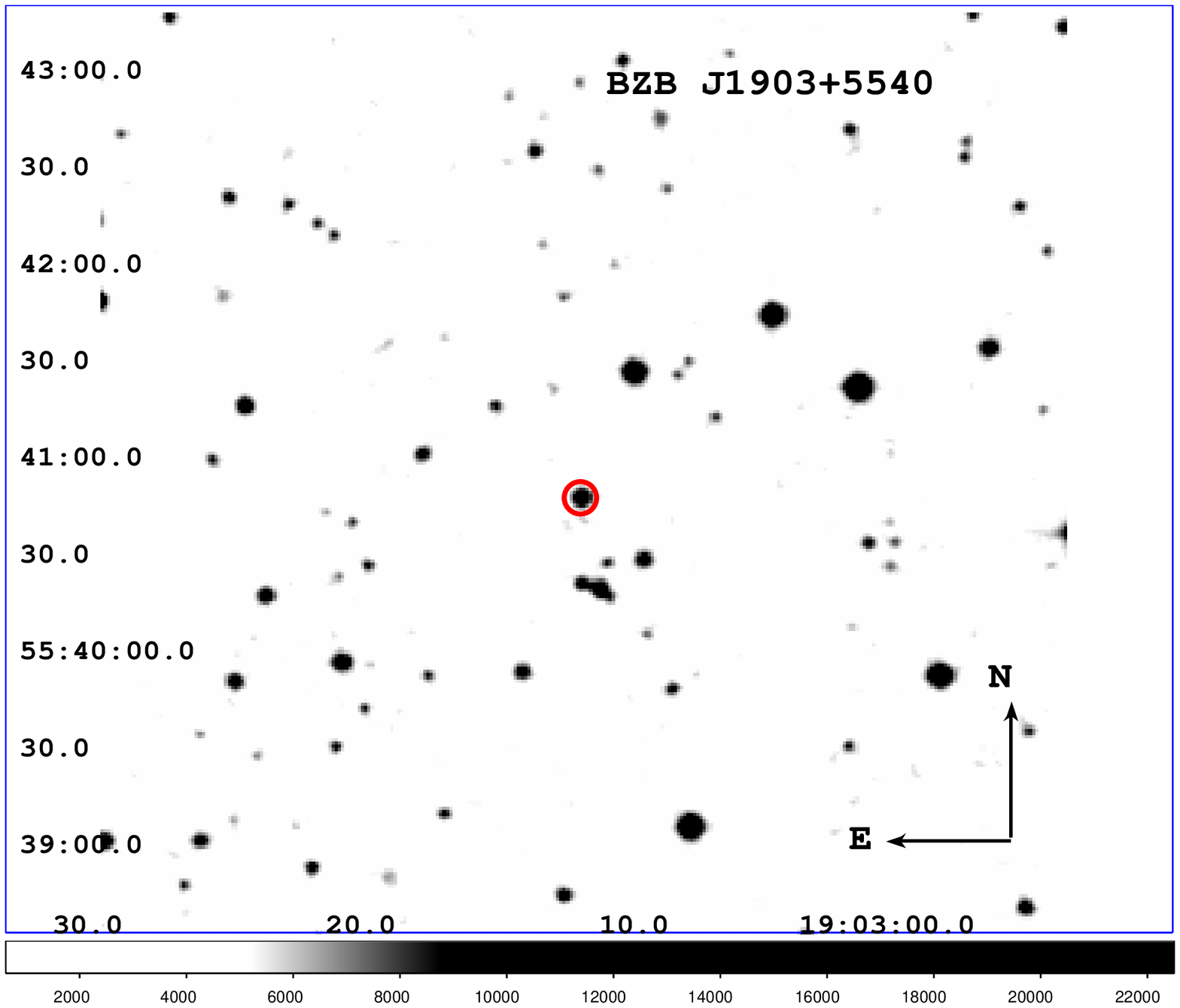}
\end{center}
\caption{Upper panel: optical spectra of the \fer\ BL Lac candidate
BZB J1903+5540 listed in the \bzcat\ v4.1.
Our observation clearly shows a featureless continuum and allows us to verify its classification.
The average S/N is also indicated.
Middle panel: normalized spectrum.
Lower panel: 5\arcmin\,x\,5\arcmin\ finding chart from the Digital Sky Survey (red filter). 
The source  is indicated by the red circle.}
\label{fig:J1903}
\end{figure}

\begin{figure}[]
\begin{center}
\includegraphics[height=12.2cm,width=12.2cm,angle=0]{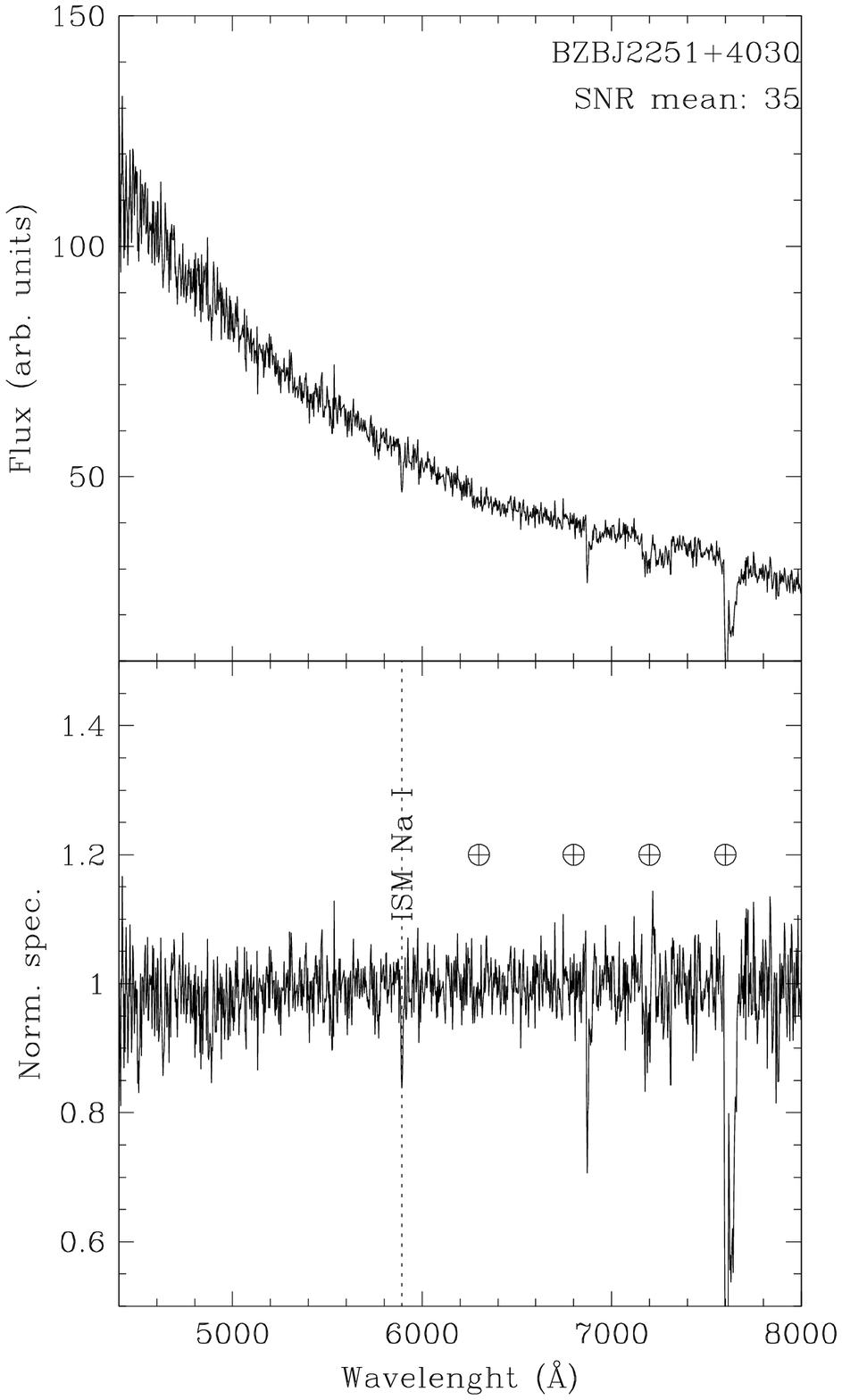}
\includegraphics[height=5.6cm,width=5.6cm,angle=0]{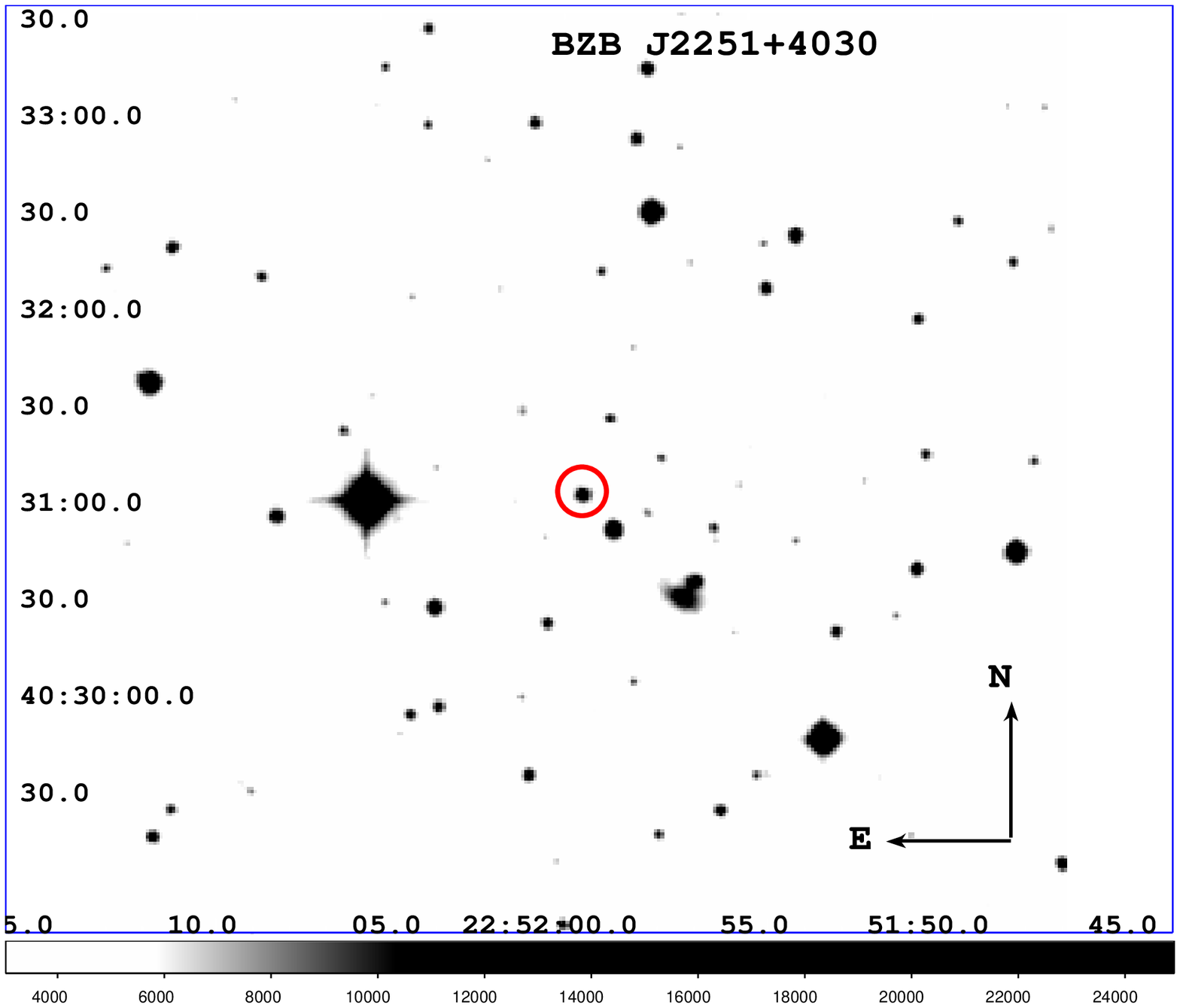}
\end{center}
\caption{Upper panel: optical spectra of the \fer\ BL Lac candidate
BZB J2251+4030 listed in the \bzcat\ v4.1.
Our observation clearly shows a featureless continuum and allows us to verify its classification.
The average S/N is also indicated.
Middle panel: normalized spectrum.
Lower panel: 5\arcmin\,x\,5\arcmin\ finding chart from the Digital Sky Survey (red filter). 
The source is indicated by the red circle.}
\label{fig:J2251}
\end{figure}

\begin{figure}[]
\begin{center}
\includegraphics[height=12.2cm,width=12.2cm,angle=0]{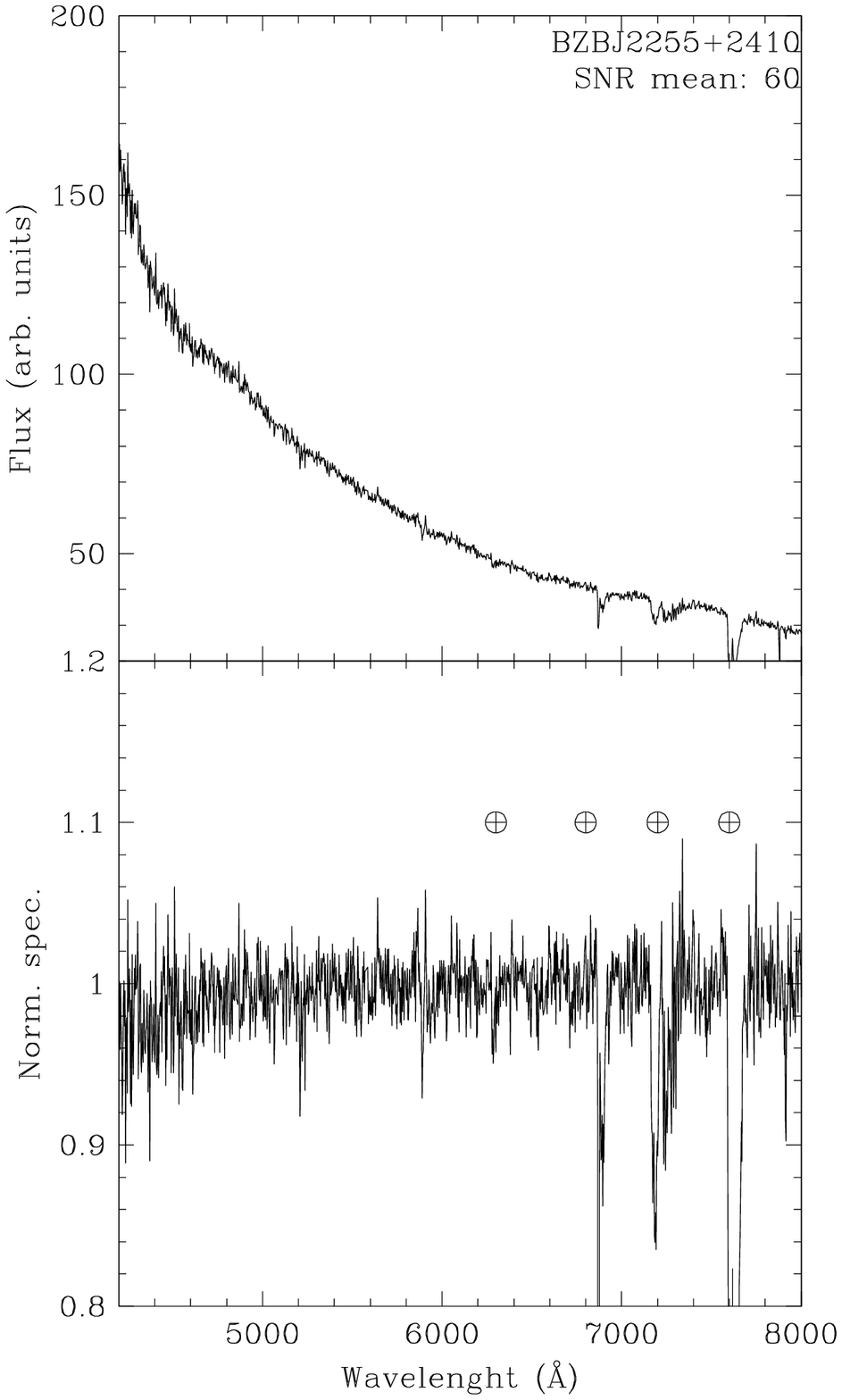}
\includegraphics[height=5.6cm,width=5.6cm,angle=0]{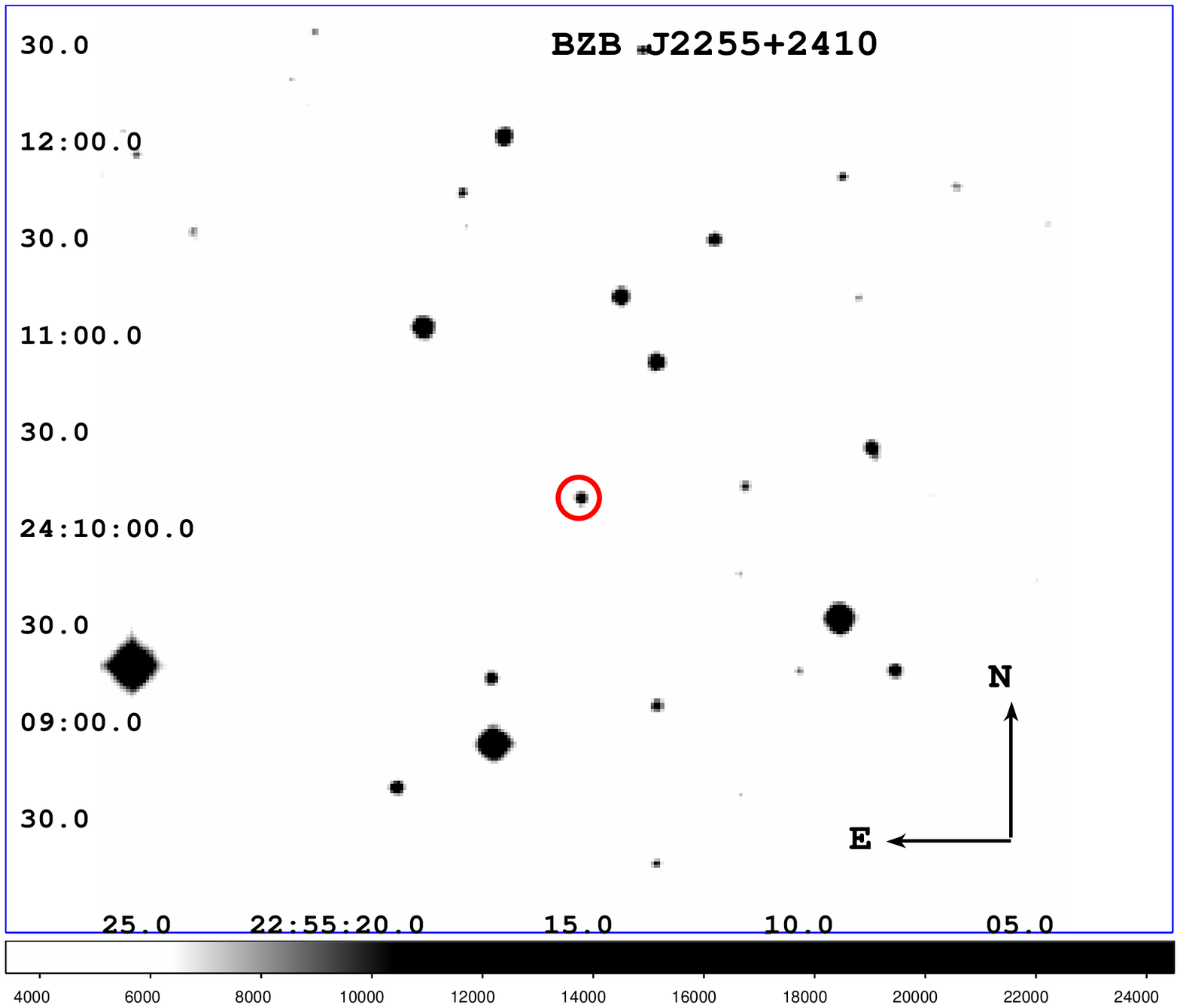}
\end{center}
\caption{Upper panel: optical spectra of the \fer\ BL Lac candidate
BZB J2255+2410 listed in the \bzcat\ v4.1.
Our observation clearly shows a featureless continuum and allows us to verify its classification.
The average S/N is also indicated.
Middle panel: normalized spectrum.
Lower panel: 5\arcmin\,x\,5\arcmin\ finding chart from the Digital Sky Survey (red filter). 
The source is indicated by the red circle.}
\label{fig:J2255}
\end{figure}

\begin{figure}[]
\begin{center}
\includegraphics[height=12.2cm,width=12.2cm,angle=0]{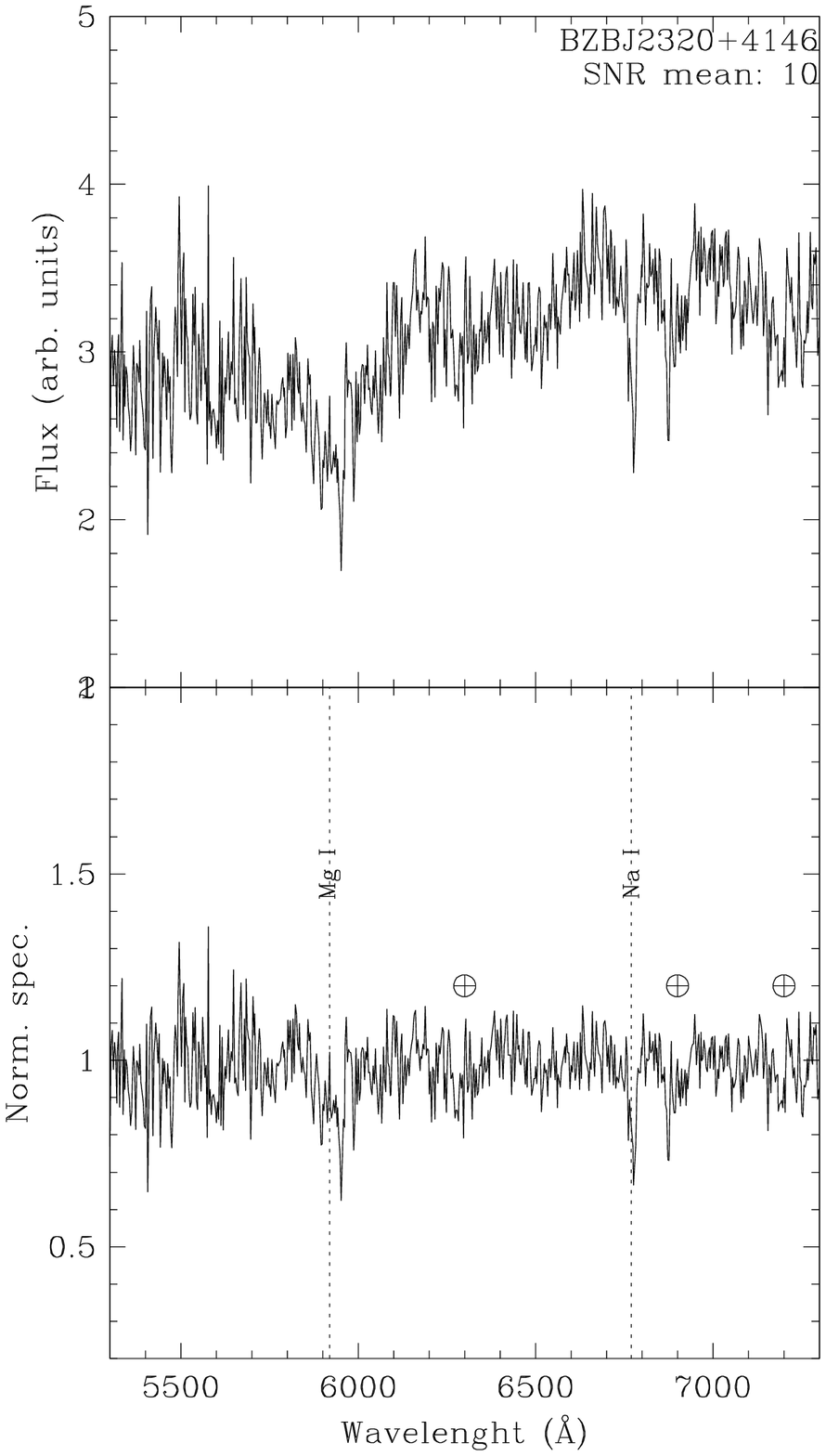}
\includegraphics[height=5.6cm,width=5.6cm,angle=0]{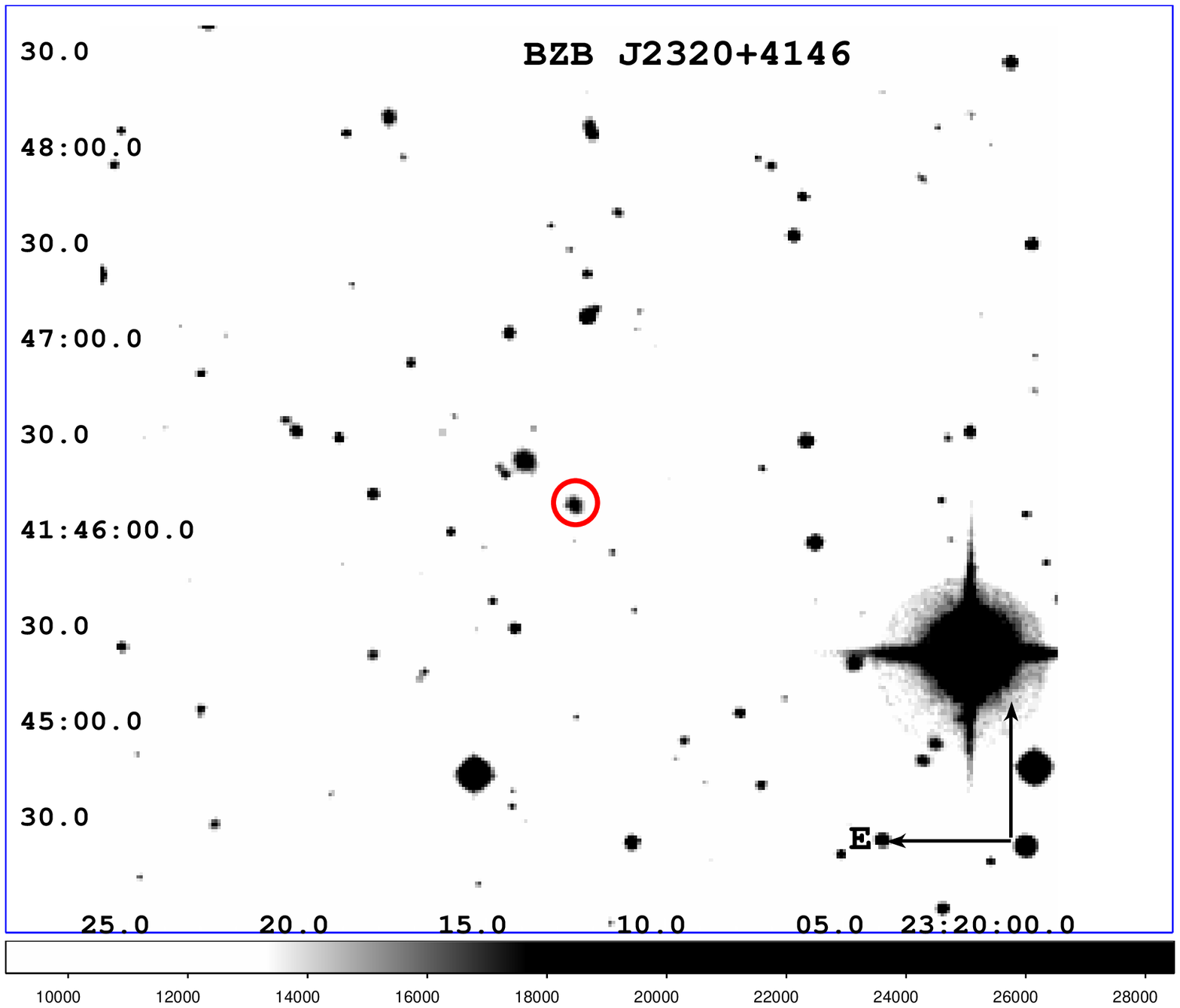}
\end{center}
\caption{Upper panel: optical spectra of the BL Lac candidate
BZB J2320+4146 listed in the \bzcat\ v4.1.
Our observation clearly shows two absorption features allowing us to verify its classification
and its redshfit estimate.
The average S/N is also indicated.
Middle panel: normalized spectrum.
Lower panel: 5\arcmin\,x\,5\arcmin\ finding chart from the Digital Sky Survey (red filter). 
The source is indicated by the red circle.}
\label{fig:J2320}
\end{figure}

\begin{figure}[]
\begin{center}
\includegraphics[height=12.2cm,width=12.2cm,angle=0]{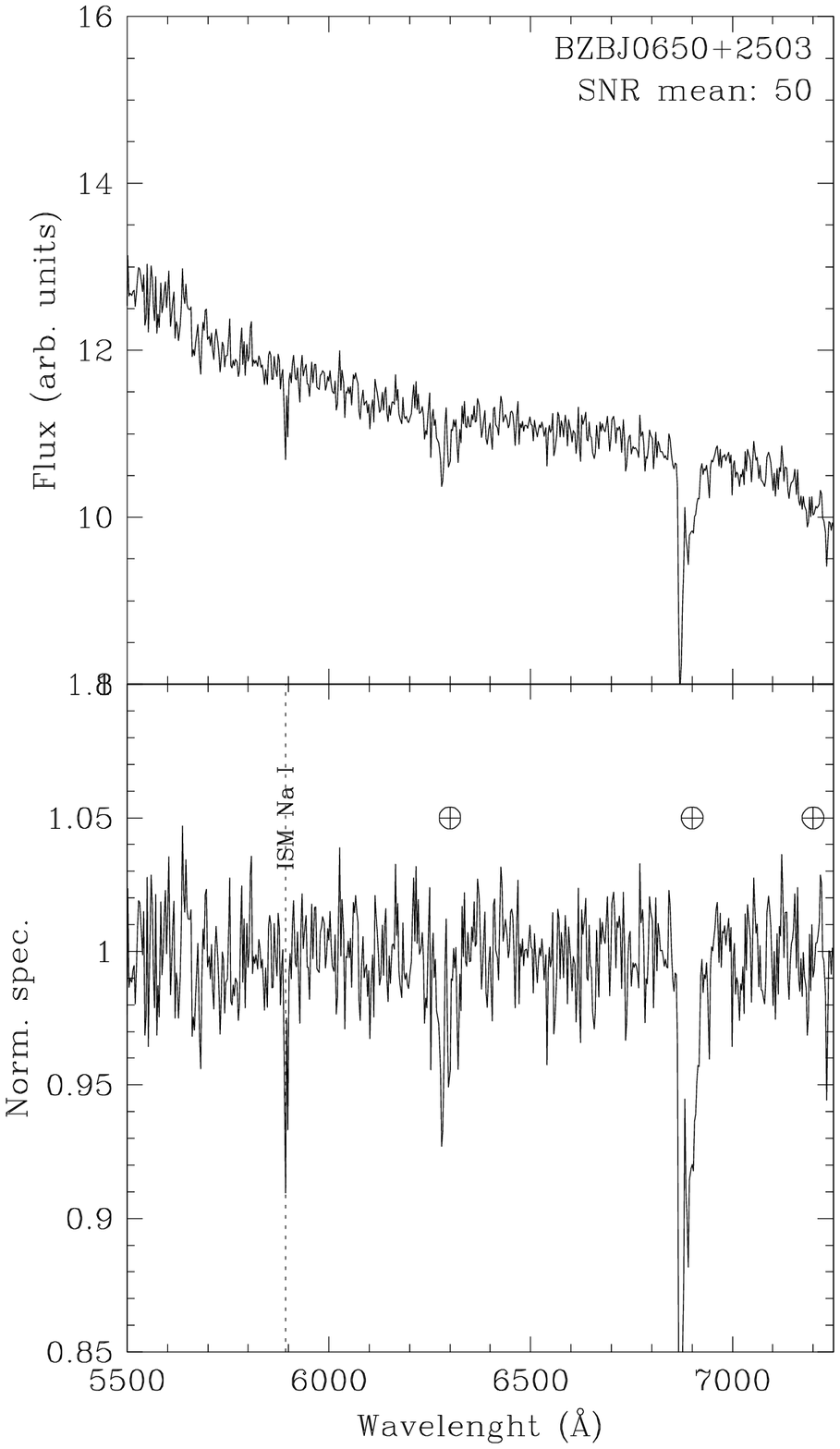}
\includegraphics[height=5.6cm,width=5.6cm,angle=0]{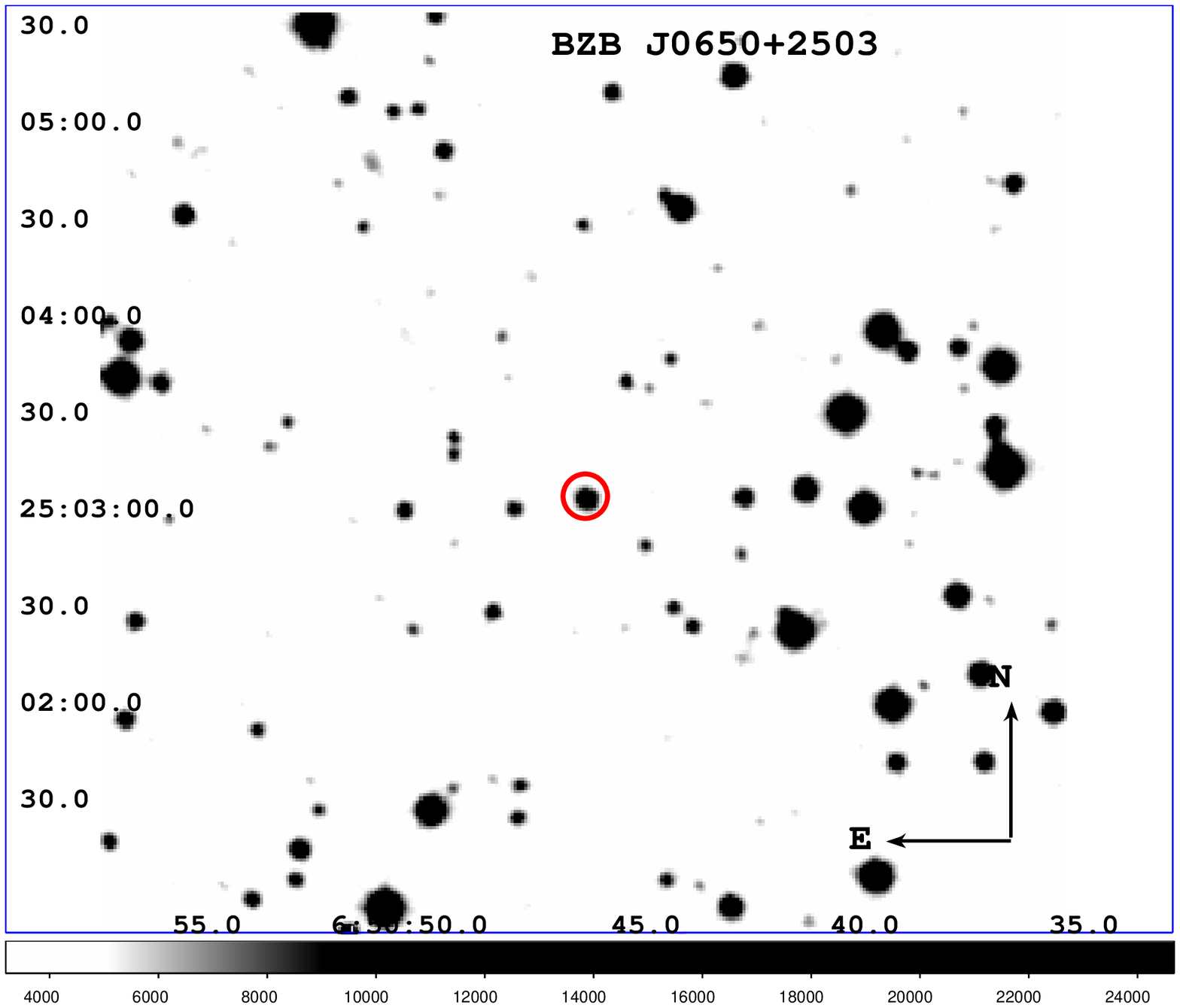}
\end{center}
\caption{Upper panel: optical spectra of the \fer\ source: BZB J0650+2503 listed in the \bzcat\ v4.1.
Our observation clearly shows a featureless continuum and allows us to confirm its classification, but
unfortunately, we were unable to confirm the redshift estimate reported in the \bzcat.
The average S/N is also indicated.
Middle panel: normalized spectrum.
Lower panel: 5\arcmin\,x\,5\arcmin\ finding chart from the Digital Sky Survey (red filter). 
The source  is indicated by the red circle.}
\label{fig:J0650}
\end{figure}

\begin{figure}[]
\begin{center}
\includegraphics[height=12.2cm,width=12.2cm,angle=0]{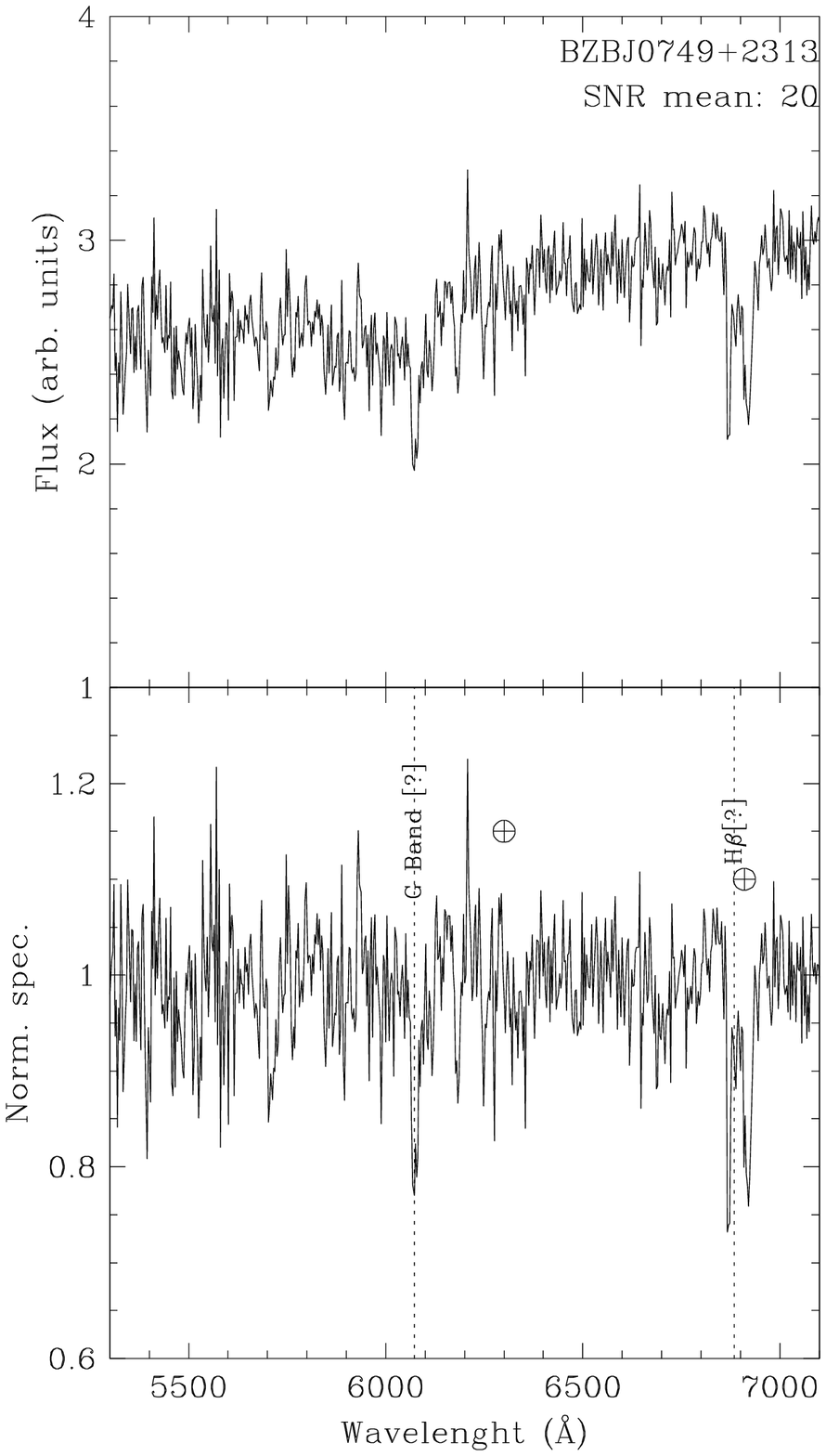}
\includegraphics[height=5.6cm,width=5.6cm,angle=0]{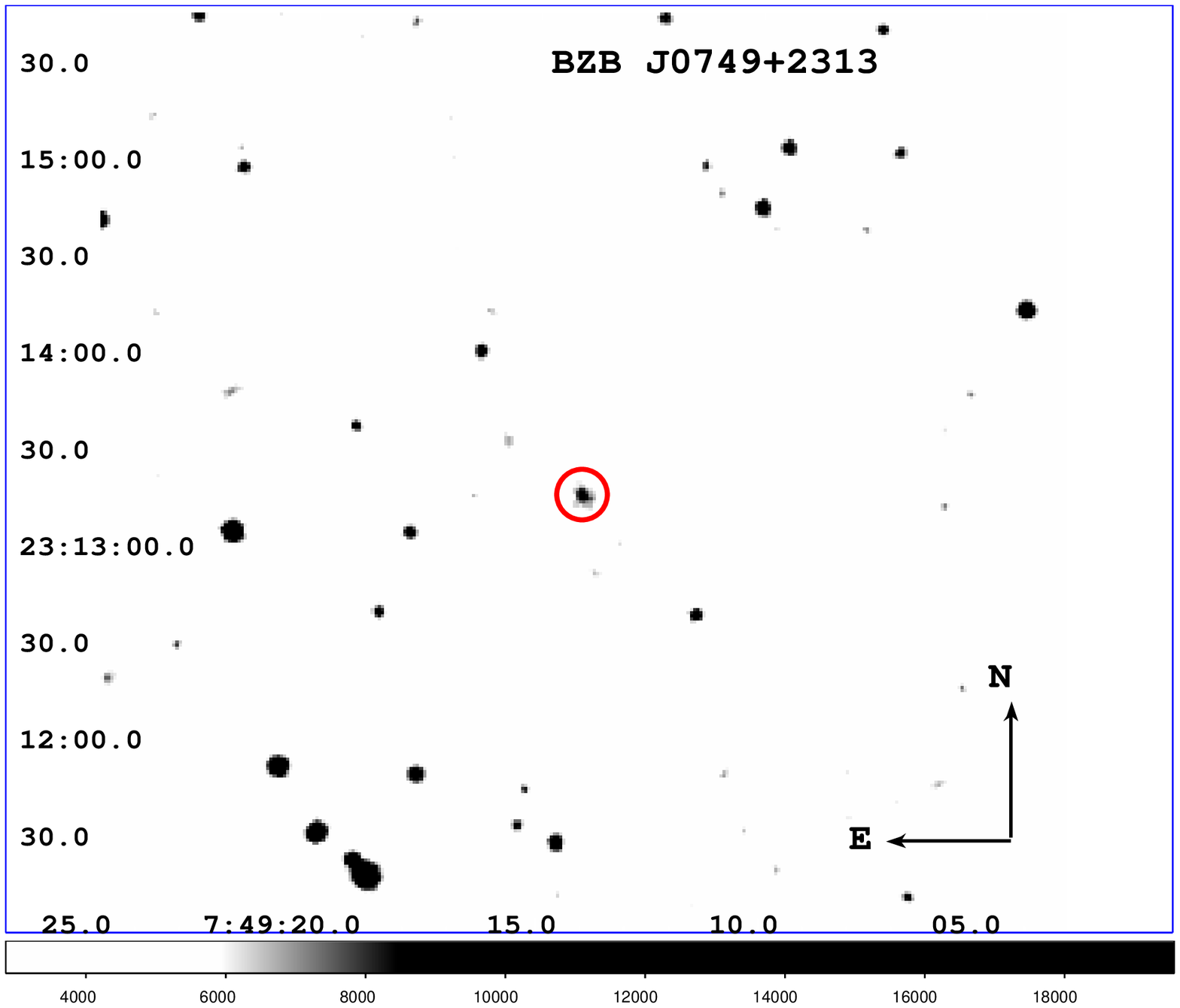}
\end{center}
\caption{Upper panel: optical spectra of the BZB J0749+2313 listed in the \bzcat\ v4.1.
Our observation clearly shows a featureless continuum and allows us to confirm its classification, but
unfortunately, we were unable to confirm the redshift estimate reported in the \bzcat.
The average S/N is also indicated.
Middle panel: normalized spectrum.
Lower panel: 5\arcmin\,x\,5\arcmin\ finding chart from the Digital Sky Survey (red filter). 
The source is indicated by the red circle.}
\label{fig:J0749}
\end{figure}

\begin{figure}[]
\begin{center}
\includegraphics[height=12.2cm,width=12.2cm,angle=0]{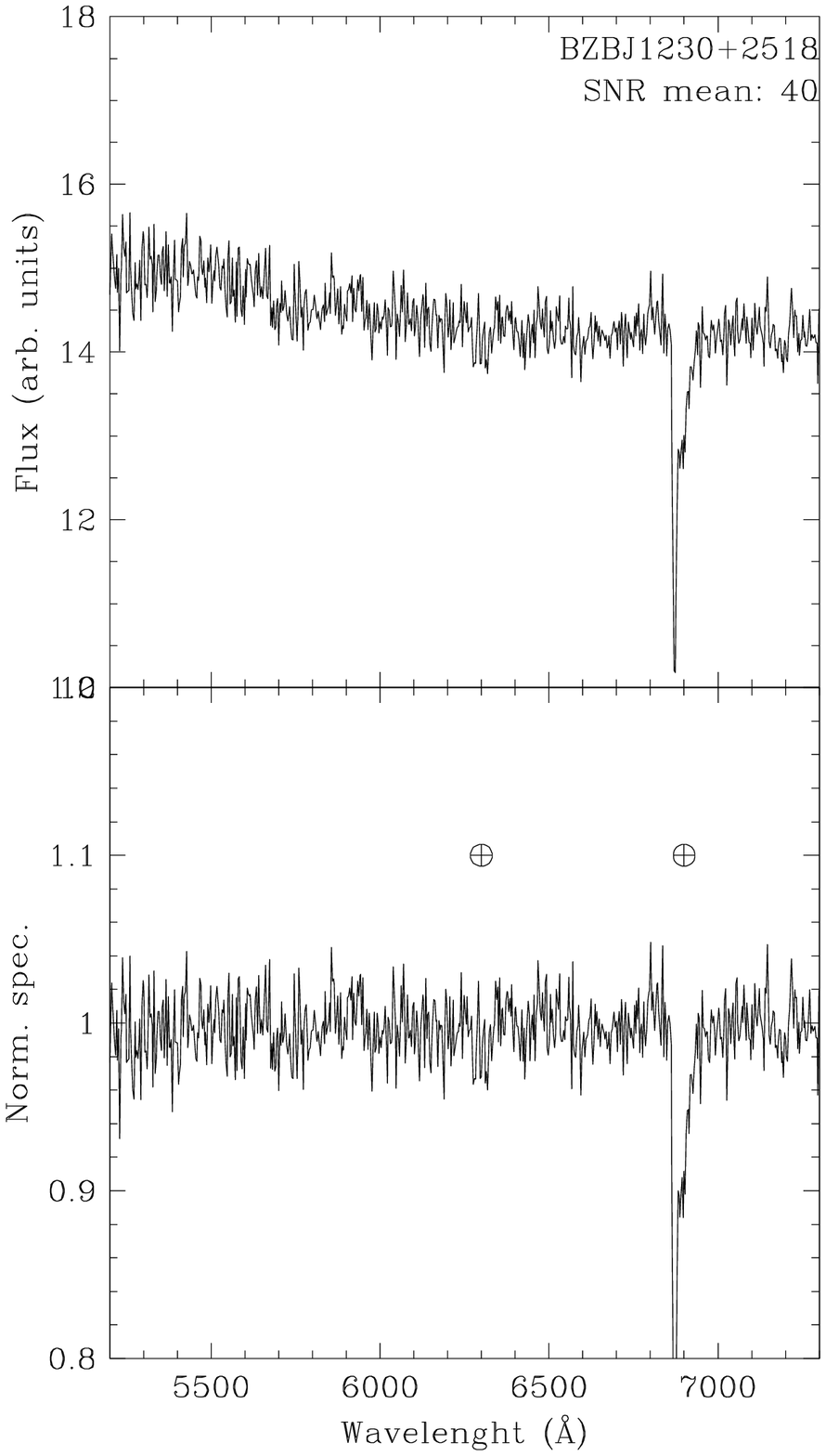}
\includegraphics[height=5.6cm,width=5.6cm,angle=0]{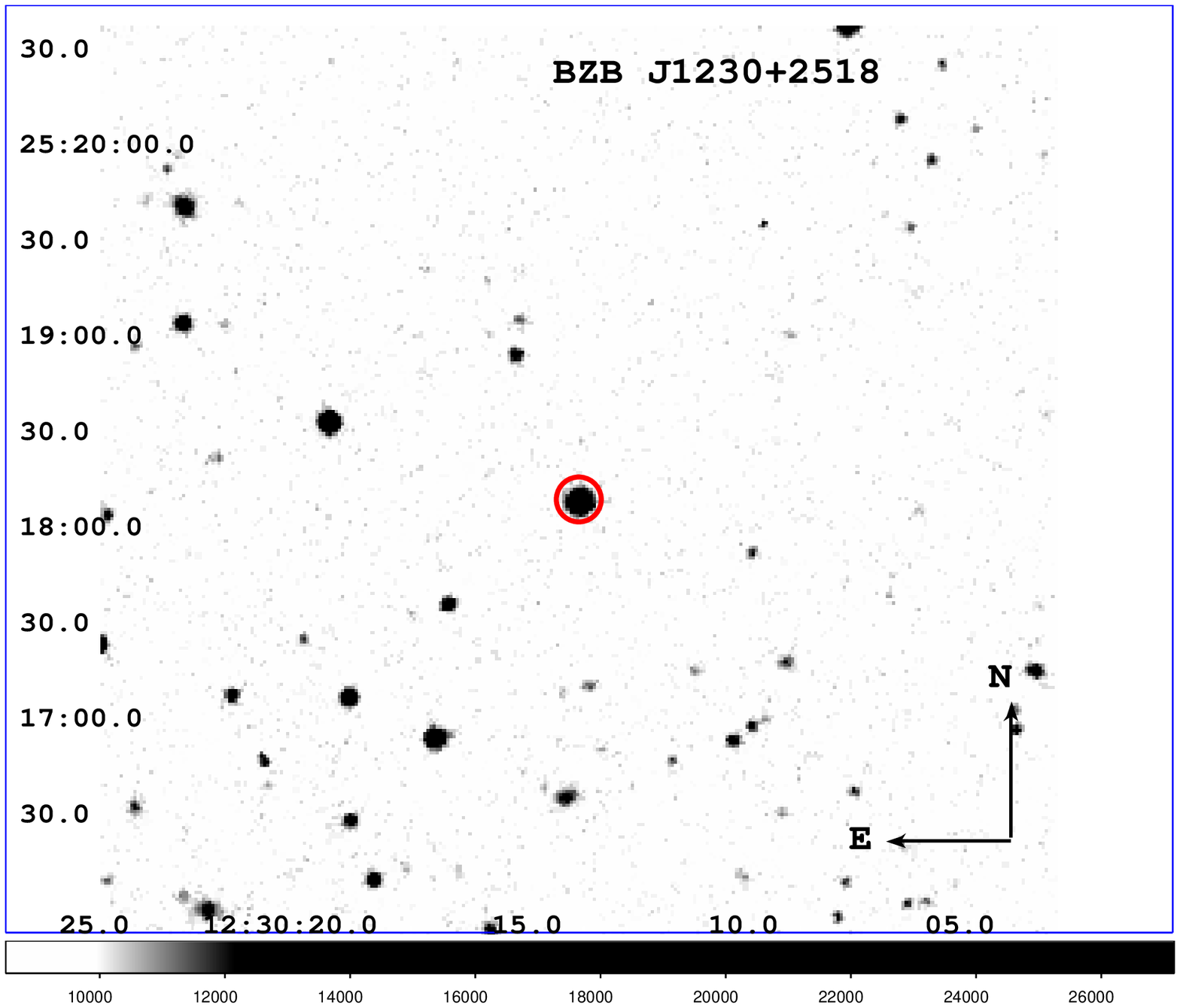}
\end{center}
\caption{Upper panel: optical spectra of the \fer\ source: BZB J1230+2518 listed in the \bzcat\ v4.1.
Our observation clearly shows a featureless continuum and allows us to confirm its classification, but
unfortunately, we were unable to confirm the redshift estimate reported in the \bzcat.
The average S/N is also indicated.
Middle panel: normalized spectrum.
Lower panel: 5\arcmin\,x\,5\arcmin\ finding chart from the Digital Sky Survey (red filter). 
The source is indicated by the red circle.}
\label{fig:J1230}
\end{figure}

\begin{figure}[]
\begin{center}
\includegraphics[height=12.2cm,width=12.2cm,angle=0]{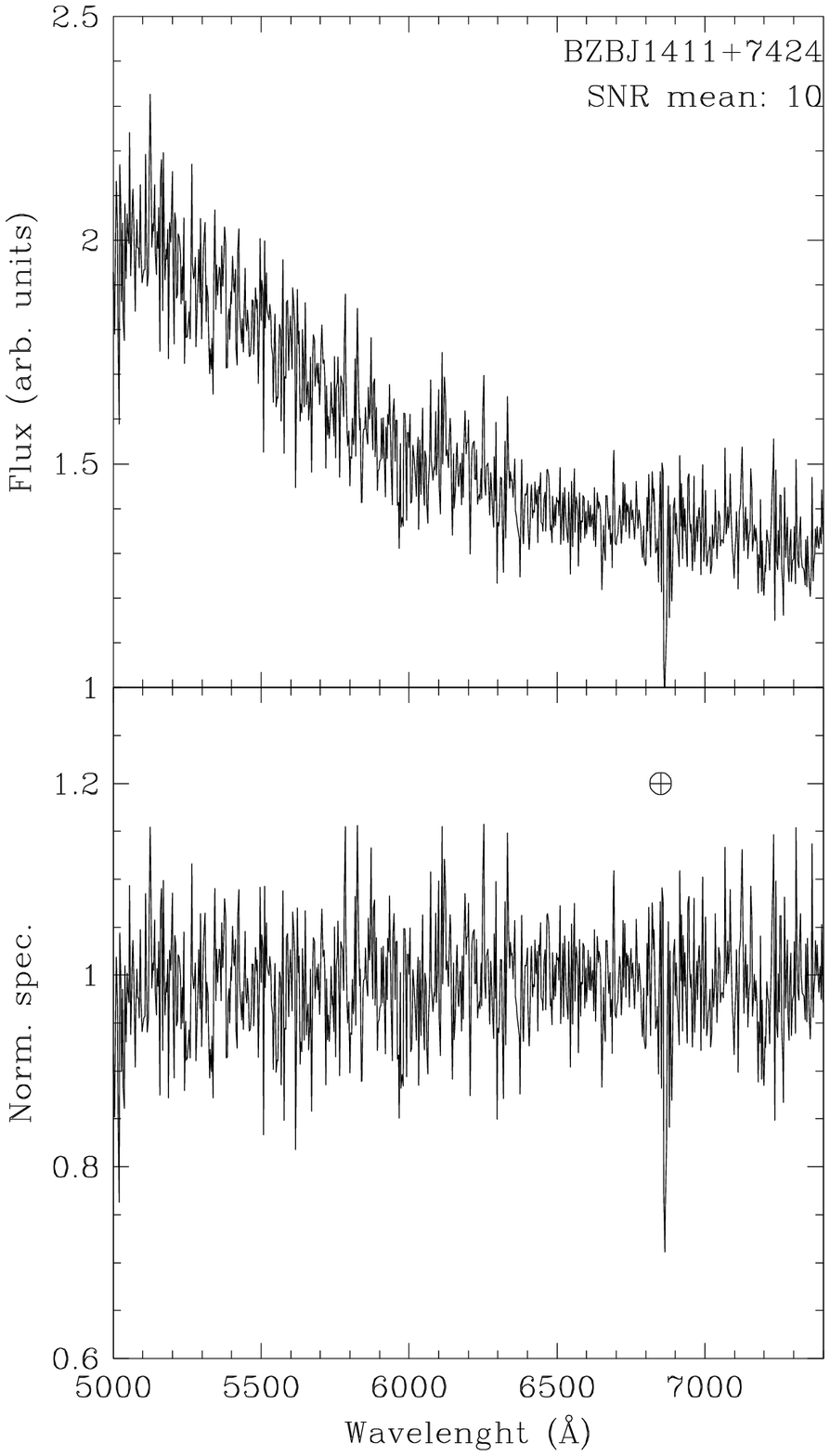}
\includegraphics[height=5.6cm,width=5.6cm,angle=0]{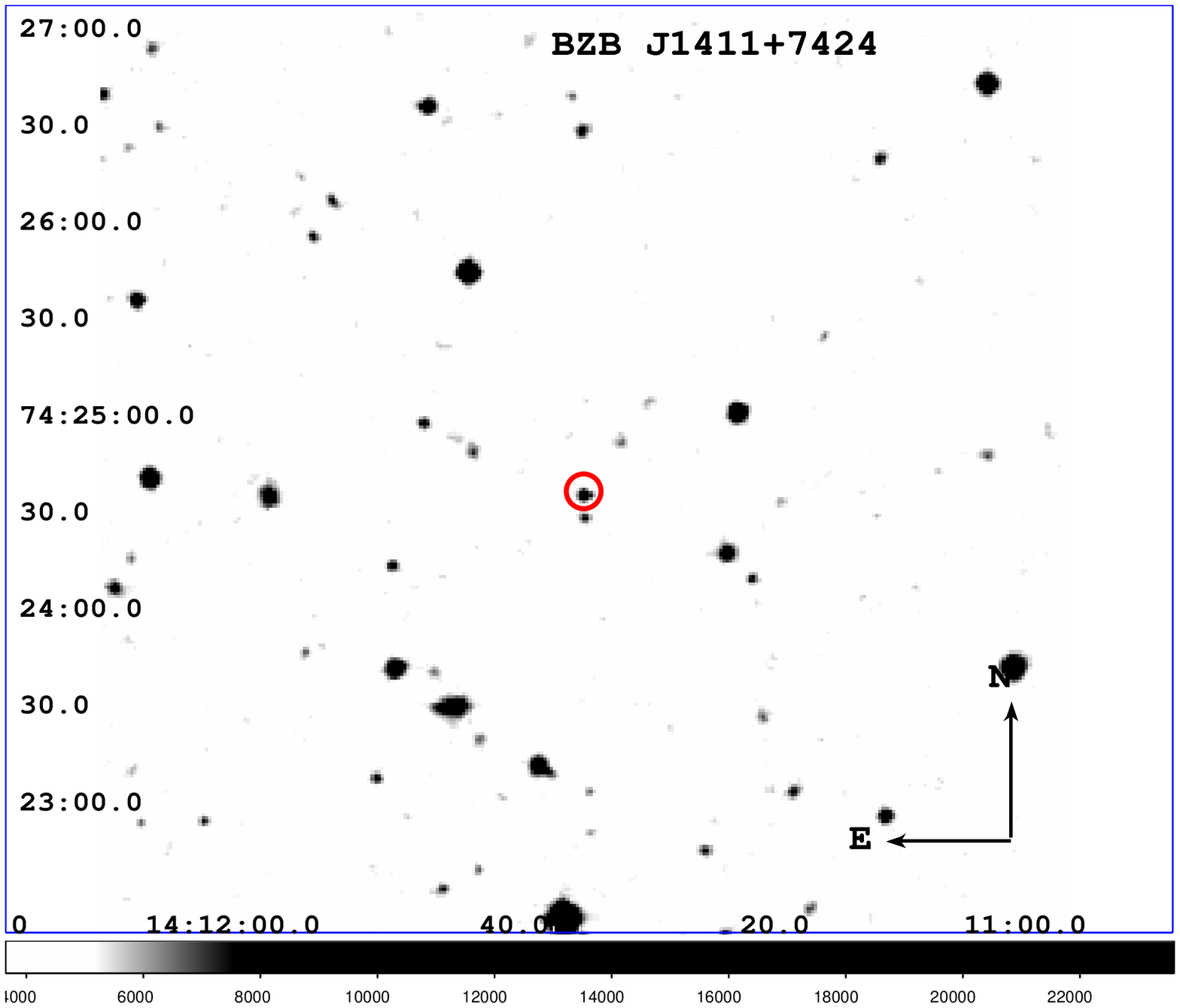}
\end{center}
\caption{Upper panel: optical spectra of the BZB J1411+7424 listed in the \bzcat\ v4.1.
Our observation clearly shows a featureless continuum and allows us to confirm its classification, but
unfortunately, we were unable to confirm the redshift estimate reported in the \bzcat.
The average S/N is also indicated.
Middle panel: normalized spectrum.
Lower panel: 5\arcmin\,x\,5\arcmin\ finding chart from the Digital Sky Survey (red filter). 
The source  is indicated by the red circle.}
\label{fig:J1411}
\end{figure}

\begin{figure}[]
\begin{center}
\includegraphics[height=12.2cm,width=12.2cm,angle=0]{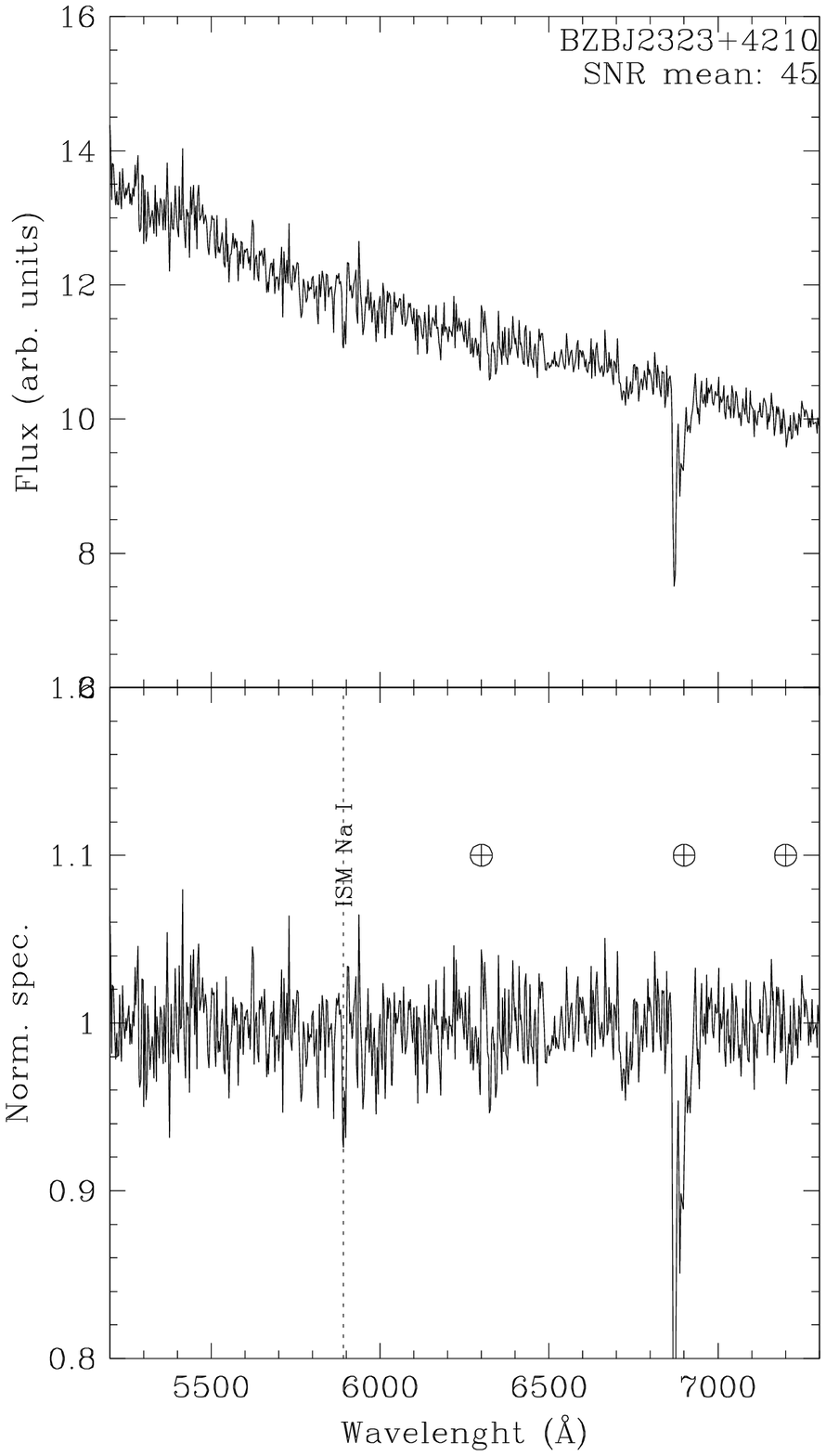}
\includegraphics[height=5.6cm,width=5.6cm,angle=0]{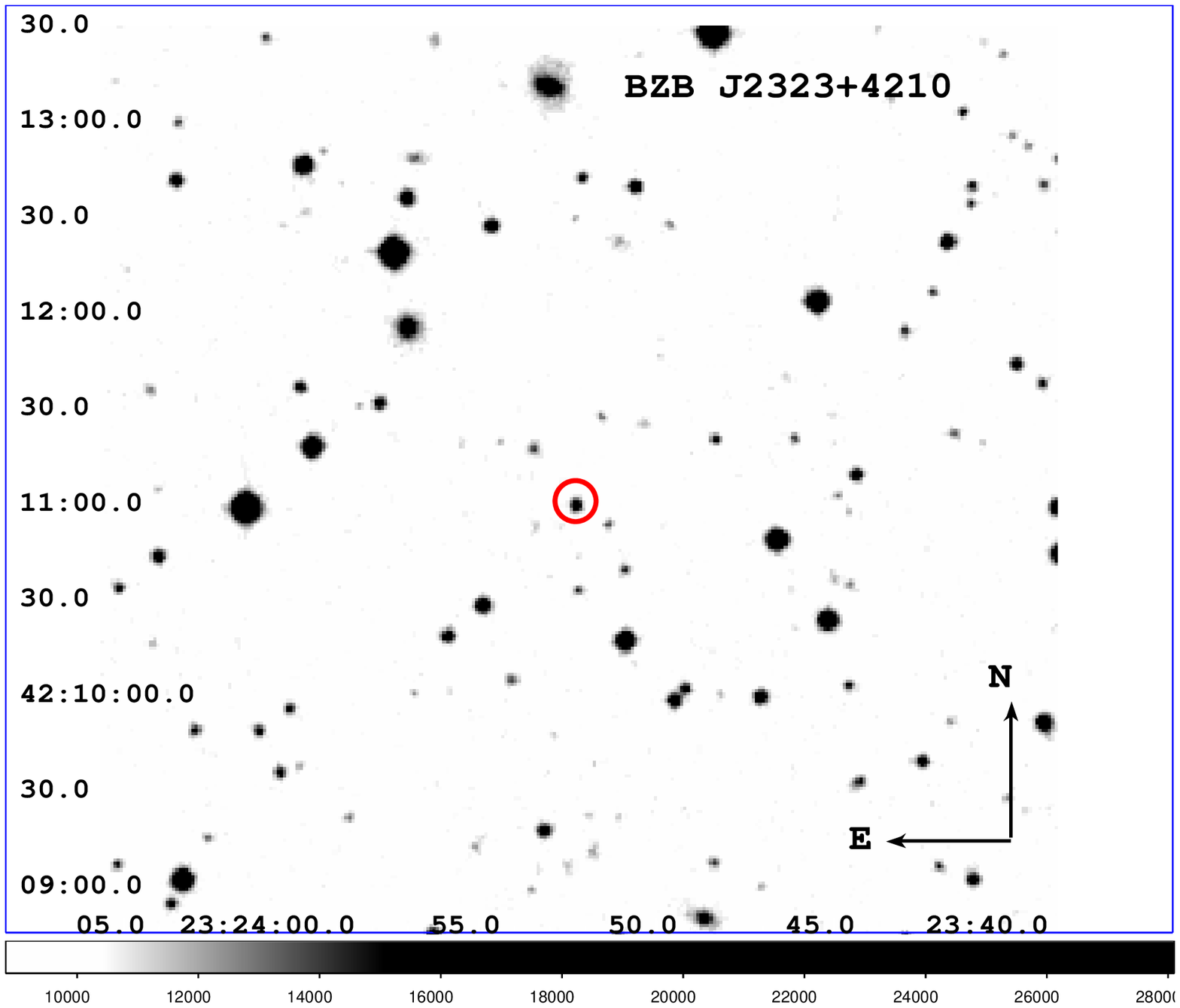}
\end{center}
\caption{Upper panel: optical spectra of the \fer\ source: BZB J2323+4210 listed in the \bzcat\ v4.1.
Our observation clearly shows a featureless continuum and allows us to confirm its classification, but
unfortunately, we were unable to confirm the redshift estimate reported in the \bzcat.
The average S/N is also indicated.
Middle panel: normalized spectrum.
Lower panel: 5\arcmin\,x\,5\arcmin\ finding chart from the Digital Sky Survey (red filter). 
The source is indicated by the red circle.}
\label{fig:J2323}
\end{figure}

\begin{figure}[]
\begin{center}
\includegraphics[height=12.2cm,width=12.2cm,angle=0]{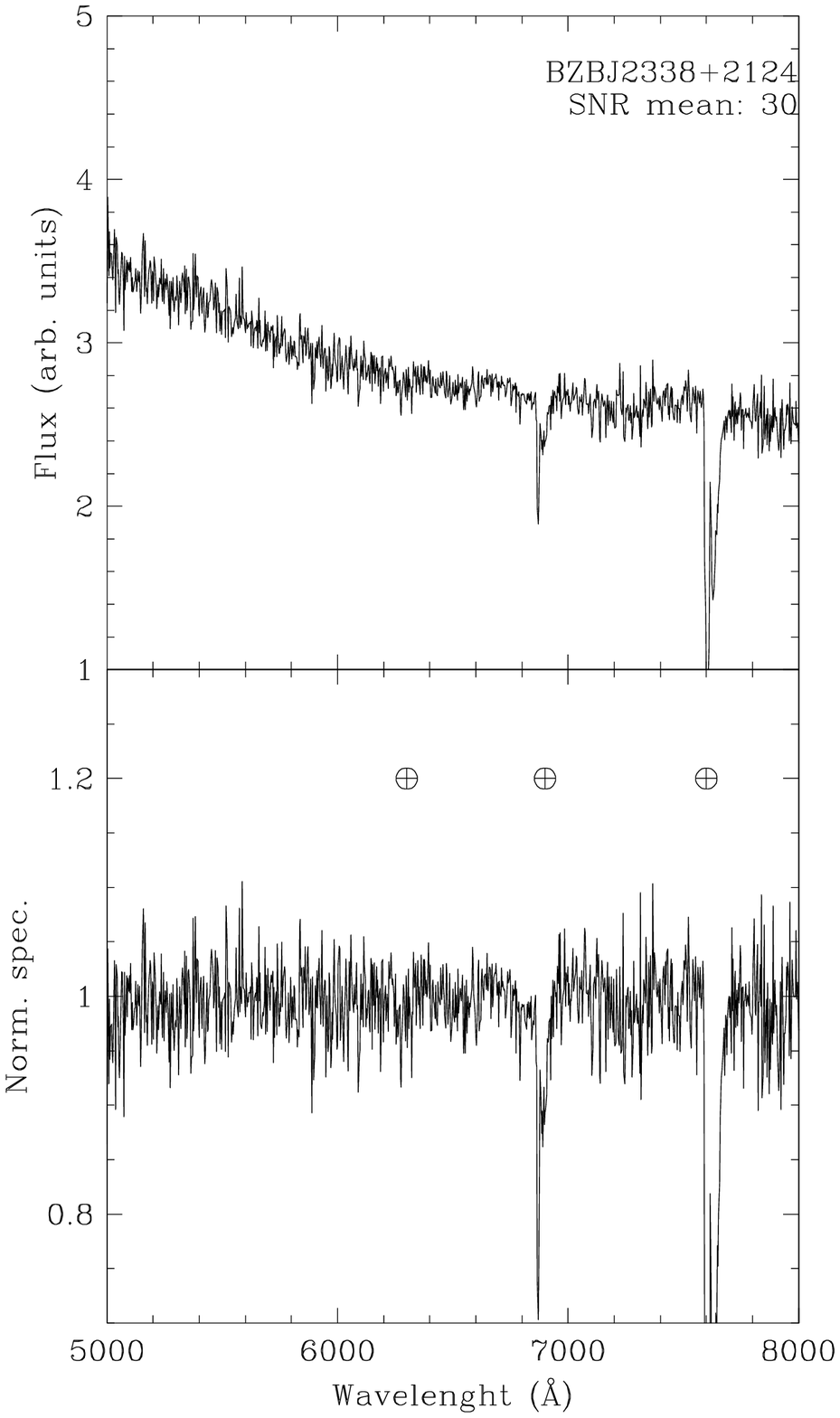}
\includegraphics[height=5.6cm,width=5.6cm,angle=0]{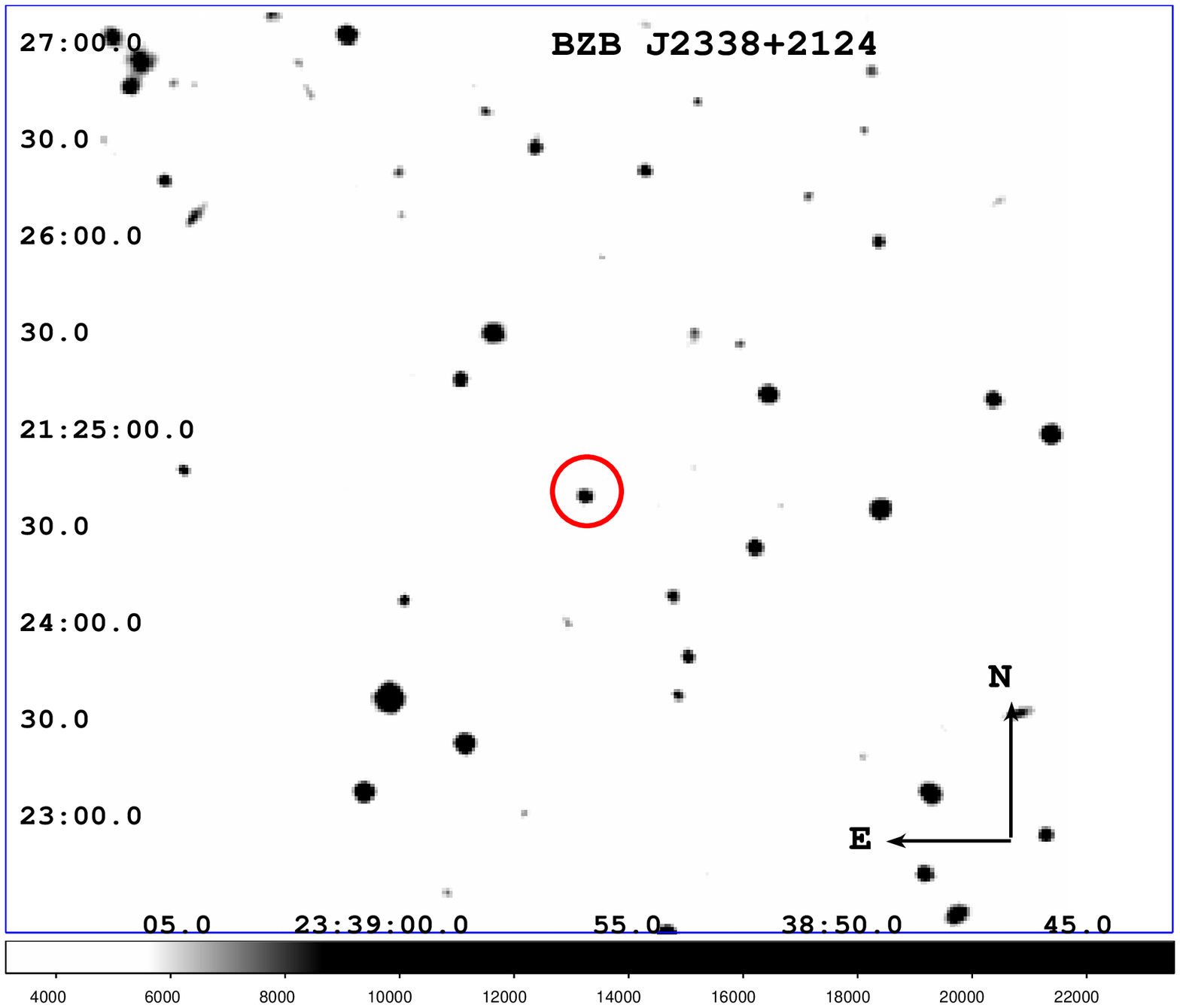}
\end{center}
\caption{Upper panel: optical spectra of the \fer\ source: BZB J2338+2124 listed in the \bzcat\ v4.1.
Our observation clearly shows a featureless continuum and allows us to confirm its classification, but
unfortunately, we were unable to confirm the redshift estimate reported in the \bzcat.
The average S/N is also indicated.
Middle panel: normalized spectrum.
Lower panel: 5\arcmin\,x\,5\arcmin\ finding chart from the Digital Sky Survey (red filter). 
The source  is indicated by the red circle.}
\label{fig:J2338}
\end{figure}

The finding charts were taken from the Digital Sky Survey\footnote{\underline{http://archive.eso.org/dss/dss}}
for all the sources investigated and they are shown below each spectrum (see Figures~\ref{fig:J0409.8} - \ref{fig:J2338} in the following).

\section{Observations and data reduction}
\label{sec:obs}
We obtained the spectra in both visiting and remote observing mode at the KPNO Mayall 4m class telescope using the R-C spectrograph. 
We adopted the recommended setup for the instrument according to our scientific goal, which is the classification of the observed targets. 
Thus we considered a slit width of 1\arcsec.2 and low-resolution gratings (KPC10A and BL181 depending on the availability at the telescope),
yielding a dispersion of 3 $\textrm{\AA}$ pixel$^{-1}$ in both cases.
We observed the selected targets during the nights  of 25 and 26 September 2013 and on 26 and 27 December 2013 during gray time.
The average seeing during both runs was about 1\arcsec\ and conditions were clear.

Data reduction was assessed using IRAF standard procedures. 
For each object we performed bias subtraction, flat-field, and cosmic rays rejection. 
Since for each target we secured two individual exposures to remove cosmic rays, we also averaged them according to their signal-to-noise ratios (S/N). 
We then exploited the availability of the two individual exposures for ambiguous detected 
spectral features to better reject spurious ones.

The wavelength calibration was achieved using the spectra of a helium-neon-argon (HeNeAr) lamp, which guarantees 
a smooth coverage of the entire range. To take into account flexures of the instruments and drift due to poor long-term stability during each night,
we took an arc frame before any target to guarantee a good wavelength solution for the scientific spectra. 
The accuracy reached is $\sim$3$\textrm{\AA}$ rms.

We corrected for the Galactic absorption assuming the $E_{B - V}$ values computed by Schlegel et al. (1998) and the relation
reported by Cardelli et al. (1989).
Although our program did not require precise photometric
precision, we observed a spectrophotometric standard star to
perform relative flux calibration on each spectrum.
To detect faint spectral features, especially because our targets might be BL Lac objects, aimed at hunting redshifts, 
we normalized each spectrum to its continuum. 
\begin{table*}
\label{tab:log}
\caption{Description of the selected sample. Our sources are divided into four subsamples including
1) potential counterparts of UGSs, 2) sources are classified as active galaxies of uncertain type according to the 
2LAC catalog, 3) BZBs classified in Ackermann et al. (2011) that had no available spectrum, (4)
BL Lac candidates, both detected and not detected by \fer,\ 
for which no optical spectroscopic information was found in the literature \citep{massaro11} or BZBs with uncertain/unknown redshift estimate.}
\resizebox{\textwidth}{!}{
\begin{tabular}{|lllllllll|}
\noalign{\smallskip}
\hline
\noalign{\smallskip}
Name & \fer\ & R.A.   & Dec.    & Obs.\,Date   & Exp. & multifrequency notes$^*$ & z & class\\
     & name & (J2000) & (J2000) & (yyyy-mm-dd) & (sec) & & & \\ 
\hline
\hline
Unidentified Gamma-ray Sources & & & & & & & & \\
\hline
\hline
   WISE J040946.58-040003.5 & 2FGL J0409.8-0357 & 04:09:46.59 & -04:00:03.6 & 2013-09-25 & 1200 & N,w,U,g,u,x - SED in Takeuchi+13 & ? & BL Lac \\
   WISE J090038.69+674223.3 & 2FGL J0900.9+6736 & 09:00:38.74 & +67:42:23.4 & 2013-12-28 & 3000 & V,T,W,N,87,GB,rf,w,g,u,U & ? & BL Lac \\
   WISE J101544.43+555100.6 & 2FGL J1016.1+5600 & 10:15:44.42 & +55:51:00.5 & 2013-12-28 & 3600 & V,T,W,N,F,87,GB,rf,w,M,s,g & 0.677 & QSO \\
   WISE J122358.06+795328.2 & 2FGL J1223.3+7954 & 12:23:58.22 & +79:53:29.0 & 2013-12-27 & 2400 & L,N,w,M,g,U & ? & BL Lac \\
   WISE J200505.97+700439.5 & 2FGL J2004.6+7004 & 20:05:06.30 & +70:04:40.7 & 2013-09-26 & 3000 & N,w,M,U & ? & BL Lac \\
   WISE J210805.46+365526.5 & 2FGL J2107.8+3652 & 21:08:09.76 & +36:56:17.9 & 2013-09-25 & 2700 & W,T,N,87,rf,w & ? & BL Lac \\
   WISE J211020.19+381659.2 & 2FGL J2110.3+3822 & 21:10:20.21 & +38:16:58.8 & 2013-09-25 & 1800 & W,N,w,M & 0.46 & QSO \\
\hline
\hline
\fer\ Active Galaxies of Uncertian type & & & & & & & & \\
\hline
\hline
   WISE J030943.23-074427.5 & 2FGL J0309.3-0743 & 03:09:43.24 & -07:44:27.5 & 2013-09-25 & 1200 & N,A,c,rfw,M,s,g & ? & BL Lac \\
   WISE J060915.06-024754.5 & 2FGL J0609.4-0248 & 06:09:15.02 & -02:47:54.2 & 2013-09-24 & 1200 & N,w,M,U,X & ? & BL Lac \\ 
   WISE J070858.28+224135.4 & 2FGL J0709.0+2236 & 07:08:58.29 & +22:41:35.5 & 2013-12-28 & 2400 & N,87,GB,rf,w,M,g,X & ? & BL Lac \\
   WISE J081240.84+650911.1 & 2FGL J0812.6+6511 & 08:12:40.84 & +65:09:10.9 & 2013-12-27 & 2400 & L,N,87,GB,rf,w,M,g,X & ? & BL Lac \\
   WISE J184450.96+570938.6 & 2FGL J1844.7+5716 & 18:44:51.19 & +57:09:40.6 & 2013-09-26 & 1800 & T,L,N,87,c,rf,w,M & ? & BL Lac \\ 
   WISE J224753.22+441315.5 & 2FGL J2247.8+4412 & 22:47:53.21 & +44:13:15.3 & 2013-09-26 & 1200 & L,N,rf,w,M,X & ? & BL Lac \\ 
   WISE J232445.32+080206.1 & 2FGL J2324.6+0801 & 23:24:45.31 & +08:02:05.9 & 2013-09-25 & 1800 & Pm,N,87,rf,w,M & ? & BL Lac \\
   WISE J232538.11+164642.7 & 2FGL J2325.4+1650 & 23:25:38.12 & +16:46:42.7 & 2013-09-25 & 1800 & N,w,M & ? & BL Lac \\
\hline
\hline
\fer\ BL Lacs with no optical spectra & & & & & & & & \\
\hline
\hline
   WISE J212743.03+361305.7 & 2FGL J2127.8+3614 & 21:27:43.03 & +36:13:05.8 & 2013-12-28 & 2400 & V,T,W,N,87,c,rf,w,M & ? & BL Lac \\ 
   WISE J231101.29+020505.3 & 2FGL J2310.9+0204 & 23:11:01.31 & +02:05:04.2 & 2013-09-25 & 1200 & N,w,M,U,g & ? & BL Lac \\ 
   WISE J235205.84+174913.7 & 2FGL J2352.0+1753 & 23:52:05.88 & +17:49:14.3 & 2013-12-27 & 2400 & N,87,GB,rf,w,M,U,g,u,x - SED in Takeuchi+13 & ? & BL Lac \\ 
   WISE J235612.70+403646.8 & 2FGL J2356.1+4034 & 23:56:12.68 & +40:36:48.5 & 2013-12-28 &  600 & L,N,w,M,U & 0.131 & BL Lac/galaxy \\ 
\hline
\hline
BZB candidates in the \bzcat\ & & & & & & & & \\
\hline
\hline
   BZB J0103+4322       &                   & 01:03:28.80 & +43:22:59.5 & 2013-09-26 & 1200 & & ? & BL Lac \\
   BZB J0607+4739       & 2FGL J0607.4+4739 & 06:07:23.25 & +47:39:46.9 & 2013-12-27 & 1800 & & ? & BL Lac \\ 
   BZB J0612+4122       & 2FGL J0612.8+4122 & 06:12:51.18 & +41:22:37.4 & 2013-12-27 & 1800 & & ? & BL Lac \\ 
   BZB J0814+6431       & 2FGL J0814.7+6429 & 08:14:39.19 & +64:31:22.0 & 2013-09-25 & 1200 & & ? & BL Lac \\ 
   BZB J0848+6606       & 2FGL J0849.2+6606 & 08:48:54.60 & +66:06:09.3 & 2013-12-28 & 2400 & & ? & BL Lac \\ 
   BZB J1123+7230       &                   & 11:23:49.20 & +72:30:00.0 & 2013-12-27 & 3000 & & ? & BL Lac \\
   BZB J1143+7304       &                   & 11:43:04.73 & +73:04:09.3 & 2013-12-27 & 1200 & & 0.123 & BL Lac/galaxy\\ 
   BZB J1330+7001       & 2FGL J1330.9+7001 & 13:30:25.81 & +70:01:38.7 & 2013-12-27 & 2400 & & ? & BL Lac \\ 
   BZB J1435+5815       &                   & 14:35:45.96 & +58:15:24.7 & 2013-12-28 & 2400 & & 0.299 & BL Lac/galaxy\\ 
   BZB J1836+3136       & 2FGL J1836.2+3137 & 18:36:21.24 & +31:36:26.8 & 2013-09-25 & 1800 & & ? & BL Lac \\ 
   BZB J1903+5540       & 2FGL J1903.3+5539 & 19:03:11.61 & +55:40:38.4 & 2013-09-26 &  600 & & ? & BL Lac \\ 
   BZB J2251+4030       & 2FGL J2251.9+4032 & 22:51:59.77 & +40:30:58.2 & 2013-09-25 &  600 & & ? & BL Lac \\ 
   BZB J2255+2410       & 2FGL J2255.2+2408 & 22:55:15.34 & +24:10:12.4 & 2013-09-25 & 1800 & & ? & BL Lac \\
   BZB J2320+4146       &                   & 23:20:12.20 & +41:46:05.3 & 2013-12-28 &  600 & & 0.152 & BL Lac \\ 
\hline
\hline
BZBs listed in the \bzcat\ with uncertain $z$ & & & & & & & & \\
\hline
\hline
   BZB J0650+2503       & 2FGL J0650.7+2505 & 06:50:46.52 & +25:03:00.3 & 2013-12-28 & 1800 & & ? & BL Lac \\ 
   BZB J0749+2313       &                   & 07:49:14.00 & +23:13:16.7 & 2013-12-28 & 1200 & & ? & BL Lac \\ 
   BZB J1230+2518       & 2FGL J1230.2+2517 & 12:30:14.09 & +25:18:07.1 & 2013-12-28 & 1800 & & ? & BL Lac \\ 
   BZB J1411+7424       &                   & 14:11:34.75 & +74:24:29.2 & 2013-12-28 & 1800 & & ? & BL Lac \\ 
   BZB J2323+4210       & 2FGL J2323.8+4212 & 23:23:52.10 & +42:10:58.7 & 2013-12-28 & 1200 & & ? & BL Lac \\ 
   BZB J2338+2124       & 2FGL J2339.0+2125 & 23:38:56.38 & +21:24:41.4 & 2013-12-27 & 1800 & & ? & BL Lac \\ 
\hline
\hline
\noalign{\smallskip}
\end{tabular}}
($^*$) Symbols used for the multifrequency notes are all reported in Sect.~\ref{sec:crossmatches} 
together with the references of the catalogs/surveys.
\label{tab:log}
\end{table*}

\section{Results}
\label{sec:results}

\subsection{Unidentified gamma-ray sources}
Five out of seven candidate counterparts for seven UGSs listed in the 2FGL catalog and observed in our sample are classified
as BL Lac objects.
Despite the good quality of the spectra obtained at the KPNO Mayall 4m telescope,  we were unfortunately unable to establish the redshift for any of them.
The remaining two sources are classified as quasars (QSOs): the one potentially associated with 2FGL J1016.1+5600
lies at $z=$0.677, and given its flat radio spectrum, clearly appears to be a BZQ, while the lack of radio information for
the counterpart of 2FGL J2110.3+3822, which lies at $z=$0.46, does not yet guarantee its BZQ classification.
All the spectra are shown in Figs.~\ref{fig:J0409.8} - \ref{fig:J2110.3}, together with their finding charts.

\subsection{Gamma-ray active galaxies of uncertain type}
In the AGU sample we confirmed the BL Lac nature of eight sources observed with our spectroscopic campaign.
In particular, we reobserved WISE J060915.06-024754.5 associated with 2FGL J0609.4-0248 
to obtain a redshift estimate for this source,
but we were only able to confirm the previous results by Shaw et al. (2013a).
The same is true for the  WISE J184450.96+570938.6 counterpart of 2FGL J1844.7+5716 and WISE J224753.22+441315.5
associated with 2FGL J2247.8+4412, for which we did not detect any spectral features to derive a lower limit of the redshifts.
No redshifts were determined for this AGU sample.
All the spectra are shown in Figs.~\ref{fig:J0309.3} - \ref{fig:J2325.4} together with their finding charts.

In addition to these eight AGUs, we also found four sources listed in the 2LAC as BZBs that do not belong to the \bzcat\ 
(version 4.1, the same as used to search for the associations of the 2FGL sources) and for which no data and/or spectra
are available in the \fer\ catalog \citep{abdo10,nolan12}.
For 2FGL J2127.8+3614, 2FGL J2310.9+0204, and 2FGL J2352.0+1753, the optical spectra were also published by Shaw et al. (2013a),
but we are not able to confirm their lower limit on the redshift estimates because our observations show only a featureless continuum.
Our campaign confirms that three out of four sources are BZBs, as previously mentioned. 
The only difference with respect to the 2LAC catalog concerns the remaining source 2FGL J2356.1+4034, which according to our analysis appears as an elliptical galaxy at $z=$0.131 in the optical band, 
not dominated by a nonthermal continuum in the optical band,
as might be expected for a BZB at $z=$0.331 as reported in Ackermann et al. (2011).
The spectra for these additional four sources are shown in 
Figs.~\ref{fig:J2127.8} - \ref{fig:J2356.1} together with their finding charts.

\subsection{\bzcat\ sources with uncertain nature or unknown redshifts}
Details for the BZBs that belong to the \bzcat\ and were observed in our northern campaign are listed below.
Table~\ref{tab:log} reports their \bzcat\ name, that of the \fer\ counterpart, when associated with a $\gamma$-ray source in the 2FGL, together with the coordinates.
There are no multifrequency notes this table since they are already discussed in the \bzcat\ \citep{massaro09}.

There are 14 sources listed in \bzcat\ as BL Lac candidates, 9 of which are also associated with \fer\ sources, as reported in Table~\ref{tab:log}, 
with no optical spectra present in literature that permit a firm classification.
Our spectroscopic follow-up observations confirm the BL Lac nature for all of them, but no redshifts were estimated 
except for
BZB J2320+4146, for which we obtain a $z$ value in agreement with that published in the literature and reported in the \bzcat,
BZB J1143+7304 lying at $z=$0.123, and BZB J1435+5815 
that instead lies at $z=$0.299. However, these two sources appear to have an optical spectrum 
dominated by the emission of their host elliptical galaxies instead
of by nonthermal continuum arising from their jet.
To establish whether the optical emission is dominated by the host galaxy or by the jet emission, we adopted the same criterion as defined by Massaro et al. (2012d) based on the calcium break (Ca, H, and K) contrast adapted to the SDSS optical colors. We considered the optical spectrum dominated by nonthermal emission if the flux emitted at frequencies above the Ca, H, and K was
higher than the one radiated below. This threshold is more conservative than those proposed by March\~a et al. (1996) and Landt, Padovani \& Giommi (2002), and it was also recently adopted in our analyses of blazar candidates \citep[e.g.,][]{sdss}.
All these spectra are shown in Figs.~\ref{fig:J0103} - \ref{fig:J2320} together with their finding charts.

Finally, We also observed an additional sample of six \bzcat\ sources, all classified as BZBs,
for which the redshift estimate reported in the literature is uncertain.
Unfortunately, we did not obtain $z$ estimates for these known BL Lacs.
These additional six spectra, together with the finding charts, are shown in Figs.~\ref{fig:J0650} - \ref{fig:J2338}.

\section{Summary and conclusions}
\label{sec:conclusions}
We presented the results of our 2013 optical spectroscopic campaign carried out in the Northern Hemisphere with
the Kitt Peak National Observatory (KPNO) Mayall 4m class telescope.
The main aim of our program is to use optical spectroscopy to confirm the nature of sources selected 
for their IR colors or low radio frequency spectra (i.e., below $\sim$1GHz) similar to the known \fer-detected blazars that lie within the positional uncertainty regions of the unidentified gamma-ray sources (UGSs).
Identifying blazars among these objects will improve and refine future associations for the \fer\ catalogs.
Our spectroscopic campaign will also allow us to search for redshift estimates of the potential UGS counterparts.

During our campaign we also observed several active galaxies of uncertain type, as defined according to the \fer\ 
catalogs \citep[see, e.g.,][]{ackermann11a,nolan12}, to verify whether they are blazars.
In addition, we observed several sources that already belong to the \bzcat\ but were classified
as BL Lac candidates as a result of the lack of optical spectra available in literature, 
or are BZBs with uncertain redshift estimates.

Thirty-nine sources are observed in the Northern Hemisphere.
The results of this campaign, which is complementary to that carried out in the Southern Hemisphere \citep{landoni14},
can be summarized as follows:

\begin{enumerate}
\item In the sample of the potential counterpart for the UGSs, selected with IR colors \citep{paper1,ugs1,ugs2} 
and on the basis of the flat radio spectrum below $\sim$1GHz \citep{ugs3,ugs6,lorcat}, we confirmed the blazar-like nature of six 
out of seven sources. Five are clearly BZBs with a classical featureless optical spectrum, while the remaining two are QSOs.
In particular, for WISE J101544.43+555100.6, potentially associated with 2FGL J1016.1+5600, the radio data available above $\sim$1.4GHz
allowed us to classify this source as a flat-spectrum radio quasar at $z=$0.677, while we cannot claim the same for the
WISE J211020.19+381659.2 candidate counterpart for 2FGL J2110.3+3822 because we lack additional radio observations.
\item In the AGU sample observed during our campaign we found that all eight sources are classified as BZBs.
\item We also obtained the spectra of four BZBs, classified in the 2LAC catalog, but with no spectra published at the time of the observations.
All of them are indeed BZBs, two also previously confirmed by Shaw et al. (2013a), with the exception of the 2FGL J2356.1+4034 counterpart. 
According to our analysis, 2FGL J2356.1+4034 seems to be associated to an ``uncertain'' BL Lac that in the optical 
appears to be dominated by its host galaxy emission (see Sect.~\ref{sec:results} for more details).
\item Within the \bzcat\ sources we found 14 BL Lac candidates that were
all confirmed as genuine BZBs via optical spectroscopy. For three
of these we confirmed a z estimate, and for two the optical spectra
were dominated by the host galaxy more than the expected nonthermal
continuum. For the remaining six BZBs listed in the \bzcat\ with uncertain redshift estimates, we were unable to 
obtain any $z$ value with our observations. 
\end{enumerate}

We thank the referee Y. Tanaka for useful comments that led to improvements in the paper.
We are grateful to D. Hammer for her help to schedule, prepare, and perform the KPNO observations.
This investigation is supported by the NASA grants NNX12AO97G and NNX13AP20G.
H. A. Smith acknowledges partial support from NASA/JPL grant RSA 1369566.
The work by G. Tosti is supported by the ASI/INAF contract I/005/12/0.
Part of this work is based on archival data, software or on-line services provided by the ASI Science Data Center.
This research has made use of data obtained from the high-energy Astrophysics Science Archive
Research Center (HEASARC) provided by NASA's Goddard Space Flight Center; 
the SIMBAD database operated at CDS,
Strasbourg, France; the NASA/IPAC Extragalactic Database
(NED) operated by the Jet Propulsion Laboratory, California
Institute of Technology, under contract with the National Aeronautics and Space Administration.
Part of this work is based on the NVSS (NRAO VLA Sky Survey):
The National Radio Astronomy Observatory is operated by Associated Universities,
Inc., under contract with the National Science Foundation and on the VLA low-frequency Sky Survey (VLSS).
The Molonglo Observatory site manager, Duncan Campbell-Wilson, and the staff, Jeff Webb, Michael White and John Barry, 
are responsible for the smooth operation of Molonglo Observatory Synthesis Telescope (MOST) and the day-to-day observing programme of SUMSS. 
The SUMSS survey is dedicated to Michael Large whose expertise and vision made the project possible. 
The MOST is operated by the School of Physics with the support of the Australian Research Council and the Science Foundation for Physics within the University of Sydney.
This publication makes use of data products from the Wide-field Infrared Survey Explorer, 
which is a joint project of the University of California, Los Angeles, and 
the Jet Propulsion Laboratory/California Institute of Technology, 
funded by the National Aeronautics and Space Administration.
This publication makes use of data products from the Two Micron All Sky Survey, which is a joint project of the University of 
Massachusetts and the Infrared Processing and Analysis Center/California Institute of Technology, funded by the National Aeronautics 
and Space Administration and the National Science Foundation.
This research has made use of the USNOFS Image and Catalogue Archive
operated by the United States Naval Observatory, Flagstaff Station
(http://www.nofs.navy.mil/data/fchpix/).
Funding for the SDSS and SDSS-II has been provided by the Alfred P. Sloan Foundation, 
the Participating Institutions, the National Science Foundation, the U.S. Department of Energy, 
the National Aeronautics and Space Administration, the Japanese Monbukagakusho, 
the Max Planck Society, and the Higher Education Funding Council for England. 
The SDSS Web Site is http://www.sdss.org/.
The SDSS is managed by the Astrophysical Research Consortium for the Participating Institutions. 
The Participating Institutions are the American Museum of Natural History, 
Astrophysical Institute Potsdam, University of Basel, University of Cambridge, 
Case Western Reserve University, University of Chicago, Drexel University, 
Fermilab, the Institute for Advanced Study, the Japan Participation Group, 
Johns Hopkins University, the Joint Institute for Nuclear Astrophysics, 
the Kavli Institute for Particle Astrophysics and Cosmology, the Korean Scientist Group, 
the Chinese Academy of Sciences (LAMOST), Los Alamos National Laboratory, 
the Max-Planck-Institute for Astronomy (MPIA), the Max-Planck-Institute for Astrophysics (MPA), 
New Mexico State University, Ohio State University, University of Pittsburgh, 
University of Portsmouth, Princeton University, the United States Naval Observatory, 
and the University of Washington.
The WENSS project was a collaboration between the Netherlands Foundation 
for Research in Astronomy and the Leiden Observatory. 
We acknowledge the WENSS team consisted of Ger de Bruyn, Yuan Tang, 
Roeland Rengelink, George Miley, Huub Rottgering, Malcolm Bremer, 
Martin Bremer, Wim Brouw, Ernst Raimond and David Fullagar 
for the extensive work aimed at producing the WENSS catalog.
TOPCAT\footnote{\underline{http://www.star.bris.ac.uk/$\sim$mbt/topcat/}} 
\citep{taylor05} for the preparation and manipulation of the tabular data and the images.
The Aladin Java applet\footnote{\underline{http://aladin.u-strasbg.fr/aladin.gml}}
was used to create the finding charts reported in this paper \citep{bonnarell00}. 
It can be started from the CDS (Strasbourg - France), from the CFA (Harvard - USA), from the ADAC (Tokyo - Japan), 
from the IUCAA (Pune - India), from the UKADC (Cambridge - UK), or from the CADC (Victoria - Canada).
\clearpage

\clearpage

\end{document}